\documentclass[
amsmath,amssymb,
prd,nofootinbib,floatfix,12pt,
]{revtex4}

\usepackage{amssymb,amsmath,amsbsy}
\usepackage[dvips]{color}
\usepackage{xcolor}
\usepackage{dsfont}
\usepackage{mathrsfs}
\usepackage{float}

\usepackage{tikz}
\usetikzlibrary{tikzmark}

\definecolor{Red}{cmyk}{0,1,1,0}
\definecolor{BrickRed}{cmyk}{0,0.89,0.94,0.28}
\definecolor{Blue}{cmyk}{1,1,0,0}
\definecolor{Green}{cmyk}{1,0,1,0 }

\newcommand\beq{\begin{eqnarray}}
\newcommand\eeq{\end{eqnarray}}
\newcommand{\RomanNumeralCaps}[1]{\MakeUppercase{\romannumeral #1}}

\def\lsim{\mathrel{\rlap{\lower4pt\hbox{$\sim$}}
    \raise1pt\hbox{$<$}}}                
\def\gsim{\mathrel{\rlap{\lower4pt\hbox{$\sim$}}
    \raise1pt\hbox{$>$}}}            

\allowdisplaybreaks
\usepackage{graphicx}
\usepackage{graphics}
\usepackage{rotating} 
\usepackage{pdflscape} 
\usepackage{longtable}
\usepackage{setspace}

\newcommand{\nocontentsline}[3]{}
\newcommand{\tocless}[2]{\bgroup\let\addcontentsline=\nocontentsline#1{#2}\egroup}

\begin{document}
\renewcommand{\theequation}{\arabic{section}.\arabic{equation}}
\renewcommand{\thefigure}{\arabic{section}.\arabic{figure}}
\renewcommand{\thetable}{\arabic{section}.\arabic{table}}

\title{\large \baselineskip=20pt 
High-quality axions in solutions to the $\mu$ problem}

\author{Prudhvi N.~Bhattiprolu and Stephen P.~Martin}
\affiliation{\it
Department of Physics, Northern Illinois University, DeKalb IL 60115}

\begin{abstract}\normalsize \baselineskip=15.5pt
Solutions to the $\mu$ problem in supersymmetry based on the Kim-Nilles mechanism naturally feature a Dine-Fischler-Srednicki-Zhitnitsky (DFSZ) axion with decay constant of order the geometric mean of the Planck and TeV scales, consistent with astrophysical limits. We investigate minimal models of this type with two gauge-singlet fields that break a Peccei-Quinn symmetry, and extensions with extra vectorlike quark and lepton supermultiplets consistent with gauge coupling unification. We show that there are many anomaly-free discrete symmetries, depending on the vectorlike matter content, that protect the Peccei-Quinn symmetry to sufficiently high order to solve the strong CP problem. We study the axion couplings in this class of models.  Models of this type that are  automatically free of the domain wall problem require at least one pair of strongly interacting vectorlike multiplets with mass at the intermediate scale, and predict axion couplings that are greatly enhanced compared to the minimal supersymmetric DFSZ models, putting them within reach of proposed axion searches.
\end{abstract}

\maketitle
\vspace{-1.1cm}

\makeatletter
\def\l@subsubsection#1#2{}
\makeatother

\baselineskip=18pt

\tableofcontents

\baselineskip=14.55pt

\setcounter{footnote}{1}
\setcounter{figure}{0}
\setcounter{table}{0}

\newpage

\section{Introduction\label{sec:introduction}}
\setcounter{equation}{0}
\setcounter{figure}{0}
\setcounter{table}{0}
\setcounter{footnote}{1}

Extensions of the Standard Model (SM) of particle physics are plagued by several apparent hierarchy problems, which can be viewed as hints towards the ultimate completion of the theory. The vacuum energy, as manifested in the observed expansion rate of the universe due to the cosmological constant, is approximately 120 orders of magnitude smaller than its naive
dimensional analysis estimate $M_P^4$, where $M_P = 2.4 \times 10^{18}$ GeV is the reduced Planck mass scale. The ``big hierarchy problem" is that the electroweak scale set by the Higgs field squared mass parameter is 32 orders of magnitude smaller than $M_P^2$. The strong CP problem is that the CP-odd angle $\theta$ in the QCD Lagrangian could be of order 1, but is constrained to be smaller than $9 \times 10^{-11}$ by the measured value \cite{Abel:2020gbr} of the electric dipole moment of the neutron.

In this paper, we will be concerned with a class of models that simultaneously address the latter two problems. The big hierarchy problem of the Higgs field is addressed by supersymmetry, with supersymmetry breaking terms characterized by the TeV scale,\footnote{In this paper, we will loosely refer to the mass scales associated with superpartner masses and supersymmetry breaking as the TeV scale. Given the negative search results so far at the Large Hadron Collider (LHC), the masses of the strongly interacting superpartners are evidently somewhat larger than 1 TeV. The fact that their masses exceed the Higgs vacuum expectation value by an order of magnitude is the ``little hierarchy problem", which we do not address in this paper; it is obviously much less severe than the big hierarchy problem, for which the LHC results have not favored any competing hypotheses.}
while the strong CP problem is addressed by including a global Peccei-Quinn (PQ) $U(1)$ symmetry \cite{Peccei:1977hh,Peccei:1977ur} that is explicitly broken by a QCD anomaly but also spontaneously broken, minimizing the effective value of $|\theta|$ and giving rise to a light, very weakly coupled axion \cite{Weinberg:1977ma}-\cite{Dine:1982ah}. For reviews of the axion solution to the strong CP problem from various points of view, see refs.~\cite{Turner:1989vc}-\cite{Semertzidis:2021rxs}.

The ``$\mu$ problem" of supersymmetry relates the issues of supersymmetry breaking and the PQ symmetry breaking. The superpotential of the Minimal Supersymmetric Standard Model (MSSM, for a review see \cite{Martin:1997ns}) contains terms of the form
\beq
W &=& \mu H_u H_d + y_u H_u q \overline u  - y_d H_d q \overline d - y_e H_d \ell \overline e,
\label{eq:MSSMsuperpotential}
\eeq
where $H_u$ and $H_d$ are the Higgs doublet chiral superfields. The $q, \overline u, \overline d, \ell$, and $\overline e$ are quark and lepton chiral superfields, for which we suppress flavor and gauge indices and use lowercase letters to distinguish them from additional vectorlike quark and lepton superfields to be considered below. Besides the Yukawa coupling matrices $y_u$, $y_d$, and $y_e$, this superpotential contains a single dimensionful parameter $\mu$, which preserves supersymmetry but nevertheless should be roughly of the same order as the supersymmetry breaking mass scale in order to allow for electroweak symmetry breaking. To accomplish this, the Kim-Nilles mechanism \cite{Kim:1983dt} takes the $\mu$ term to be absent from eq.~(\ref{eq:MSSMsuperpotential}) in the ultraviolet theory, and in its place introduces  
non-renormalizable terms, for example of the form proposed in ref.~\cite{Murayama:1992dj}:
\beq
W_{\rm \RomanNumeralCaps{1}} &=& 
\frac{\lambda_\mu}{M_{P}} X Y H_u H_d + 
\frac{\lambda}{6 M_{P}} X^3 Y
\label{eq:MSYsuperpotential}
\eeq
where $X,Y$ are gauge-singlet chiral superfields, and $\lambda_\mu$ and $\lambda$ are dimensionless couplings. The term $\lambda$ in eq.~(\ref{eq:MSYsuperpotential}) is included to stabilize the potential at large $X,Y$ scalar field strengths. Including the effects of supersymmetry breaking, there are also contributions to the Lagrangian (in terms of scalar fields, for which we use the same symbol as the corresponding superfield):
\beq
{\cal L}_{\text{soft}} &=& 
\left (\frac{a_{\mu}}{M_P} X Y H_u H_d + \frac{a}{6 M_{P}} X^3 Y \right ) + {\text{c.c.}}
-m_X^2 |X|^2 - m_Y^2 |Y|^2 ,
\eeq
where $a_{\mu}, a$ are of order $m_{\text{soft}}$ and $m_X^2, m_Y^2$ are of order $m_{\text{soft}}^2$, with $m_{\text{soft}}$ roughly at the TeV scale. If $m_X^2$ and $m_Y^2$ are negative, or just sufficiently small compared to $|a/\lambda|^2$, then the resulting scalar potential has a local minimum for vacuum expectation values (VEVs) parametrically of order\footnote{Note that it is important to include the contribution of the allowed holomorphic supersymmetry-breaking coupling $a$ here. Omitting it would seemingly lead to 
$\langle Y \rangle \ll \langle X \rangle \sim M_{\text{int}}$.}
\beq
\langle X \rangle \sim \langle Y \rangle \sim \sqrt{m_{\text{soft}} M_{P}}
\equiv M_{\text{int}} ,
\label{eq:defineMint}
\eeq
a scale intermediate between the Planck and TeV scales.
This can always occur if $m_X^2$ and $m_Y^2$ are negative, but this is not a necessary condition.
For example, if $m_X^2 = m_Y^2 = m^2$, then there is a non-trivial local minimum if $|a/\lambda|^2 - 12 m^2 > 0$, and it will be a global minimum if $|a/\lambda|^2 - 16 m^2 > 0$. Symmetry breaking along such flat directions is stable against radiative corrections \cite{Martin:1999hc}. In the low energy theory, the $\mu$ and $b$ terms of the MSSM Lagrangian are
\beq
\mu &=& \frac{\lambda_\mu}{M_P} \langle X Y \rangle \sim m_{\text{soft}},
\\
b &=& \frac{a_\mu}{M_P} \langle X Y \rangle \sim m_{\text{soft}}^2,
\eeq
solving the $\mu$ problem. 

If $m_{\rm soft}$ is of order the TeV scale, then 
\beq 
10^9\>{\text{GeV}} \,\lsim\, M_{\text{int}} \,\lsim\, 10^{12}\>\text{GeV},
\label{eq:Mintrange}
\eeq 
depending on the dimensionless parameters involved. 
Equation~(\ref{eq:Mintrange}) roughly coincides with the preferred range for the decay constant of the axion from astrophysical constraints on the low end and from cosmological dark matter density on the upper end. It is therefore notable that $X$ and $Y$ carry non-zero charges for a PQ symmetry, which the VEVs spontaneously break, so that the axion could be a linear combination of the pseudo-scalar components of the fields $X$ and $Y$ (with a very small admixture of $H_u$ and
$H_d$). Since the MSSM quarks also carry non-zero PQ charge, this is an example of a
Dine-Fischler-Srednicki-Zhitnitsky (DFSZ) \cite{Dine:1981rt,Zhitnitsky:1980tq} axion model. In this way, the solution of the $\mu$ problem in supersymmetry can also be the solution of the strong CP problem.

Besides the axion, the other components of the supermultiplets 
$X$ and $Y$ all get TeV scale masses. These include
a scalar saxion, and gauge-singlet axino fermions, 
one of which could be the lightest supersymmetric particle.
The lightest MSSM superpartner could decay to the axino with a macroscopic proper decay length, which can be much larger than the size of a collider detector. However some of the decays can occur within the detector, providing a striking search signal \cite{Martin:2000eq} for the Large Hadron Collider.

In addition to eq.~(\ref{eq:MSYsuperpotential}), there are three other similar but distinct superpotential structures involving two PQ-breaking superfields $X$ and $Y$. 
(For simplicity and economy, we restrict our attention to only two such fields. Although it is
possible to have more than two, this would seem to make it harder to find solutions to the axion quality problem to be discussed shortly.) They are:
\beq
W_{\rm \RomanNumeralCaps{2}}&=&\frac{\lambda_\mu}{2M_{P}} X^2 H_u H_d + 
\frac{\lambda}{6 M_{P}} X^3 Y
,
\label{eq:CCKsuperpotential}
\\
W_{\rm \RomanNumeralCaps{3}}&=&\frac{\lambda_\mu}{2M_{P}} Y^2 H_u H_d + 
\frac{\lambda}{6 M_{P}} X^3 Y
,
\label{eq:SPMasuperpotential}
\\
W_{\rm \RomanNumeralCaps{4}}&=&\frac{\lambda_\mu}{2M_{P}} X^2 H_u H_d + 
\frac{\lambda}{4 M_{P}} X^2 Y^2
,
\label{eq:SPMbsuperpotential}
\eeq
with corresponding holomorphic soft supersymmetry-breaking parameters $a_\mu$ and $a$ in each case, in the obvious way. The structure in eq.~(\ref{eq:CCKsuperpotential}) was proposed in 
ref.~\cite{Choi:1996vz}, and those in eqs.~(\ref{eq:SPMasuperpotential}) and (\ref{eq:SPMbsuperpotential}) in ref.~\cite{Martin:2000eq}. 
For a review and further elucidation, see \cite{Bae:2019dgg}.
In the following, 
we will refer to the models defined by eqs.~(\ref{eq:MSYsuperpotential}),
(\ref{eq:CCKsuperpotential}), (\ref{eq:SPMasuperpotential}), and
(\ref{eq:SPMbsuperpotential})
as base models $\text{B}_{\rm \RomanNumeralCaps{1}}, \text{B}_{\rm \RomanNumeralCaps{2}}, \text{B}_{\rm \RomanNumeralCaps{3}}$, and $\text{B}_{\rm \RomanNumeralCaps{4}}$, respectively, since we will be interested in extensions of them. Each of them implies a different assignment of PQ charges,
which are summarized in Table \ref{tab:base models}.
We choose the normalization of the PQ charges so that $H_u H_d$ has charge $-2$.
\begin{table}
\begin{minipage}[]{0.95\linewidth}
\caption{The four base models. 
Each model has two gauge-singlet chiral superfields  $X,Y$ whose
scalar component VEVs spontaneously break the Peccei-Quinn symmetry and generate the $\mu$ term through the schematic superpotential terms shown. The PQ charges are normalized so that $H_u H_d$ has charge $-2$.\label{tab:base models}}
\end{minipage}
\begin{center}
\begin{tabular}{|c | c c|}
\hline
~Base model~ & ~Superpotential terms~  &  ~PQ charges of $(X, Y)$~\\
\hline
\hline
&&
\\[-13pt]
~$\text{B}_{\rm \RomanNumeralCaps{1}}$~ 
& 
~$X Y H_u H_d + X^3 Y$~ 
&  
~$(-1, 3)$~\\[1pt]
~$\text{B}_{\rm \RomanNumeralCaps{2}}$~ 
& 
~$X^2 H_u H_d + X^3 Y$~ 
&  
~$(1, -3)$~\\[1pt]
~$\text{B}_{\rm \RomanNumeralCaps{3}}$~ 
& 
~$Y^2 H_u H_d + X^3 Y$~  
&  
~$(-1/3, 1)$~\\[1pt]
~$\text{B}_{\rm \RomanNumeralCaps{4}}$~ 
& 
~$X^2 H_u H_d + X^2 Y^2$~
&  
~$(1, -1)$~\\[1pt]
\hline
\end{tabular}
\end{center}
\end{table}

\baselineskip=15pt

The axion quality problem \cite{Georgi:1981pu}-\cite{Holman:1992us} results from the possible presence of higher dimensional contributions to the Lagrangian that explicitly violate the $U(1)$ PQ symmetry, since these can displace the QCD $\theta$ parameter away from 0 at the minimum of the scalar potential, spoiling the solution to the strong CP problem. Such contributions are expected to be allowed if they are not forbidden, because ungauged symmetries are not respected by quantum gravitational effects, at least in thought experiments such as black hole evaporation processes. In supersymmetry, we therefore consider superpotential operators of the form
\beq
W &=& \frac{\kappa}{M_P^{p-3}} X^{j} Y^{p-j},
\label{eq:XaYbsuperpotential}
\eeq
with dimensionless $\kappa$.
Together with the $\lambda$ term in the superpotential, this gives rise to scalar potential terms (from $|F_X|^2$ and $|F_Y|^2$) that have $p+2$ powers of $X$ and $Y$ and are suppressed by $1/M_P^{p-2}$. These are also accompanied by holomorphic supersymmetry-breaking terms in the scalar potential
\beq
V &=& \frac{a_{\kappa}}{M_P^{p-3}} X^j Y^{p-j} + {\text{c.c.}},
\label{eq:XaYbsoft}
\eeq
where $a_{\kappa}$ should be of order the TeV scale. For generic phases of $\kappa$ and $a_{\kappa}$, both of these types of contributions result in tadpoles for the axion field $A$ (a linear combination of the imaginary parts of $X$ and $Y$ 
expanded around the CP-conserving vacuum) that are parametrically of the same form
\beq
V &=& -\delta \frac{f_A^{p+1}}{M_P^{p-2}} A ,
\label{eq:deltaApotential}
\eeq
where we have identified the axion decay constant $f_A$ with the intermediate scale VEVs $M_{\text{int}}$, and the dimensionless quantity $\delta$ depends on the parameters (including the magnitudes and phases of $\lambda$, $\kappa$, and $a_{\kappa}$ and the integer $k$) in a complicated way. The axion potential also includes the usual squared mass term, approximately for small $A$,
\beq
V &=& 
\frac{{\cal M}_{\text{QCD}}^4}{2 f_A^2 }A^2
\label{eq:approxA2potential}
\eeq
where
\beq
{\cal M}_{\text{QCD}}^4 
&\approx& 
\frac{m_u m_d m_\pi^2 f_\pi^2}{(m_u + m_d)^2}
\>\approx\> (\mbox{0.0754 GeV})^4.
\eeq
The combination of eqs.~(\ref{eq:deltaApotential}) and (\ref{eq:approxA2potential}) gives rise to
\beq
|\theta_{\text{eff}}| &=& \langle A \rangle/f_A = \delta \frac{f_A^{p+2}}{{\cal M}_{\text{QCD}}^4 M_P^{p-2}} .
\eeq
Now by requiring that the $\theta_{\text{eff}}$ parameter at the minimum of the potential is less than $10^{-10}$ from the experimental bound on the neutron electric dipole moment, one therefore finds \cite{Barr:1992qq,Kamionkowski:1992mf,Holman:1992us}  that one should have
\beq
p+2 &>& \frac{88 + \log_{10}(\delta)}{9.4 - \log_{10}(f_A/10^{9} {\text{GeV}})} .
\eeq
Note that the bound on $p$ becomes weaker for smaller $f_A$. With the naive $\delta \approx 1$, if PQ-violating superpotential terms with $p = (8, 9, 10, 11,$ or $12)$ are present one should have $f_A \lsim (4 \times 10^9,\> 3 \times 10^{10},\> 10^{11},\> 4 \times 10^{11},$ or $10^{12})$ GeV, respectively. However, like all naturalness criteria, this one is inherently fuzzy, as $\delta$ could be significantly less than 1, for example because the corresponding coupling(s) happen to have a small magnitude and/or a phase alignment with the bare $\theta$.  We will therefore not commit to a specific requirement for $p$,  with the understanding that larger $p$ is safer in some sense.

Previous works exploring solutions to the axion quality problem have invoked composite axions models \cite{Kim:1984pt}-\cite{Yin:2020dfn}, additional continuous gauge symmetries 
\cite{Barr:1992qq,Holman:1992us}, \cite{Cheng:2001ys}-\cite{Chen:2021haa}, and 
discrete gauged symmetries in non-supersymmetric \cite{Dias:2002gg}-\cite{Bjorkeroth:2017tsz}
and supersymmetric \cite{Chun:1992bn}-\cite{Nakai:2021nyf} models.
Here, we will be interested in the latter type of idea, in which the $U(1)$ PQ symmetry arises as an approximate accidental consequence of a discrete symmetry imposed on superfields. We will consider models in which discrete symmetries forbid PQ-violating Lagrangian terms up to some mass dimension,  and refer to the smallest exponent allowed in a particular model for the PQ-violating terms of the types in eqs.~(\ref{eq:XaYbsuperpotential}) and (\ref{eq:XaYbsoft}) as the PQ-violation suppression $p$. Note that if the suppression $p$ for a particular model is odd, then it can be bumped up to the next integer by simply imposing another $Z_2$ symmetry under which both $X$ and $Y$ are odd, since this forbids all terms of the forms eq.~(\ref{eq:XaYbsuperpotential}) and (\ref{eq:XaYbsoft}) with odd $p$. We therefore consider as potentially viable any models whose discrete symmetries predict that $p$ should be 7 or more, and consider models with $p\geq 12$ for all PQ-violating terms involving only $X$ and $Y$ as presumptively high-quality. 

Since the Kim-Nilles mechanism provides a $\mu$ term for the $H_u$ and $H_d$ fields at the TeV scale, it is reasonable to suppose that the same mechanism can give masses to other vectorlike pairs of chiral superfields as well, some of which could therefore be at the TeV scale just like the MSSM Higgs and higgsino particles. For each additional pair $\Phi + \overline \Phi$ of chiral superfields, supersymmetric mass terms near the TeV scale can arise in three possible ways, due to non-renormalizable superpotential terms of the forms:

\vspace{-0.9cm}

\beq
W_{\text{mass}} &=& 
\begin{cases}
 \frac{\lambda_\Phi}{M_P} X Y \Phi \overline{\Phi}, \\
 \frac{\lambda_\Phi}{2 M_P} X^2 \Phi \overline{\Phi}, \\
 \frac{\lambda_\Phi}{2 M_P} Y^2 \Phi \overline{\Phi},
\end{cases}
\label{mass_possibilitiesNONREN}
\eeq
\vspace{-0.5cm}

\noindent assuming $\lambda_\Phi$ is of order one. Alternatively, masses at the intermediate scale can be achieved by renormalizable superpotential terms of the forms:
\beq
W_{\text{mass}} &=& 
\begin{cases}
 \lambda_\Phi X \Phi \overline{\Phi}, \\
 \lambda_\Phi Y \Phi \overline{\Phi}.
\end{cases}
\label{mass_possibilitiesREN}
\eeq
The net PQ charges of $\Phi\overline{\Phi}$ are important for understanding the low-energy axion couplings, as discussed in the next section. Table~\ref{tab:PQchargesofPhiPhibar} shows the possible values of PQ charges $Q_{\Phi \overline{\Phi}} \equiv (Q_\Phi + Q_{\overline{\Phi}})$ in the extensions of the four base models, for the various possible superpotential mass terms.
\begin{table}
\begin{minipage}[]{0.95\linewidth}
\caption{Possible values of net PQ charges $Q_{\Phi \overline{\Phi}} \equiv (Q_\Phi + Q_{\overline{\Phi}})$ for vector-like chiral superfield pairs, in the extensions of the four base models, depending on the possible superpotential terms that provide masses due to the scalar components of $X$ and $Y$ obtaining VEVs of order $M_{\text{int}}$. The first three mass terms provide for a TeV scale mass, and the last two provide for an intermediate scale mass.
\label{tab:PQchargesofPhiPhibar}}
\end{minipage}
\begin{center}
\begin{tabular}{|c | r r r r |}
\hline
~Mass terms~ & ~$\text{B}_{\rm \RomanNumeralCaps{1}}$~  &  ~$\text{B}_{\rm \RomanNumeralCaps{2}}$~  &  ~$\text{B}_{\rm \RomanNumeralCaps{3}}$~  &  ~$\text{B}_{\rm \RomanNumeralCaps{4}}$~\\
\hline
\hline
~$X Y \Phi \overline{\Phi}$~ & ~$-2$~  &  ~$2$~  &  ~$-2/3$~  &  ~$0$~\\
~$X^2 \Phi \overline{\Phi}$~ & ~$2$~  &  ~$-2$~  &  ~$2/3$~  &  ~$-2$~\\
~$Y^2 \Phi \overline{\Phi}$~ & ~$-6$~  &  ~$6$~  &  ~$-2$~  &  ~$2$~\\
\hline
\hline
~$X \Phi \overline{\Phi}$~ & ~$1$~  &  ~$-1$~  &  ~$1/3$~  &  ~$-1$~\\
~$Y \Phi \overline{\Phi}$~ & ~$-3$~  &  ~$3$~  &  ~$-1$~  &  ~$1$~\\
\hline
\end{tabular}
\end{center}
\end{table}
 
\begin{figure}[!tb]
 \begin{center}  
  \begin{minipage}[]{0.495\linewidth}
    \includegraphics[width=8.0cm,angle=0]{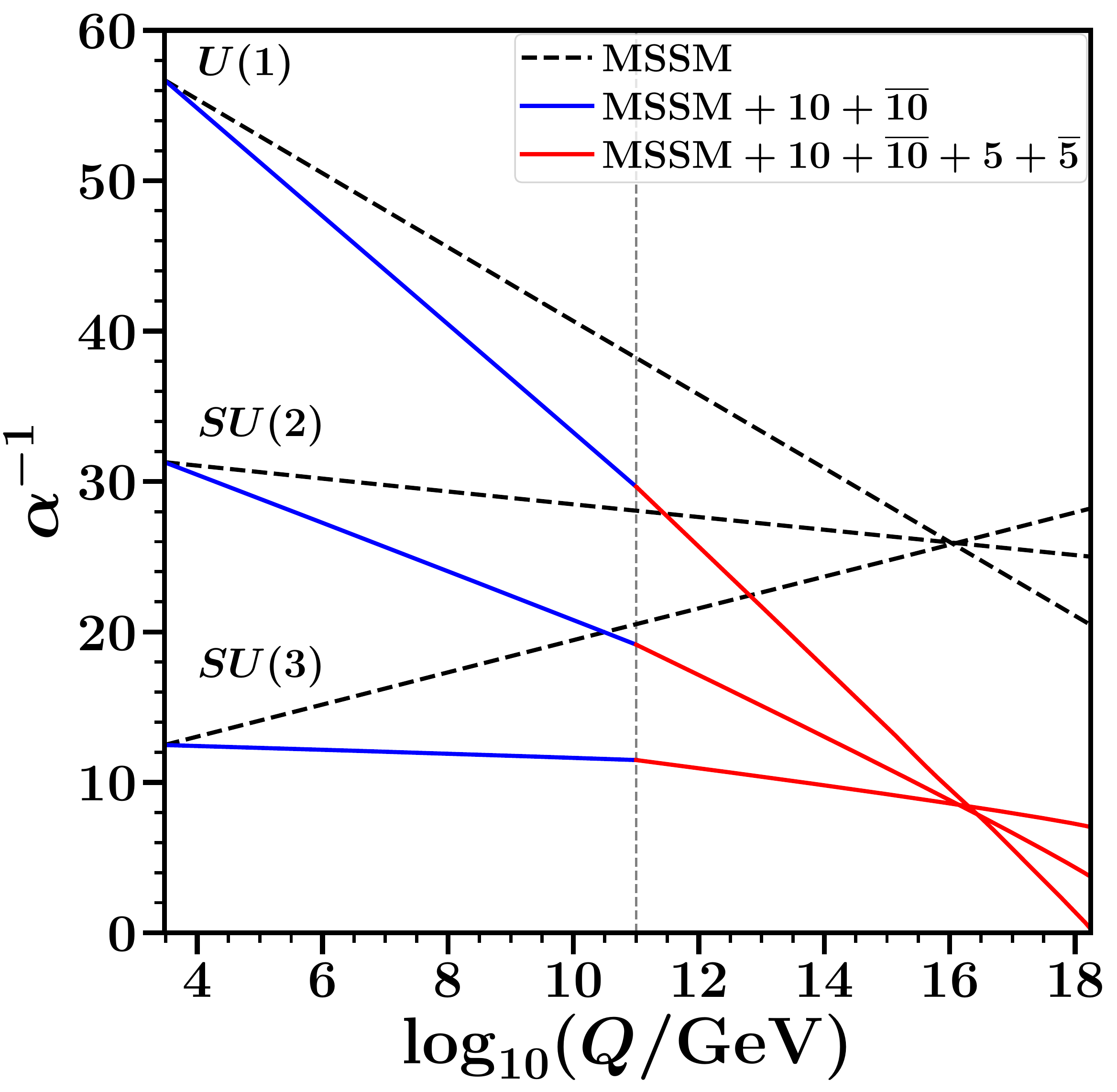}
  \end{minipage}
  \begin{minipage}[]{0.495\linewidth}
    \includegraphics[width=8.0cm,angle=0]{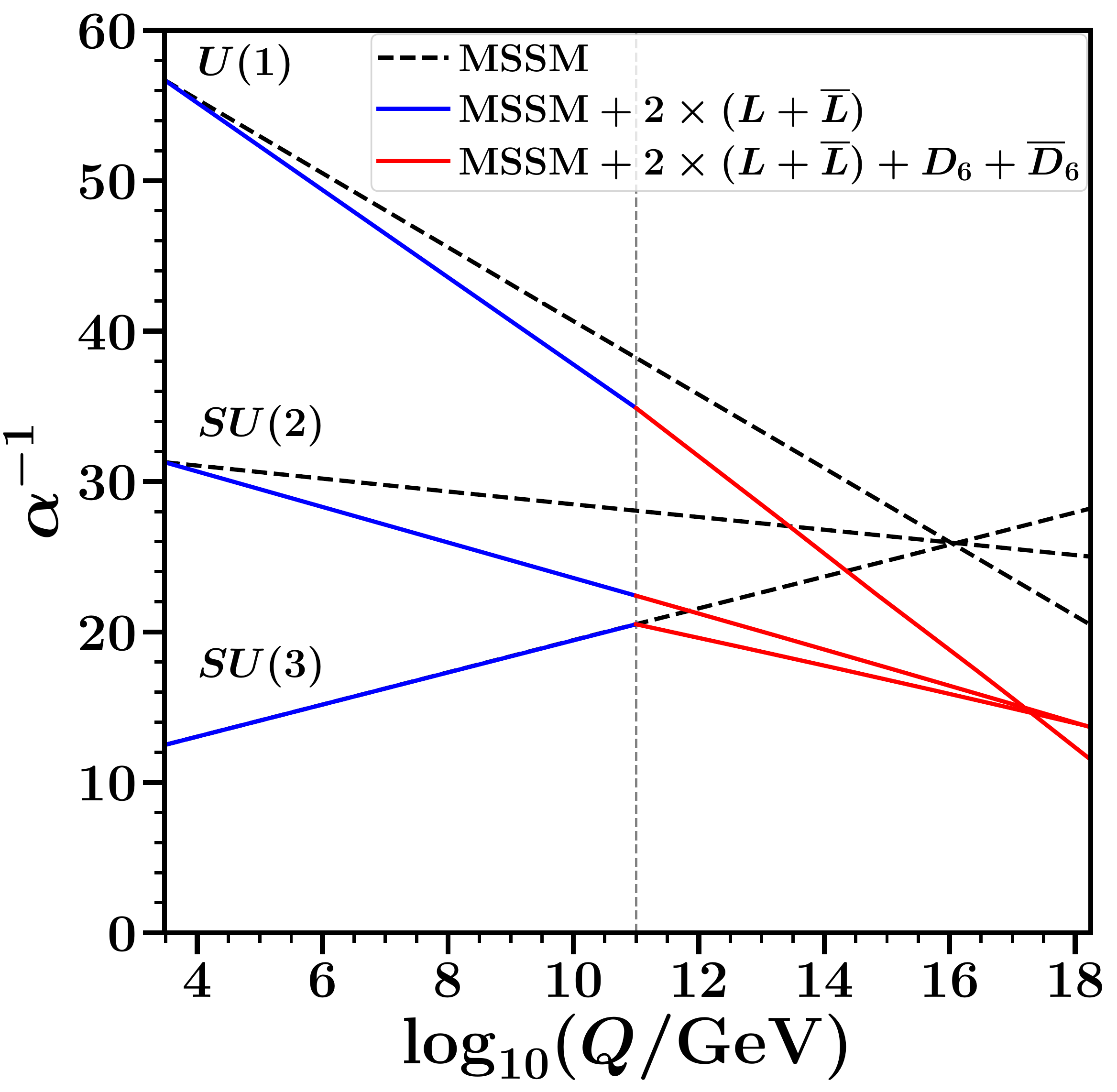} 
  \end{minipage}   
  \begin{minipage}[]{0.95\linewidth}
\caption{Examples of gauge coupling unification with extra chiral supermultiplets added to the MSSM at both the TeV and $M_{\text{int}}$ scales. The left panel shows a typical example with extra vectorlike chiral supermultiplets in complete $SU(5)$ representations, in this case a ${\bf 10} + {\bf \overline{10}}$ near the TeV scale and a ${\bf 5}+{\bf \overline 5}$ at the intermediate scale. The right panel shows a more exotic example with a two pairs of vectorlike doublet leptons at the TeV scale, and a pair of quixes at the intermediate scale $M_{\text{int}} = 10^{11}$ GeV, leading
to unification of couplings above $10^{17}$ GeV.
\label{fig:unification}}
\end{minipage}\end{center}
\end{figure}

We will consider pairs of additional vectorlike quark or lepton superfields $\Phi + \overline \Phi$ chosen from among those in Table \ref{tab:additionalfields}. 
\begin{table}
\begin{center}
\begin{minipage}[]{0.95\linewidth}
\caption{Vectorlike pairs of chiral superfields $\Phi + \overline \Phi$ 
that can be added to the base models, and their Standard Model gauge transformation properties.
These pairs will carry non-zero net PQ charges as shown in Table \ref{tab:PQchargesofPhiPhibar}, depending on the source of their mass.\label{tab:additionalfields}}
\end{minipage}

\vspace{0.2cm}

\begin{tabular}{|c c|}
\hline
~Superfields~  &  ~$SU(3)_c \times SU(2)_L \times U(1)_Y$~\\
\hline
\hline
~$Q + \overline{Q}$~  &  ~$({\bf 3}, {\bf 2}, 1/6)$ + $({\bf \overline{3}}, {\bf 2}, -1/6)$~\\
~$U + \overline{U}$~  &  ~$({\bf 3}, {\bf 1}, 2/3)$ + $({\bf \overline{3}}, {\bf 1}, -2/3)$~\\
~$E + \overline{E}$~  &  ~$({\bf 1}, {\bf 1}, -1)$ + $({\bf 1}, {\bf 1}, 1)$~\\
~$D + \overline{D}$~  &  ~$({\bf 3}, {\bf 1}, -1/3)$ + $({\bf \overline{3}}, {\bf 1}, 1/3)$~\\
~$L + \overline{L}$~  &  ~$({\bf 1}, {\bf 2}, -1/2)$ + $({\bf 1}, {\bf 2}, 1/2)$~\\
~$D_6 + \overline{D}_6$~  &  ~$({\bf 6}, {\bf 1}, 1/3)$ + $({\bf \overline{6}}, {\bf 1}, -1/3)$~\\
\hline
\end{tabular}
\end{center}
\end{table}
The first 5 pairs each include one field with the same color and electroweak quantum numbers as the MSSM chiral superfields, from which they are distinguished by the use of capital letters. While any of the possible mass terms could be used independently for each of types of fields, it is well-known that nearly degenerate sets of chiral superfields in 
${\bf 5}+{\bf \overline 5} = D + \overline D + L + \overline L$ 
or 
${\bf 10} + {\bf \overline{10}} = Q + \overline Q + U + \overline U + E + \overline E$ representations of the $SU(5)$ grand unified theory \cite{Georgi:1974sy} will preserve the apparent unification of gauge couplings observed in the MSSM, as illustrated for example in the left panel of Figure \ref{fig:unification}. Note that we do not assume that $SU(5)$ is actually the unbroken gauge group in the ultraviolet, and we allow for different components of the ${\bf 5}+{\bf \overline 5}$ and/or ${\bf 10} + {\bf \overline{10}}$ to have different mass source terms and therefore different PQ charges. In Table \ref{tab:additionalfields} we have also included a more exotic pair, an $SU(3)_c$ color sextet (quix) and its conjugate with electric charges  $\pm 1/3$, denoted $D_6 + \overline D_6$. If these have masses at an intermediate scale $M_{\text{int}} \approx 10^{11}$ GeV and there are two $L+ \overline L$ pairs near the TeV scale, then one can again have gauge coupling unification at a scale somewhat higher than in the MSSM, as shown in the right panel of Figure \ref{fig:unification}. Although this combination might seem somewhat of an ad hoc choice, it is of interest because it provides an example of an even more enhanced coupling of the axion to photons, as we will see.
%

There are other classes of quixotic models that can have gauge coupling unification, for example $D_6 + \overline{D}_6$ or $U_6 + \overline{U}_6 = ({\bf 6}, {\bf 1}, -2/3)$ + $({\bf \overline{6}}, {\bf 1}, 2/3)$ quixes at the intermediate scale that have one or two weak isotriplets $T \sim ({\bf 1}, {\bf 3}, 0)$ at the same scale. Such models do not predict any new particles at the TeV scale, and therefore are omitted from our discussion.

The rest of this paper is organized as follows. In section \ref{sec:axion couplings}, we review the general properties of axion models, in particular the relation between PQ charge assignments and
axion couplings, the domain wall problem, and the axion quality problem. The relevant relations
are obtained for supersymmetric DFSZ axion models of the type described above. In section \ref{sec:discrete}, we explore possible discrete symmetries that can solve the axion quality problem, and establish that there are many available for all of the base models and their extensions. In section \ref{sec:results}, we discuss axion signals and detection prospects for our models, including both present constraints and future detection prospects. Section \ref{sec:conclusion} contains some summarizing remarks. 

\clearpage

\section{Peccei-Quinn charges, anomalies and axion couplings\label{sec:axion couplings}}
\setcounter{equation}{0}
\setcounter{figure}{0}
\setcounter{table}{0}
\setcounter{footnote}{1}

\subsection{Axion properties in general}

In this section we review the general properties of QCD axion models, following the discussion in \cite{DiLuzio:2020wdo}. The anomalous divergence of the Peccei-Quinn current in a model with each left-handed Weyl fermion $\psi_f$ transforming under $SU(3)_c$ in the representation $R_{f}$ with a PQ charge $Q_{f}$ (such that $\psi_f \rightarrow e^{i Q_f \alpha} \psi_f$) is given by:
\beq
\partial_{\mu}j^{\mu}_{\text{PQ}} &=& 
\frac{g_s^2}{16 \pi^2} N G^a_{\mu \nu} \Tilde{G}^{a \mu \nu} 
+ \frac{e^2}{16 \pi^2} E F_{\mu \nu} \Tilde{F}^{\mu \nu},
\label{eq:PQdivergence}
\eeq
where the $U(1)_{\text{PQ}}$-$[SU(3)_c]^2$ and $U(1)_{\text{PQ}}$-$[U(1)_{\rm EM}]^2$ anomaly coefficients, which play an important role in the low-energy phenomenology, are
\beq
N &=& 
\textrm{Tr} \left [ Q_f T(R_{f}) \right ]
,
\label{eq:QCDanomaly}
\\
E &=& 
\textrm{Tr} \left [ Q_f q_f^2 \right ]
,
\label{eq:EManomaly}
\eeq
respectively. Here, the trace in $N$ is over all left-handed fermion representations with $SU(3)_c$ index $T(R_f)$, while the trace in $E$ is over all left-handed fermion fields with electromagnetic (EM) charge $q_f$. The PQ current can be explicitly written as:
\beq
j^{\rm \mu}_{\text{PQ}} = - i \smash[b]{\sum_s} Q_s 
\bigl ( \varphi_s^\dag \partial^\mu \varphi_s - \varphi_s \partial^\mu \varphi_s^\dag \bigr )
- \smash[b]{\sum_f} Q_f \psi_f^\dag \overline{\sigma}^\mu \psi_f,
\eeq
where $\varphi_s$ are complex scalar fields, each with a PQ charge $Q_s$ (such that $\varphi_s \rightarrow e^{i Q_s \alpha} \varphi_s$).

The PQ symmetry can be spontaneously broken by non-zero vacuum expectation values (VEVs) of some scalars $\langle \varphi_s \rangle \equiv v_s$ (without breaking color or EM) that can be written as:
\beq
\varphi_s = \frac{1}{\sqrt{2}} (v_s + \rho_s) e^{i a_s/v_s}.
\label{eq:scalarVEV}
\eeq
Here, $\rho_s$ and $a_s$ are the canonically normalized radial and angular fields, respectively. In our case, the scalars with VEVs will be either doublets or singlets under $SU(2)_L$. The PQ current becomes:
\beq
j^{\rm \mu}_{\text{PQ}} = \smash[b]{\sum_{s}} Q_s v_s \partial^\mu a_s
- \smash[b]{\sum_f} Q_f \psi_f^\dag \overline{\sigma}^\mu \psi_f,
\eeq
in the decoupling limit, i.e.~ignoring the terms with heavy radial fields $\rho_s$, which are PQ invariant.
Now, defining
\beq
v_A^2 = \smash[b]{\sum_{s}} Q_s^2 v_s^2
,
\label{eq:definevA}
\eeq
the pseudo-Nambu-Goldstone boson axion field $A$ is given by:
\beq
A = \frac{1}{v_A} \sum_{s} Q_s v_s a_s
. 
\eeq
This results in
\beq
j^{\rm \mu}_{\text{PQ}} = v_A \partial^\mu A
- \smash[b]{\sum_f} Q_f \psi_f^\dag \overline{\sigma}^\mu \psi_f.
\label{eq:PQ current}
\eeq
We also note that under a $U(1)_{\text{PQ}}$ transformation parameterized by $\alpha$,
\beq
a_s &\rightarrow& a_s + \alpha Q_s v_s ,
\\
A  &\rightarrow& A + \alpha v_A.
\label{eq:axionshift}
\eeq

Assuming the PQ symmetry is not explicitly broken,
the PQ charges in a theory are determined by the terms in the Lagrangian.
However, there is an ambiguity in the charges due to the freedom to add multiples of the weak hypercharge and any other non-anomalous
$U(1)$ symmetries that might be present in the theory.
The ``physical" PQ charges can be obtained by imposing an orthogonality condition:
\beq
\smash[b]{\sum_{s}} Y_s Q_s v_s^2 = 0,
\label{eq:Z_ortho}
\eeq
to ensure that the axion and the $Z$ boson do not mix.
Here, $Y_s$ are the $U(1)_Y$ weak hypercharges of the scalars that get VEVs.
The above condition can be obtained by requiring the axion $A$ to be invariant
under $U(1)_Y$ gauge transformations parameterized by $\alpha_Y$:
\beq
a_s &\rightarrow& a_s +  \alpha_Y Y_s v_s,
\\[3pt]
A  &\rightarrow& A + \frac{\alpha_Y}{v_A} \smash[b]{\sum_s} Y_s Q_s v_s^2 = A.
\eeq
The would-be Nambu-Goldstone boson $G$ that becomes the longitudinal component of the $Z$-boson is:
\beq
G &=& \frac{1}{v^\prime} \smash[b]{\sum_{s}} Y_s v_s a_s,
\qquad\quad
{v^\prime}^2 \>=\> \smash[b]{\sum_s} Y_s^2 v_s^2.
\eeq

The anomalous divergence of the PQ current in eq.~(\ref{eq:PQdivergence}), with $j^{\rm \mu}_{\text{PQ}}$ in eq.~(\ref{eq:PQ current}), now implies:
\beq
\partial_\mu \partial^\mu A
- \frac{1}{v_A} \smash[b]{\sum_f Q_f} \partial_\mu \left ( \psi_f^\dag \overline{\sigma}^\mu \psi_f \right )
&=&
\frac{N}{v_A} \frac{g_s^2}{16 \pi^2}  G^a_{\mu \nu} \Tilde{G}^{a \mu \nu} +
\frac{E}{v_A} \frac{e^2}{16 \pi^2}  F_{\mu \nu} \Tilde{F}^{\mu \nu}.
\eeq

This equation can be obtained as the Euler-Lagrange equation of the following effective Lagrangian:
\beq
\mathcal{L}_{A} &\supset&
\frac{1}{2} \partial_\mu A \partial^\mu A 
+ \frac{g_s^2}{32 \pi^2} \frac{A}{f_A} G^a_{\mu \nu} \Tilde{G}^{a \mu \nu}
+ \frac{E}{N} \frac{e^2}{32 \pi^2} \frac{A}{f_A} F_{\mu \nu} \tilde{F}^{\mu \nu}
- \frac{1}{N} \frac{\partial_\mu A}{2 f_A}  \sum_f Q_f \psi_f^\dag \overline{\sigma}^\mu \psi_f,
\phantom{xxxx}
\eeq
where, provided that the PQ symmetry has a non-zero QCD anomaly $N$,  the axion decay constant is defined by
\beq
f_A &\equiv& \frac{v_A}{2 N}.
\eeq
As a convention, we choose to define $f_A$ as always positive. In some models $N$ is negative, in which case $v_A$ is negative in our convention.

In terms of four-component Dirac fermions $\Psi_f$ with left-handed and conjugate of right-handed components carrying PQ charges $Q_f$ and $Q_{\overline{f}}$, respectively, 
\beq
\mathcal{L}_{A} &\supset&
\frac{1}{2} \partial_\mu A \partial^\mu A 
+ \frac{g_s^2}{32 \pi^2}  \frac{A}{f_A} G^a_{\mu \nu} \Tilde{G}^{a \mu \nu}
+  c_{\gamma} \frac{e^2}{32 \pi^2} \frac{A}{f_A} F_{\mu \nu} \tilde{F}^{\mu \nu}\notag\\
&&+ \frac{\partial_\mu A}{2 f_A} \sum_f c_f \overline{\Psi}_f \gamma^\mu \gamma_5 \Psi_f
+ \frac{\partial_\mu A}{2 f_A} \sum_f c_f^\prime \overline{\Psi}_f \gamma^\mu \Psi_f,
\label{eq:L_eff_dirac}
\eeq
with
\beq
c_{\gamma} &=& \frac{E}{N},\label{eq:coeff_photon}\\
c_{f} &=& \frac{Q_f + Q_{\overline{f}}}{2 N}\label{eq:coeff_fermion_a}\\
c_{f}^\prime &=& \frac{Q_{\overline{f}} - Q_f}{2 N}\label{eq:coeff_fermion_v}.
\eeq
One can now express the axion-fermion couplings in terms of axial currents only, eliminating the vector currents. This can be achieved by redefining Dirac fermions $\Psi_f$ by the following $U(1)$ vector transformation (as opposed to the axial transformation that contributes to the PQ anomaly):
\beq
\Psi_f \rightarrow e^{i c^\prime_f A/2 f_A} \Psi_f, \label{eq:U1-vector-transf}
\eeq
under which the kinetic terms of the fermions give an additional contribution that cancels the axion-fermion couplings involving the vector currents,
\beq
\sum_f i \Psi_f \gamma^\mu \partial_\mu \Psi_f \rightarrow \sum_f i \Psi_f \gamma^\mu \partial_\mu \Psi_f
- \frac{\partial_\mu A}{2 f_A} \sum_f c_f^\prime \overline{\Psi}_f \gamma^\mu \Psi_f.
\eeq
After evaluating the coefficients of axion-photon and axion-fermion couplings given in
eqs.~(\ref{eq:coeff_photon}) and (\ref{eq:coeff_fermion_a}), respectively,
the resulting low-energy axion interaction Lagrangian can be obtained
as
\beq
\mathcal{L}_{A} &=& \frac{1}{2} \partial^\mu A \partial_\mu A - \frac{1}{2} m_A^2 A^2 +
 C_{A \gamma} \frac{e^2}{32 \pi^2} \frac{A}{f_A} F_{\mu \nu} \tilde{F}^{\mu \nu}
+ \frac{\partial_\mu A}{2 f_A} \sum_{f = p, n, e} C_{A f} \overline{\Psi}_f \gamma^\mu \gamma_5 \Psi_f
,\phantom{xxx}
\label{eq:LA_int}
\eeq
where  
the axion mass is given in terms of the axion decay constant $f_A$ by \cite{Borsanyi:2016ksw}
\beq
m_A = 5.691(51) \frac{10^{12} \, {\text{GeV}}}{f_A} \mu{\rm eV},
\label{eq:mAintermsoffA}
\eeq
and
\cite{diCortona:2015ldu}
\beq
C_{A \gamma} &=& c_\gamma - 1.92(4),\label{eq:axion_couplings_begin}\\
C_{A p} &=& -0.47(3) + 0.88(3) \, c_u - 0.39(2) \, c_d - C_{A, \, {\rm sea}},\\
C_{A n} &=& -0.02(3) + 0.88(3) \, c_d - 0.39(2) \, c_u - C_{A, \, {\rm sea}},\\
C_{A, \, {\rm sea}} &=& 0.038(5) \, c_s + 0.012(5) \, c_c + 0.009(2) \, c_b + 0.0035(4) \, c_t,\\
C_{A e} &=& c_e
\label{eq:axion_couplings_end}
\eeq
The tree-level couplings to fermions above do not include the effects of renormalization group running \cite{Srednicki:1985xd}-\cite{Choi:2021kuy}, which are usually small compared to experimental uncertainties. A notable exception is the coupling
$C_{Ae}$ when $c_e$ vanishes or is very small in hadronic axion models, which we will include as non-supersymmetric benchmarks below. In those cases we include the leading logarithmic contribution \cite{Srednicki:1985xd,Chang:1993gm} 
\beq
\Delta C_{A e} &=& \frac{3 \alpha^2}{4 \pi^2} \left ( c_\gamma \ln{\frac{f_A}{m_e}} - 1.92(4) \ln{\frac{{\text{GeV}}}{m_e}} \right ) .
\label{eq:DeltaCae}
\eeq
Finally, the effective Lagrangian which provides the axion coupling to photons and on-shell electrons, protons, and neutrons can be written as
\beq
\mathcal{L}_A^{\text{int}} &=&
\frac{1}{4} g_{A \gamma} A F_{\mu \nu} \tilde{F}^{\mu \nu}
- \sum_{f = p, n, e} i g_{A f} A \overline{\Psi}_f \gamma_5 \Psi_f
,
\label{eq:LA_int_alt}
\eeq
with the definitions
\beq
g_{A \gamma} = \frac{\alpha}{2 \pi} \frac{C_{A \gamma}}{f_A},
\qquad
g_{A f} = m_f \frac{C_{A f}}{f_A}
.
\label{eq:low-energy axion couplings}
\eeq
Here we note that the contribution from the CP-violating terms where axion derivatively couples to
vector fermion current in eq.~(\ref{eq:L_eff_dirac}) vanish after integrating by parts and using the Dirac equation (without having to redefine the fermion fields as done in eq.~(\ref{eq:U1-vector-transf})).

After the PQ symmetry breaking, a discrete subgroup $Z_{N_{\rm DW}}
(= e^{2 k \pi i/N_{\rm DW}}$, $k = 0, 1, \ldots, N_{\rm DW} - 1)$
is left  unbroken. $N_{\rm DW}$ is the Domain Wall (DW) number
that corresponds to the number of discrete set of inequivalent
degenerate minima of the axion potential \cite{Sikivie:1982qv}.
With the above definition of PQ-QCD-QCD anomaly coefficient $N$, the DW number can be computed as \cite{Ernst:2018bib}:
\beq
N_\textrm{DW} \equiv \textrm{minimum integer} \left ( 
2 N \sum_{s} \frac{n_s Q_s v_s^2}{v_A^2}
\right ),
\label{eq:DWnumber}
\eeq
where
$n_s \in \mathbb{Z}$.
The domain wall number must be invariant under the rescalings of the PQ charges, as is reflected in the above formula.

Models with $N_\textrm{DW} > 1$ may have non-trivial cosmological implications. In particular, formation of topological defects such as stable domain walls, due to degenerate vacua with different possible phases of the axion, can dominate the early universe \cite{Sikivie:1982qv}. However, the cosmological domain wall problem may not arise if PQ symmetry is broken in the pre-inflationary era and is not restored after inflation ends, so that the observable universe today consists of a single patch that initially had a common value of $\theta$. Another possibility \cite{Lazarides:1982tw} is that the apparent $Z_{N_{\text{DW}}}$ discrete symmetry relating different vacua are actually embedded within a continuous gauged symmetry that is spontaneously broken at high energies. The $N_{\text{DW}}$ apparently distinct vacua are then actually the same, being connected by the continuous gauge symmetry. In this paper we will concentrate instead on the possibility that safety is achieved by $N_\textrm{DW} = 1$, in which case any domain walls that form are bounded by strings on their edges, and are highly unstable \cite{Vilenkin:1982ks}. This can be achieved using a strategy similar to that in \cite{Georgi:1982ph}, by introducing heavy fermions charged under the PQ symmetry, in our case components of chiral supermultiplets that are vectorlike under the Standard Model gauge group.

\subsection{Application to DFSZ axions in supersymmetry}

In the following, we consider supersymmetric DFSZ-type axion models with two PQ-breaking gauge-singlet fields $X,Y$ as discussed in the Introduction, consisting of the four  base models $\text{B}_{\rm \RomanNumeralCaps{1}}$, $\text{B}_{\rm \RomanNumeralCaps{2}}$, $\text{B}_{\rm \RomanNumeralCaps{3}}$, and $\text{B}_{\rm \RomanNumeralCaps{4}}$ summarized in Table \ref{tab:base models}, with possible extensions by additional vectorlike superfields 
that maintain approximate gauge coupling unification. The additional chiral superfield content in the models considered below are listed in Table \ref{tab:additionalfields}, with mass terms as in Table \ref{tab:PQchargesofPhiPhibar}.

From our choice of normalization in Table \ref{tab:base models}, 
we get the following condition on the PQ charges of the MSSM Higgses:
\beq
Q_{H_u} + Q_{H_d} &=& -2.
\label{eq:Higgs PQ charge condition in base models}
\eeq
Requiring that the Yukawa terms in the MSSM superpotential are $U(1)_{\text{PQ}}$ invariant, we can express the following PQ charge combinations$-$the only combinations that enter the low-energy axion-fermion couplings$-$in terms of $Q_{H_u}$:
\beq
Q_{\overline{u}} + Q_q &=& -Q_{H_u}\label{eq:MSSM ubar-q-Hu invariance},\\
Q_{\overline{d}} + Q_q &=& 2 + Q_{H_u},\\
Q_{\overline{e}} + Q_{\ell} &=& 2 + Q_{H_u}.
\label{eq:MSSM e-l-Hd invariance}
\eeq
We have used eq.~(\ref{eq:Higgs PQ charge condition in base models}) in the last two equations.
Using the above constraints, the PQ charges of all the MSSM fields can be expressed in terms
of only three numbers $Q_{H_u}$, $Q_{q}$, and $Q_{\ell}$, which
at this point  are still unconstrained and therefore can be treated as free parameters. This corresponds to the freedom to add multiples of other $U(1)$ charges to the PQ symmetry charges. 

In order to obtain the low-energy axion couplings and the DW number, the scalars (represented by the same symbols as their respective supermultiplets) that acquire a VEV are parameterized as:
\beq
X \supset \frac{v_x}{\sqrt{2}} e^{i a_x/v_x}
,
\quad
Y \supset \frac{v_y}{\sqrt{2}} e^{i a_y/v_y}
,
\quad
H^0_u \supset \frac{v_u}{\sqrt{2}} e^{i a_u/v_u}
,
\quad
H^0_d \supset \frac{v_d}{\sqrt{2}} e^{i a_d/v_d}
,
\eeq
where $H^0_u$ and $H^0_d$ are the neutral MSSM Higgs scalars, $a_s = \{ a_x, a_y, a_u, a_d \}$ are the pseudo-scalar bosons that contribute to the axion, and $v_x, v_y \gg v_u, v_d$ in an invisible axion model.

Imposing the orthogonality condition eq.~(\ref{eq:Z_ortho}) in each of the base models yields:
\beq
\frac{Q_{H_d}}{Q_{H_u}} = \frac{v_u^2}{v_d^2} \equiv \tan^2 \beta,
\eeq
with $s_\beta \equiv \sin \beta = v_u/v$, $c_\beta \equiv \cos \beta = v_d/v$, and $v^2 = v_u^2 + v_d^2$. The orthogonality condition has essentially used the freedom to add arbitrary multiples of the $U(1)_Y$ charges to the fields. The PQ charges for the MSSM Higgs supermultiplets using eq.~(\ref{eq:Higgs PQ charge condition in base models}), consistent with the Higgs VEVs and our normalization convention $Q_{H_u} + Q_{H_d} = -2$, are:
\beq
Q_{H_u} = - 2 c_\beta^2,
\qquad
Q_{H_d} = - 2 s_\beta^2,
\eeq
and we have, from eq.~(\ref{eq:definevA}),
\beq
2 N f_A \>=\> v_A &=& 
\begin{cases}
\left[ v_x^2 + 9 v_y^2 + 4 s_\beta^2 c_\beta^2 v^2\right ]^{1/2} \qquad (\text{B}_{\rm \RomanNumeralCaps{1}}, \text{B}_{\rm \RomanNumeralCaps{2}}),
\\
\left[\frac{1}{9} v_x^2 + v_y^2 + 4 s_\beta^2 c_\beta^2 v^2\right ]^{1/2}\qquad (\text{B}_{\rm \RomanNumeralCaps{3}}), 
\\
\left [v_x^2 +  v_y^2 + 4 s_\beta^2 c_\beta^2 v^2\right ]^{1/2}\phantom{9}\qquad (\text{B}_{\rm \RomanNumeralCaps{4}}),
\end{cases}
\label{eq:VintermsoffA}
\eeq
with $N$ to be given soon below. The contribution $4 s_\beta^2 c_\beta^2 v^2$ is numerically negligible.

Using eqs.~(\ref{eq:MSSM ubar-q-Hu invariance})-(\ref{eq:MSSM e-l-Hd invariance}) the PQ charges of the rest of the  MSSM chiral superfields can be fixed in terms of $Q_q$ and $Q_{\ell}$.
We can also require that neutrino masses are provided by the superpotential version of the Weinberg operator $(H_u\ell)(H_u\ell)$. Equivalently from the point of view of anomalies, we can introduce a gauge-singlet neutrino superfield $\overline \nu$ which has superpotential terms
\beq
W_{\rm seesaw} &=& \frac{1}{2} \lambda_{\overline \nu} S \overline\nu\hspace{1pt} \overline\nu + y_{\overline\nu} H_u \ell \overline\nu,
\eeq
where $S$ is a $SU(3)_c \times SU(2)_L \times U(1)_Y$ singlet, for example either a bare mass term or $X$ or $Y$.
This then gives us the further constraint
\beq
Q_{\overline{\nu}} \,=\, -Q_S/2,\qquad
Q_{\ell} &=& - Q_{H_u} + Q_{S}/2,
\label{eq:seesaw PQ charge of doublet leptons}
\eeq
where $Q_S = 0$ in the case of a bare seesaw mass ($\lambda_{\overline \nu} S \rightarrow M_{\overline \nu}$).

The resulting PQ charge assignments for the MSSM chiral superfields are shown in Table~\ref{tab:MSSM_PQ}. 
These physical PQ charge assignments also hold for any extensions of the base models, 
as long as there are no additional scalars with non-zero VEVs that can potentially feed into the orthogonality condition in eq.~(\ref{eq:Z_ortho}). Also included for future reference are the charges for two non-anomalous family-independent $U(1)$ symmetries; one is $6Y$, 
the weak hypercharge normalized to integer charges for all fields, and the second
is $2 T_R^3 = 2Y - (B-L)$, also with integer charges.
\begin{table}
\begin{center}
\begin{minipage}[]{0.95\linewidth}\caption{The superfield charges of the Peccei-Quinn symmetry and two linearly independent anomaly-free $U(1)$ symmetries of the dimensionless part of the MSSM superpotential. Here, $\overline\nu$ represents either a gauge-singlet neutrino or the combination $-H_u \ell$ appearing in the non-renormalizable Weinberg operator for neutrino masses. The physical Peccei-Quinn charges given here are obtained by imposing the orthogonality condition eq.~(\ref{eq:Z_ortho}) and the normalization condition $Q_{H_u} + Q_{H_d} = -2$, and are given in terms of two free parameters $Q_q$, $Q_\ell$, and the ratio of Higgs expectation values $\tan\beta = s_\beta/c_\beta$. If a seesaw neutrino mass term $S \hspace{1pt} \overline \nu\hspace{1pt}\overline\nu$ is included in the superpotential for some gauge-singlet field $S$ whose scalar component gets a VEV, then $2 T_R^3 = 2 Y - (B-L)$ is explicitly broken and $Q_\ell = 2 c_\beta^2 + Q_S/2$ is fixed. \label{tab:MSSM_PQ}}
\end{minipage}

\vspace{0.3cm}

\begin{tabular}{|c | c | c | c | c | c | c | c| c|}
\hline
& ~$H_u$~ & ~$H_d$~ & ~$q$~ & ~$\ell$~ & ~$\overline{u}$~ & ~$\overline{d}$~ & ~$\overline{e}$~ & ~$\overline\nu$~ 
\\
\hline\hline
~PQ~ & ~$-2 c_\beta^2$~ & ~$-2 s_\beta^2$~ & ~$Q_q$~ & ~$Q_{\ell}$~ & ~$2 c_\beta^2 - Q_q$~ & ~$2 s_\beta^2 - Q_q$~ & ~$2 s_\beta^2 - Q_{\ell}$~ & ~$2 c_\beta^2 - Q_{\ell}$~
\\
\hline
~$6Y$~ & ~$3$~ & ~$-3$~ & ~$1$~ & ~$-3$~ & ~$-4$~ & ~$2$~ & ~$6$~ & ~$0$~
\\
\hline
~$2{T^3_R}$~ & ~$1$~ & ~$-1$~ & ~$0$~ & ~$0$~ & ~$-1$~ & ~$1$~ & ~$1$~ & ~$-1$~
\\
\hline
\end{tabular}
\end{center}
\end{table}

Using eqs.~(\ref{eq:QCDanomaly}) and (\ref{eq:EManomaly}), the PQ anomaly coefficients in the base models and their extensions to include vectorlike fields are:
\beq
N &=& 
3 + {\textstyle\sum} Q_{Q\overline Q} + \frac{1}{2} {\textstyle\sum} Q_{U \overline U} + 
\frac{1}{2} {\textstyle\sum} Q_{D \overline D} + \frac{5}{2} {\textstyle\sum} Q_{D_6 \overline D_6},
\label{eq:Nforextendedmodels}
\\
E &=& 
6 + \frac{5}{3} {\textstyle\sum} Q_{Q\overline Q} + \frac{4}{3} {\textstyle\sum} Q_{U \overline U} 
+ \frac{1}{3} {\textstyle\sum} Q_{D \overline D} 
+ {\textstyle\sum} Q_{L \overline L} 
+ {\textstyle\sum} Q_{E \overline E} 
+ \frac{2}{3} {\textstyle\sum} Q_{D_6 \overline D_6},
\phantom{xxxx}
\label{eq:Eforextendedmodels}
\eeq
where the contributions ${\textstyle\sum} Q_{\Phi\overline \Phi}$ can be read from Table \ref{tab:PQchargesofPhiPhibar}, depending on the origins of the $\Phi\overline \Phi$ mass terms. 
Note that for all four base models, we have $N=3$ and $E=6$.
The coefficients $c_\gamma, c_f$ defined in eqs.~(\ref{eq:coeff_photon})-(\ref{eq:coeff_fermion_a})
in the base models and their extensions are then:
\beq
c_\gamma = \frac{E}{N}, \quad 
c_{u} = \frac{c_\beta^2}{N} , \quad
c_{d} = \frac{s_\beta^2}{N} , \quad
c_{e} = \frac{s_\beta^2}{N}.
\label{eq:ZerothOrderExtcoeff}
\eeq
Note that all of $N$, $E$, $c_\gamma$, $c_u$, $c_d$, and $c_e$,
and therefore the low-energy effective Lagrangian parameters 
eqs.~(\ref{eq:axion_couplings_begin})-(\ref{eq:axion_couplings_end}) and (\ref{eq:low-energy axion couplings}),
are independent of the choices of $Q_{q}$ and $Q_{\ell}$, and we arrive at
\beq
g_{A\gamma} &=& 
\frac{\alpha}{2\pi f_A} (E/N - 1.92(4)),
\label{eq:gAgamma}
\\
g_{Ae} &=& 
\frac{m_e}{f_A} \frac{s_\beta^2}{N},
\label{eq:gAe}
\\
g_{An} &=& 
\frac{m_n}{f_A} \left ( -0.02(3) + [0.833(30) - 1.239(37) c_\beta^2]/N \right),
\label{eq:gAn}
\\
g_{Ap} &=& \frac{m_p}{f_A} \left ( -0.47(3) + [-0.437(21) + 1.302(37) c_\beta^2]/N \right)
\phantom{xxx}
\label{eq:gAp}
\eeq
for the central value with parenthetical uncertainty estimates in our base models and extensions, parameterized only by $N$ and $E$. Note that the model-independent ($N$-independent) part of the neutron coupling $g_{An}$ suffers from an accidental cancellation \cite{diCortona:2015ldu}.

We now compute the domain wall number (i.e. the number of independent degenerate minima
of the axion potential) for each of the four base
models listed in Table~\ref{tab:base models} using the formula given in eq.~(\ref{eq:DWnumber}). 

\subparagraph*{Model $\text{B}_{\rm \RomanNumeralCaps{1}}$:} Using eq.~(\ref{eq:DWnumber}),
we have
\beq
\sum_s \frac{n_s Q_s v_s^2}{v_A^2} = \frac{- n_x v^2_x + 3 n_y v_y^2  - 
(n_u + n_d) 2 s_\beta^2 c_\beta^2 v^2}{v_x^2 + 9 v_y^2 + 4 s_\beta^2 c_\beta^2  v^2},
\label{eq:Z1_NDW_step1}
\eeq
where $n_x$, $n_y$, $n_u$, and $n_d$ are integers accompanying the scalars
$X$, $Y$, $H_u$, and $H_d$, respectively. In the above expression, our goal
is to arrange each $n_s$ such that eq.~(\ref{eq:Z1_NDW_step1}) is an integer. This is obtained provided that
$n_y = -3 n_x$ and $n_u + n_d =  2 n_x$. Then from eqs.~(\ref{eq:DWnumber}) and (\ref{eq:Z1_NDW_step1}) we find:
\beq
N_{\rm DW} =  \textrm{minimum integer} \left | 2 N n_x \right |, 
\label{eq:Z1_NDW}
\eeq
where $n_x$ is a non-zero integer.
Therefore, since $N=3$ for the MSSM field content, the domain wall number of the base model $\text{B}_{\rm \RomanNumeralCaps{1}}$
is $N_{\rm DW} = 6$.

\subparagraph*{Model $\text{B}_{\rm \RomanNumeralCaps{2}}$:} Following a similar procedure as above,
we have:
\beq
\sum_s \frac{n_s Q_s v_s^2}{v_A^2} = \frac{n_x v_x^2 - 3 n_y v_y^2 - (n_u + n_d) 2 s_\beta^2 c_\beta^2 v^2}{v_x^2 + 9 v_y^2 + 4 s_\beta^2 c_\beta^2 v^2},
\label{eq:Z2_NDW_step1}
\eeq
which is again an integer if
$n_y = -3 n_x$ but this time with
$n_u + n_d = -2 n_x$. The domain wall number is therefore given by:
\beq
N_{\rm DW} =  \textrm{minimum integer} | 2 N n_x|, 
\label{eq:Z2_NDW}
\eeq
where again $n_x$ is a non-zero integer. For the base model $\text{B}_{\rm \RomanNumeralCaps{2}}$ with $N=3$, this again amounts to $N_{\rm DW} = 6$.

\subparagraph*{Model $\text{B}_{\rm \RomanNumeralCaps{3}}$:} Proceeding as above,
we begin with:
\beq
\sum_s \frac{n_s Q_s v_s^2}{v_A^2} = \frac{- \frac{1}{3} n_x v_x^2 + n_y v_y^2 
- (n_u + n_d) 2 s_\beta^2 c_\beta^2 v^2}{\frac{1}{9} v_x^2 + v_y^2 + 
4 s_\beta^2 c_\beta^2 v^2}.
\label{eq:Z3_NDW_step1}
\eeq
This will be an integer if $n_y = -3 n_x$ and $n_u + n_d = 6 n_x$, resulting in
\beq
N_{\rm DW} =  \textrm{minimum integer} \left | 6 N n_x \right|,
\label{eq:Z3_NDW}
\eeq
for $n_x$ a non-zero integer.
For the base model $\text{B}_{\rm \RomanNumeralCaps{3}}$ with $N=3$, we obtain $N_{\rm DW} = 18$ .

\subparagraph*{Model $\text{B}_{\rm \RomanNumeralCaps{4}}$:} To calculate the domain wall number for this model, we start with:
\beq
\sum_s \frac{n_s Q_s v_s^2}{v_A^2} = \frac{n_x v_x^2 - n_y v_y^2 - (n_u + n_d) 2 s_\beta^2 c_\beta^2 v^2}{v_x^2 + v_y^2 + 4 s_\beta^2 c_\beta^2 v^2},
\label{eq:Z4_NDW_step1}
\eeq
This will be an integer if $n_y = -n_x$ and $n_u + n_d = -2 n_x$,
with the result
\beq
N_{\rm DW} =  \textrm{minimum integer} \left | 2 N n_x \right |, 
\label{eq:Z4_NDW}
\eeq
where again $n_x$ is a non-zero integer. This gives $N_{\rm DW} = 6$ for the base model $\text{B}_{\rm \RomanNumeralCaps{4}}$.

The formulas for $N_{\rm DW}$ in eqs.~(\ref{eq:Z1_NDW}), (\ref{eq:Z2_NDW}), (\ref{eq:Z3_NDW}), and (\ref{eq:Z4_NDW}) in terms of $N$ are also applicable to any extensions of the models
$\text{B}_{\rm \RomanNumeralCaps{1}}$, 
$\text{B}_{\rm \RomanNumeralCaps{2}}$,
$\text{B}_{\rm \RomanNumeralCaps{3}}$, and 
$\text{B}_{\rm \RomanNumeralCaps{4}}$, respectively, provided that there are no additional scalars getting VEVs. In each case, the value of the anomaly $N$ should be computed by including contributions from the other strongly interacting 
chiral superfields present in the model extension, as in eqs.~(\ref{eq:Nforextendedmodels}). Then $N_{\rm DW}$ is the smallest 
integer $|2 N n_x|$ (for extensions of base models 
$\text{B}_{\rm \RomanNumeralCaps{1}}$, 
$\text{B}_{\rm \RomanNumeralCaps{2}}$, and 
$\text{B}_{\rm \RomanNumeralCaps{4}}$), or the smallest integer $|6 N n_x|$ (for extensions of base model 
$\text{B}_{\rm \RomanNumeralCaps{3}}$), where $n_x$ is a non-zero integer.

Although $N_{\rm DW} \ne 1$ in the base models and many of their extensions, it follows from the preceding that it is possible to achieve $N_{\rm DW}=1$ models and avoid the cosmological domain wall problem in certain extensions. This requires:
\beq
\textrm{For $N_{\rm DW} = 1$:}
\quad
N = 
\begin{cases}
 \textrm{$\pm \frac{1}{2}$ in model extensions of $\text{B}_{\rm \RomanNumeralCaps{1}}$, $\text{B}_{\rm \RomanNumeralCaps{2}}$, and $\text{B}_{\rm \RomanNumeralCaps{4}}$}, \\
 \textrm{$\pm \frac{1}{6}$ in model extensions of $\text{B}_{\rm \RomanNumeralCaps{3}}$}.
\end{cases}
\label{eq:NDW=1requirements}
\eeq
This occurs in a variety of extensions consistent with gauge coupling unification, all of which must have the property that the total index of $SU(3)_c$ chiral supermultiplets (i.e.~the equivalent number of quark+antiquark supermultiplets) with mass at the intermediate scale $M_{\text{int}} \sim f_A$ is odd. 
However, this is of course a necessary but not sufficient condition. In the particular case that there is only a ${\bf 5}+{\bf \overline 5}$ at $M_{\text{int}}$ and no other additional vectorlike quarks, then
from eqs.~(\ref{eq:Nforextendedmodels}) and (\ref{eq:NDW=1requirements}) it follows that $N_{\rm DW}=1$ cannot be constructed with
any of the possible $Q_{D \overline{D}}$ shown in Table~\ref{tab:PQchargesofPhiPhibar}.
Some of the model extensions with $N_{\rm DW}=1$ are listed in Table \ref{tab:NDW=1 models with two SU5 pairs}.
\begin{table}
\caption{Models with domain wall number $N_{\rm DW} = 1$, obtained by extending the base models
to include vectorlike superfield combinations consistent with gauge coupling unification. Here ${\bf 5}$ and ${\bf 10}$ refer to chiral superfields with the same content as the corresponding $SU(5)$ representations. Note that at least one vectorlike pair at the intermediate scale is required. Only the mass terms of the strongly interacting fields, which affect the PQ-QCD-QCD anomaly $N$ and therefore $N_{\rm DW}$, are shown explicitly. Different combinations of mass terms for the $L,\overline L$ and/or $E,\overline E$ fields give rise to the possible $3E/N$ values shown. For brevity, some models in the category with ${\bf 10 + \overline{10}}$ pairs at both the TeV and $M_{\text{int}}$ scales are omitted when they have the same anomaly coefficients as ones shown. The cases in the last two lines include quixes at the intermediate scale, which can provide large values of $E/N$. \label{tab:NDW=1 models with two SU5 pairs}}
\begin{center}
\begin{tabular}{| l | c | c | c | c |}
\hline
~Model extension~  &  ~Base~  &  ~Mass terms~ & $1/N$ & $3E/N$\\[1pt]
\hline
\hline
~${\bf 10 + \overline{10}}$ at $M_{\text{int}}$~  &  
~$\text{B}_{\rm \RomanNumeralCaps{1}}$~  &
~$Y Q \overline{Q} + X U \overline{U}$~
& \phantom{--.}2 & $20,\, -4$
\\
\hline
~${\bf 5 + \overline{5}}$ at TeV,~  &  ~$\text{B}_{\rm \RomanNumeralCaps{1}}$~  &
~$X Y D \overline{D} + Y D^\prime \overline{D}^\prime$~
& $\phantom{-}2$ & $44,\,20,\,-4,\,-28$
\\
~${\bf 5 + \overline{5}}$ at $M_{\text{int}}$~  &  ~~  &
~$Y^2 D \overline{D} + X D^\prime \overline{D}^\prime$~
& $\phantom{-}2$ & $44,\,20,\,-4,\,-28$
\\
\hline
~${\bf 5 + \overline{5}}$ at TeV,~  &  
~$\text{B}_{\rm \RomanNumeralCaps{1}}$~  &
~$Y Q \overline{Q} + Y U \overline{U} + X^2 D \overline{D}$~
& $-2$ & $68,\,44,\,20,\,-4$
\\
~${\bf 10 + \overline{10}}$ at $M_{\text{int}}$~  &  ~~  &
~$Y Q \overline{Q} + X U \overline{U} + X Y D \overline{D}$~
& $-2$ & $44,\,20,\,-4,\,-28$
\\
~~  &  ~~  &
~$X Q \overline{Q} + Y U \overline{U} + Y^2 D \overline{D}$~
& $-2$ & $44,\,20,\,-4,\,-28$
\\
\cline{2-5}
~~  &  ~$\text{B}_{\rm \RomanNumeralCaps{2}}$~  &
~$X Q \overline{Q} + X U \overline{U} + X^2 D \overline{D}$~
& $\phantom{-}2$ & $68,\,44,\,20,\,-4$
\\
\cline{2-5}
~~  &  ~$\text{B}_{\rm \RomanNumeralCaps{4}}$~  &
~$X Q \overline{Q} + X U \overline{U} + X^2 D \overline{D}$~
& $\phantom{-}2$ & $32,\, 20,\, 8,\, -4$
\\
\hline
~${\bf 10 + \overline{10}}$ at TeV,~
&  
~$\text{B}_{\rm \RomanNumeralCaps{1}}$~  &
~$X Y Q \overline{Q} + X Y U \overline{U} + X D \overline{D}$~
& $\phantom{-}2$ & $20,\, -4,\, -28,\, -52$
\\
~${\bf 5 + \overline{5}}$ at $M_{\text{int}}$~    
&  ~~  &
~$X Y Q \overline{Q} + X^2 U \overline{U} + Y D \overline{D}$~
& $\phantom{-}2$ & $44,\, 20,\, -4,\, -28$
\\
~~  &  ~~  &
~$X^2 Q \overline{Q} + Y^2 U \overline{U} + Y D \overline{D}$~
& $\phantom{-}2$ & ~$20,\, -4,\, -28,\, -52$~
\\
\cline{2-5}
~~  &  ~$\text{B}_{\rm \RomanNumeralCaps{2}}$~  &
~$X^2 Q \overline{Q} + X^2 U \overline{U} + X D \overline{D}$~
& $-2$ & $20,\, -4,\, -28,\, -52$
\\
\cline{2-5}
~~  &  ~$\text{B}_{\rm \RomanNumeralCaps{3}}$~  &
~$Y^2 Q \overline{Q} + X Y U \overline{U} + Y D \overline{D}$~
& $\phantom{-}6$ & $44,\, 20,\, -4,\, -28$
\\
~~  &  ~~  &
~$Y^2 Q \overline{Q} + Y^2 U \overline{U} + X D \overline{D}$~
& $\phantom{-}6$ & $20,\, -4,\, -28,\, -52$
\\
\cline{2-5}
~~  &  ~$\text{B}_{\rm \RomanNumeralCaps{4}}$~  &
~$X^2 Q \overline{Q} + X Y U \overline{U} + X D \overline{D}$~
& $\phantom{-}2$ & $32,\, 20,\, 8,\, -4$
\\
~~  &  ~~  &
~$X^2 Q \overline{Q} + X^2 U \overline{U} + X D \overline{D}$~
& $-2$ & $20,\, 8,\, -4,\, -16$
\\
~~  &  ~~  &
~$X^2 Q \overline{Q} + X^2 U \overline{U} + Y D \overline{D}$~
& $\phantom{-}2$ & $20,\, 8,\, -4,\, -16$
\\
\hline
~${\bf 10 + \overline{10}}$ at TeV,~  &  ~$\text{B}_{\rm \RomanNumeralCaps{1}}$~  &
~$X Y Q \overline{Q} + X Y U \overline{U} + X Q^\prime \overline{Q}^\prime + Y U^\prime \overline{U}^\prime$~
& $-2$ & $68,\, 44,\, 20,\, -4$
\\
~${\bf 10 + \overline{10}}$ at $M_{\text{int}}$~  &  ~~  &
~$X Y Q \overline{Q} + X^2 U \overline{U} + Y Q^\prime \overline{Q}^\prime + X U^\prime \overline{U}^\prime$~
& $-2$ & $44,\, 20,\, -4,\, -28$ 
\\
\cline{2-5}
~~  &  ~$\text{B}_{\rm \RomanNumeralCaps{2}}$~  &
~$X^2 Q \overline{Q} + X Y U \overline{U} + X Q^\prime \overline{Q}^\prime + X U^\prime \overline{U}^\prime$~
& $\phantom{-}2$ & $68,\, 44,\, 20,\, -4$
\\
\cline{2-5}
~~  &  ~$\text{B}_{\rm \RomanNumeralCaps{3}}$~  &
~$X Y Q \overline{Q} + Y^2 U \overline{U} + Y Q^\prime \overline{Q}^\prime + Y U^\prime \overline{U}^\prime$~
& $-6$ & $68,\, 44,\, 20,\, -4$
\\
~~  &  ~~  &
~$Y^2 Q \overline{Q} + X Y U \overline{U} + Y Q^\prime \overline{Q}^\prime + X U^\prime \overline{U}^\prime$~
& $-6$ & $44,\, 20,\, -4,\, -28$
\\
\cline{2-5}
~~  &  ~$\text{B}_{\rm \RomanNumeralCaps{4}}$~  &
~$X Y Q \overline{Q} + X^2 U \overline{U} + X Q^\prime \overline{Q}^\prime + X U^\prime \overline{U}^\prime$~
& $\phantom{-}2$ & $20,\, 8,\, -4,\, -16$
\\
~~  &  ~~  &
~$X^2 Q \overline{Q} + X Y U \overline{U} + X Q^\prime \overline{Q}^\prime + X U^\prime \overline{U}^\prime$~
& $-2$ & $20,\, 8,\, -4,\, -16$
\\
~~  &  ~~  &
~$X^2 Q \overline{Q} + X Y U \overline{U} + X Q^\prime \overline{Q}^\prime + Y U^\prime \overline{U}^\prime$~
& $\phantom{-}2$ & $32,\, 20,\, 8,\, -4$
\\
\hline
~$2\times (L + \overline{L})$ at TeV,~ &  ~$\text{B}_{\rm \RomanNumeralCaps{2}}$~  &
~$X D_6 \overline{D}_6$~
& $\phantom{-}2$ & ~$104,\, 80,\, 56,\, 32,\, 8$~
\\
\cline{2-5}
~$D_6 + \overline D_6$ at $M_{\text{int}}$~    &  ~$\text{B}_{\rm \RomanNumeralCaps{4}}$~  &
~$X D_6 \overline{D}_6$~
& $\phantom{-}2$ & $56,\, 44,\, 32,\, 20,\, 8$
\\
\hline
\end{tabular}
\end{center}
\end{table}
They include a model with a quix-antiquix pair at an intermediate scale near $M_{\text{int}} = 10^{11}$ GeV.

It should also be noted that the new heavy vectorlike particles in the extended models must be allowed to decay to Standard Model particles in order to avoid dangerous cosmological charged relics. This is easy to arrange, as the decays can be mediated by couplings to the Standard Model quark and lepton superfields via Yukawa terms in the superpotential, in several different ways. For example, in models with an extra ${\bf 5} + {\bf \overline 5}$, the vectorlike weak isosinglet down-type quark can decay through any of the dimensionless couplings $H_d q \overline D$ or $\ell q \overline D$ or $\overline u\hspace{0.8pt}\overline d\hspace{0.8pt} \overline D$, while the vectorlike weak isodoublet lepton can decay through any of the superpotential couplings $H_d \overline e L$ or $\ell \overline e L$ or $q \overline d L$. It is easy to construct similar couplings that allow $Q+\overline Q$, $U+\overline U$, and $E + \overline E$ to decay in models that have an extra ${\bf 10} + {\bf \overline {10}}$. In each case, in order to not overconstrain the system of equations determining the PQ charges,  or the discrete symmetry charges discussed in the next section, one may select only one of the couplings for a given $\Phi+\overline\Phi$ pair. (This also avoids tree-level violation of baryon number and/or lepton number, which could lead to proton decay.) In models with quixes (the last two rows of Table \ref{tab:NDW=1 models with two SU5 pairs}), they can decay via a superpotential term of the form $D_6 \overline{u} \overline{d}$. This fixes the PQ charge to be $Q_{D_6} = 2 Q_q - 2$. The important point for the low-energy axion phenomenology is that the existence of all such Yukawa couplings maintains the freedom to choose the net PQ charges $Q_{\Phi\overline \Phi}$ consistently with the vectorlike mass terms, and therefore does not affect the PQ anomalies.

\section{Discrete symmetries to protect $U(1)_{\text{PQ}}$\label{sec:discrete}}
\setcounter{equation}{0}
\setcounter{figure}{0}
\setcounter{table}{0}
\setcounter{footnote}{1}

\subsection{Conventions and assumptions for discrete symmetries\label{subsec:discretegeneral}}

In this section we consider discrete symmetries that can protect the Peccei-Quinn $U(1)$ symmetry up to dimension $p$ in the $X$ and $Y$ superfields in the superpotential. In equivalent language, the Peccei-Quinn symmetry is an accidental consequence of imposing the discrete symmetry. We consider as possibilities an Abelian discrete symmetry $Z_n$ or a discrete $R$-symmetry $Z_n^R$. In both cases, each chiral superfield $\Phi$ has integer charge $z_{\Phi} = 0,1,\ldots n-1$ (mod $n$). It will be convenient to treat both cases in a unified framework. We therefore take the gauginos and the anti-commuting coordinates $\theta_\alpha$ to have charge $r$, and each superpotential term is required to have total $Z_n^R$ charge $2r$ (mod $n$). For the case of ordinary (non-$R$) discrete symmetries we take $r=0$. In both cases, all Lagrangian terms have total discrete charge 0. [For continuous $R$-symmetries it is customary to take $r=1$ by a choice of normalization, but for discrete $R$-symmetries this is not always possible, since we require that all $Z_n^R$ charges for all fields are integers (mod $n$).]

In Table \ref{tab:maxsuppression}, we list the allowed\footnote{Note that $n$ must be even for non-$R$ ($r=0$) symmetries $Z_n$  for model $\text{B}_{\rm \RomanNumeralCaps{4}}$ and its extensions, in order to allow the term $X^2 Y^2$ but forbid the term $XY$.} values of $n$ that can provide for a maximum possible suppression $p$, for the base models and their extensions.
\begin{table}[b]
\begin{center}
\begin{tabular}{|c||c|c|c|c|}
\hline
Max suppression & 
$Z_n$, $\text{B}_{\rm \RomanNumeralCaps{1},\RomanNumeralCaps{2}, \RomanNumeralCaps{3}}$ 
& $Z_n$, $\text{B}_{\rm \RomanNumeralCaps{4}}$
& $Z_n^R$, $\text{B}_{\rm \RomanNumeralCaps{1}, \RomanNumeralCaps{2}, \RomanNumeralCaps{3}}$ 
& $Z_n^R$, $\text{B}_{\rm \RomanNumeralCaps{4}}$
\\
\hline\hline
$p\leq 6$ & $n\leq12$, or $14,15,18$ & 
$n\leq 10$ & $n\leq 12$, or $14$ & 
$n\leq 11$, or $14$
\\
\hline
$p=7$ & $13,\, 17,\, 21$ & $14$ & $13, 15$ & $13$ 
\\
\hline 
$p=8$ & $16,\, 20,\, 24$ & $12$ & $16, 17, 18, 19, 20, 22$ & $12, 16, 17$
\\
\hline 
$p=9$ & $19,\, 23,\, 27$ & $18$ & $21, 23$ & $15, 19, 20, 22$
\\
\hline 
$p=10$ & $22,\, 26,\, 30$ & $16$ & $24, 25, 26, 27, 28, 30, 36$ & $18, 23, 25, 26$
\\
\hline 
$p=11$ & $25,\, 29,\, 33$ & $22$ & $29, 31$ & $21, 28, 29$
\\
\hline 
$p=12$ & $28,\, 32,\, 36$ & $20$ & $32,33,34,35,38,44$ & $24, 31, 32, 34$
\\
\hline 
$p=13$ & $31,\, 35,\, 39$ & $26$ & $37, 39$ & $27, 35, 37, 38$
\\
\hline 
$p\geq 14$ & $34, 37, 38,$ or $\geq 40$ & 
$24$, or $\geq 28$ & 
$40,41,42,43$, or $\geq 45$ & 
$30, 33, 36$, or $\geq 39$
\\
\hline
\end{tabular}
\caption{Allowed $n$, for discrete symmetries $Z_n$ and discrete $R$-symmetries $Z_n^R$, that can provide a maximum possible suppression up to $p$ for the base models and their extensions. The meaning of $p$ is that the lowest dimension PQ-violating superpotential term(s) allowed by the discrete symmetry and involving only $X$ and $Y$ are of the form $X^j Y^{p-j}$. No anomaly cancellation constraints are imposed, yet. \label{tab:maxsuppression}}
\end{center}
\end{table}
This means that the lowest dimension superpotential term that is allowed by the discrete symmetry but violates the PQ symmetry has dimension $p$, and so has the form $X^j Y^{p-j}/M_P^{p-3}$. From these results, we can conclude that $n$ must be at least 13 (for models $\text{B}_{\rm \RomanNumeralCaps{1},\RomanNumeralCaps{2},\RomanNumeralCaps{3}}$) or 12 (for model $\text{B}_{\rm \RomanNumeralCaps{4}}$) even if the axion decay constant $f_A$ is as low as $10^9$ GeV so that we can have $p=7$. For higher required $p$, the smallest possible values of the order $n$ of the discrete symmetry group can be read from the table. Of course, not every discrete symmetry at order $n$ will provide the suppression listed. 

We now consider the conditions imposed by anomaly cancellation on the discrete symmetry group, which depend on the discrete symmetry charges of the superfields that are charged under the Standard Model gauge group. For simplicity, we assume that the MSSM quark and lepton chiral superfields have charges that are generation-independent. Using the symbol $z_\Phi$ for the additive $Z_n^R$ charge of the chiral superfield $\Phi$, the $Z_n^R \times G \times G$ anomalies for $G = SU(3)_c$, $SU(2)_L$, and $U(1)_Y$ are respectively:
\beq
A_3 &=& 6r + n_g (2 z_q + z_{\overline u} + z_{\overline d} - 4r) + 
2 \Delta_{Q\overline Q}+ \Delta_{U\overline U} + \Delta_{D\overline D}
+ 5 \Delta_{D_6\overline D_6}
,\phantom{xxx}
\\
A_2 &=& 4r + n_g (3 z_q + z_{l} - 4r) + z_{H_u} + z_{H_d} -2r
+ 3 \Delta_{Q\overline Q}
+ \Delta_{L\overline L}
,
\\
A_1 &=&
n_g \left (z_q + 3 z_l + 8 z_{\overline u} + 2 z_{\overline d} + 6 z_{\overline e}- 20 r \right ) + 3 (z_{H_u} + z_{H_d} - 2r) 
\nonumber \\ &&
+ \Delta_{Q\overline Q} 
+ 3 \Delta_{L\overline L} 
+ 8 \Delta_{U\overline U} 
+ 2 \Delta_{D\overline D}
+ 6 \Delta_{E\overline E}
+ 4 \Delta_{D_6\overline D_6}.
\eeq
Here, we have adopted a convenient but somewhat unusual normalization, by taking the index for the fundamental representation of the non-Abelian groups to be 1 rather than 1/2, so that $A_2$ and $A_3$ are integers defined (mod $n$). Likewise, the anomaly $A_1$ has been given in a normalization so that it is always an integer (assuming the weak hypercharge quantization of the Standard Model), and again defined (mod $n$). The first term in each of  $A_3$ and $A_2$ is the gaugino contribution. The number of chiral quark and lepton generations is $n_g = 3$. Finally, the contributions $\Delta_{\Phi\overline\Phi}$ are equal to the sum of the fermion discrete symmetry charges 
$z_{\Phi}-r$ and  $z_{\overline\Phi} - r$  for the indicated vectorlike chiral superfields, and are also equal to the negative of the sum of $z_X$ and $z_Y$ contributions for the terms in eqs.~(\ref{mass_possibilitiesNONREN}) and (\ref{mass_possibilitiesREN}) that produce their masses.

With our normalization of the $A_i$, the conditions 
\cite{Ibanez:1991hv,Ibanez:1991pr,Banks:1991xj,Ibanez:1992ji}, \cite{Lee:2011dya}
for the discrete symmetry to be anomaly free can be obtained by treating $Z_n^R$ as a subgroup of an anomaly-free continuous $U(1)$ symmetry that is spontaneously broken by a scalar field VEV of charge $n$ \cite{Krauss:1988zc}, in the presence of other possible very heavy fermions. The anomaly free condition is that one must have
\beq
\frac{A_3 + m_3 n}{k_3} 
\,=\,
\frac{A_2 + m_2 n}{k_2}
\,=\,
\frac{A_1 + m_1 n}{5 k_1}
\,=\,
\rho_{\rm GS},   
\eeq
where $\rho_{\rm GS}$ is a constant that can arise from the Green-Schwarz (GS) mechanism \cite{Green:1984sg} that may be used to cancel the anomalies for the $U(1)$ group that contains $Z_n^R$, and $m_{1,2,3}$ are integers, and $k_{2,3}$ are positive integer Kac-Moody levels, while $k_1$ in general could be arbitrary. For simplicity, we will therefore consider as a weak assumption that $k_2 = k_3$, but also consider as a stronger assumption motivated by gauge coupling unification that $k_3 = k_2 = k_1 = 1$, corresponding to the normalization in which  $SU(3)_c \times SU(2)_L \times U(1)_Y$ could be embedded in a simple GUT group, $SU(5)$ or $SO(10)$ or $E_6$.  We then have:
\beq
A_2 &=& A_3 \quad\mbox{(mod $n$)},
\eeq
for the weaker condition in which the $U(1)_Y$ normalization is considered arbitrary, with the additional stronger condition
\beq
A_1 &=& 5 A_3 \quad\mbox{(mod $n$)}
\eeq
if the gauge couplings are required to unify in the usual way.
The value $\rho_{\rm GS} = 0$ corresponds to the notable special case in which the Green-Schwarz mechanism does not play a role, in which case
\beq
A_1 = A_2 = A_3 = 0 \quad\mbox{(mod $n$)} .
\eeq

From the requirement that the MSSM superpotential couplings are allowed, we have
$
z_{\overline u} = -z_{H_u} - z_q + 2r,
$
and
$
z_{\overline d} = -z_{H_d} - z_{q} + 2r,
$ 
and
$z_{\overline e} = -z_{H_d} - z_{\ell} + 2r.$
Furthermore, $z_{H_d}$ is determined either by $z_{H_d} = -z_{H_u} - z_X - z_Y + 2r$ for base model $\text{B}_{\rm \RomanNumeralCaps{1}}$ and its extensions, or $z_{H_d} = -z_{H_u} - 2 z_X + 2r$ for base models $\text{B}_{\rm \RomanNumeralCaps{2}}$ and $\text{B}_{\rm \RomanNumeralCaps{4}}$ and their extensions, or $z_{H_d} = -z_{H_u} - 2 z_Y + 2r$ for base model $\text{B}_{\rm \RomanNumeralCaps{3}}$ and its extensions. We also require that neutrino masses are provided by the superpotential version of the Weinberg operator $(H_u\ell)(H_u\ell)$. Equivalently from the point of view of anomalies and low-energy phenomenology, we can introduce three gauge-singlet neutrino superfields $\overline \nu$ which have superpotential terms
\beq
W_{\rm seesaw} &=& \frac{1}{2} M_{\overline\nu}\, \overline\nu\hspace{1pt} \overline\nu + 
y_{\overline\nu} H_u \ell \overline\nu,
\eeq
and then we have that $z_{\overline \nu} = 2r - z_\ell - z_{H_u}$ must either be $r$, or perhaps $r+n/2$ if $n$ is even.

If we temporarily neglect the $M_{\overline \nu}$ term, the superpotential $y_u H_u q \overline u  - y_d H_d q \overline d - y_e H_d l \overline e + y_{\overline\nu} H_u \ell \overline\nu$ is invariant under two anomaly-free $U(1)$ symmetries with integer charges. One is $U(1)_{6Y}$ where $Y$ is the weak hypercharge, and the other can be taken to be either $3 (B-L)$ or $2 T^3_R = 2 Y - (B-L)$. The charges were listed in Table \ref{tab:MSSM_PQ}. It is apparent that we can always redefine the $Z_n^R$ charges by adding a multiple of  $6Y$ in order to make $q=0$, as a convention without loss of generality.  Now, if $n$ is even, we can add $(n/2) 2 T^3_R$ to every $z_\Phi$ to obtain another
discrete symmetry which has the same anomalies $A_{1,2,3}$, and 
differs from the original discrete symmetry only by adding charges $n/2$ (mod $n$)
for the fields $H_u$, $H_d$, $\overline u$, $\overline d$, $\overline e$, and $\overline \nu$. This amounts to simply toggling whether matter-parity-violating terms are allowed or forbidden by $Z_n^R$. We therefore always take $z_{\overline\nu} = r$, rather than the alternative $z_{\overline\nu} = r + n/2$ in the case that $n$ is even, with the understanding that there is always a corresponding discrete symmetry that can be obtained in this way. In the case of odd $n$, one can also always choose to impose matter parity, or not.

All of the charges and $r$ are defined (mod $n$). If $n$ is even, then any $Z_n^R$ symmetry with $r=n/2$ is equivalent to the corresponding $Z_n$ non-$R$ symmetry from the low-energy point of view, because the two symmetries forbid and allow precisely the same Lagrangian terms. Also, if $r > n/2$, then one can replace $(z_\Phi, r) \rightarrow (n - z_\Phi, n-r)$ to obtain an equivalent version of the discrete symmetry. Therefore, for the $R$-symmetries, one can take $0 < r < n/2$ without loss of generality. 

For any given discrete symmetry, equivalent versions of it can be obtained by multiplying all of the charges (including $r$) by a common integer relatively prime to $n$, and then taking the results (mod $n$). We use this to eliminate redundant low-energy descriptions of a given discrete symmetry. We also reduce every discrete symmetry to the smallest $n$ that describes it, by removing any common factors from the list of charges $z_{\Phi}$ and $r$.

It follows that the $Z_n$ or $Z_n^R$ charges of the base model chiral fields can always be written, by a conventional choice, in terms of only two independent charges $x$ and $h$ (and one binary choice $m=0,1$ for model $\text{B}_{\rm \RomanNumeralCaps{4}}$ when $n$ is even and $r \not= 0$), as summarized in Table \ref{tab:Zncharges}.
\begin{table}
\begin{center}
\begin{minipage}[]{0.98\linewidth}\caption{Charges for discrete symmetry ($Z_n$ for $r=0$, or $Z_n^R$ for non-zero integer $0 < r < n/2$), for chiral superfields in the base models, in terms of two integers $h$ and $x$, in the convention adopted here. In the case of $\text{B}_{\rm \RomanNumeralCaps{4}}$, $m=0$ for odd $n$, while $m=0,1$ for even $n$ and $r \not=0$, and $m=1$ for even $n$ if $r=0$. If vectorlike pairs are added to the model, their net discrete symmetry charges can be obtained from those for $X$ and/or $Y$, given the mass term as specified in eq.~(\ref{mass_possibilitiesNONREN}) for TeV scale masses, or (\ref{mass_possibilitiesREN}) for intermediate scale masses. \label{tab:Zncharges}}
\end{minipage}

\vspace{0.3cm}

\begin{tabular}{|c|c|c | c | c | c | c | c | c | c | c |}
\hline
~~ & ~$X$~ & ~$Y$~ & ~$H_u$~ & ~$H_d$~ & ~$q$~ & ~$\ell$~ & ~$\overline{u}$~ 
& ~$\overline{d}$~ & ~$\overline{e}$~ & ~$\overline\nu$~ 
\\
\hline\hline
~$\text{B}_{\rm \RomanNumeralCaps{1}}$~ & ~$x$~ & ~$2r\!-\!3x$~ & ~$h$~ & ~$2x\!-\!h$~ & ~$0$~ & ~$r\!-\!h$~ & ~$2r\!-\!h$~ & ~$h\!-\!2x\!+\!2r$~ & ~$2h\!-\!2x\!+\!r$~ & ~$r$~
\\
\hline
~$\text{B}_{\rm \RomanNumeralCaps{2}}$~ & ~$x$~ & ~$2r\!-\!3x$~ & ~$h$~ & ~$2r\!-\!2x\!-\!h$~ & ~$0$~ & ~$r\!-\!h$~ & ~$2r\!-\!h$~ & ~$h\!+\!2x$~ & ~$2h\!+\!2x\!-\!r$~ & ~$r$~
\\
\hline
~$\text{B}_{\rm \RomanNumeralCaps{3}}$~ & ~$x$~ & ~$2r\!-\!3x$~ & ~$h$~ & ~$6x\!-\!h\!-\!2r$~ & ~$0$~ & ~$r\!-\!h$~ & ~$2r\!-\!h$~ & ~$h\!-\!6x\!+\!4r$~ & ~$2h\!-\!6x\!+\!3r$~ & ~$r$~
\\
\hline
~$\text{B}_{\rm \RomanNumeralCaps{4}}$~ & ~$x$~ & ~$r\!-\!x\!+\!mn/2$~ & ~$h$~ & ~$2r\!-\!2x\!-\!h$~ & ~$0$~ & ~$r\!-\!h$~ & ~$2r\!-\!h$~ & ~$h\!+\!2x$~ & ~$2h\!+\!2x\!-\!r$~ & ~$r$~
\\
\hline
\end{tabular}
\end{center}
\end{table}
Therefore, for the sake of brevity, we will not list discrete symmetry charges for $H_d$ and the MSSM quark and lepton fields below, since they can be obtained from Table \ref{tab:Zncharges}. For extensions of the base models, the total discrete symmetry charges for each vectorlike pair can be obtained from those for $X$ and/or $Y$, given the mass term as specified in eq.~(\ref{mass_possibilitiesNONREN}) for TeV scale masses, or (\ref{mass_possibilitiesREN}) for intermediate scale masses. 

Gaugino masses, and holomorphic soft-supersymmetry breaking terms corresponding to terms in the superpotential, are forbidden by unbroken discrete $R$-symmetries. Since they are a phenomenological necessity, a discrete $R$-symmetry must be spontaneously broken. This can be done with a spurion with $Z_n^R$ charge 0, whose $F$-term component (with $Z_n^R$ charge equal to $-2r$) obtains a VEV. In the low-energy theory, this will give rise to the usual MSSM supersymmetry breaking terms including gaugino masses, as well as PQ-violating terms of the form in eq.~(\ref{eq:XaYbsoft}). As already noted in the Introduction, the resulting axion tadpole contribution is parametrically of the same order as the corresponding superpotential terms in eq.~(\ref{eq:XaYbsuperpotential}), and therefore has the same order-of-magnitude effect on $\theta_{\text{eff}}$.

\subsection{Discrete non-$R$ symmetries $Z_n$\label{subsec:discretenonR}}

In Table \ref{tab:Znbase}, we give a complete list of the possible distinct non-$R$ $(r=0)$ $Z_n$ symmetries for the base models, consistent with a bare Weinberg operator $(H_u\ell)(H_u\ell)$ (or equivalently $\overline\nu\hspace{1pt}\overline\nu$) for neutrino masses and the anomaly constraint $A_2 = A_3$ (mod $n$). In the second column of the table, we show the resulting suppression factor $p$. We see that there is a large selection of $Z_n$ symmetries that satisfy the weaker anomaly-free constraint. These include, for example, a $Z_{22}$ symmetry with $p=11$ for base model $\text{B}_{\rm \RomanNumeralCaps{4}}$, equivalent to one previously proposed and studied in ref.~\cite{Babu:2002ic}. In our Table \ref{tab:Znbase}, this symmetry has $(X,Y,H_u)$ charges $(2, 6k+3, 4k-2)$ with $k=1$; the charges listed in ref.~\cite{Babu:2002ic} are different but the symmetry is equivalent in the sense discussed above. There are many other discrete symmetries with equal or greater suppression $p$. For example, with the same base model $\text{B}_{\rm \RomanNumeralCaps{4}}$, we see that there are also $Z_{16}$ and $Z_{20}$ symmetries that provides protection up to $p=10$ and 12, respectively.
\begin{table}[p]
\caption{Exhaustive list of distinct $Z_n$ symmetries that satisfy the discrete anomaly cancellation constraint $A_2 = A_3$ (mod $n$) for the MSSM base models including the bare Weinberg operator $(H_u\ell)(H_u\ell)$ for neutrino masses. The meaning of $p$ is that the lowest dimension PQ-violating superpotential term(s) allowed by the discrete symmetry and involving only $X$ and $Y$ are of the form $X^j Y^{p-j}$. In this table, $k$ is any integer, and $m=0,1,2$. The middle three columns are the $Z_n$ charges for $X$, $Y$, and $H_u$. The $H_d$ and quark and lepton superfield charges are then determined as in Table \ref{tab:Zncharges}, with the charges of $q$ and $H_u\ell$ set equal to 0 without loss of generality, as discussed in the text.  Despite this infinite number of possible $Z_n$ symmetries, the only cases that satisfy the additional condition $A_1 = 5 A_3$ (mod $n$) and have adequate suppression of PQ-violating terms $(p\geq7$) are $Z_{36}$  for model $\text{B}_{\rm \RomanNumeralCaps{3}}$ with $p = k = 12$, and the $Z_{36}$ for model $\text{B}_{\rm \RomanNumeralCaps{4}}$ with $p = 8$ and $k=1$. 
\label{tab:Znbase}}
\begin{center}
\begin{tabular}{|c c |c||c|c|c||c|c|c|c|}
\hline
Model & $n$& ~$p$~ & ~$X$~ & $Y$ & $H_u$ &  $A_2\!=\!A_3$ & $A_1$ 
\\
\hline\hline
$\text{B}_{\rm \RomanNumeralCaps{1}}$  
&
${3k\!+\!1} $ & $k\!+\!3$ & $1$ & $-3$ & $\phantom{-}k\!+\!3$ &  $-6$ & $-18$ 
\\
\cline{2-8}
~~
&
${3k\!+\!2} $ & $k\!+\!2$ &$1$ & $-3$ & $-k\!+\!2$ &  $-6$ & $-18$ 
\\
\cline{2-8}
&
${9k} $ & $k$ & $3$ & $-9$ & ~$8$~ & $-18$ & $-54$ 
\\
\cline{2-8}
&
${9k\!+\!3} $ & $k\!+\!3$ & $3$ & $-9$ & ~$8$~ &  $-18$ & $-54$ 
\\
\cline{2-8}
&
${9k\!+\!6} $ & $k\!+\!2$ & $3$ & $-9$ & ~$8$~ &  $-18$ & $-54$ 
\\
\hline
\hline
$\text{B}_{\rm \RomanNumeralCaps{2}}$  
& ${3k\!+\!1} $ & $k\!+\!3$ & $1$ & $-3$ & $-k\!-\!3$ &  $\phantom{-}6$ & $\phantom{-}18$ 
\\
\cline{2-8}
~~
& ${3k\!+\!2} $ & $k\!+\!2$ & $1$ & $-3$ & $\phantom{-}k\!-\!2$ & $\phantom{-}6$ & $\phantom{-}18$ 
\\
\cline{2-8}
~~
& ${9k} $ & $k$ & $3$ & $-9$ & $\phantom{-}3k\!-\!8$ &  $\phantom{-}18$ & $\phantom{-}54$ 
\\
\cline{2-8}
~~
&
${9k\!+\!3} $ & $k\!+\!3$ & $3$ & $-9$ & $\phantom{-}3k\!-\!7$ &  $\phantom{-}18$ & $\phantom{-}54$ 
\\
\cline{2-8}
~~
&
${9k\!+\!6} $ & $k\!+\!2$ & $3$ & $-9$ & $-3k\!-\!10$ &  $\phantom{-}18$ & $\phantom{-}54$ 
\\
\hline
\hline
$\text{B}_{\rm \RomanNumeralCaps{3}}$
&
${3k} $ & $k$ & $1$ & $-3$ & $8\!+\!m k$ & $-18$ & $-54$ 
\\
\cline{2-8}
&
${3k\!+\!1} $ & $k\!+\!3$ & $1$ & $-3$ & ~$8$~ &  $-18$ & $-54$ 
\\
\cline{2-8}
&
${3k\!+\!2} $ & $k\!+\!2$ & $1$ & $-3$ & ~$8$~ & $-18$ & $-54$ 
\\
\cline{2-8}
&
${9k\!+\!3} $ & $k\!+\!3$ & $3$ & $-9$ & $3k\!+\!25$ &  $-54$ & $-162$ 
\\
\cline{2-8}
&
${9k\!+\!6} $ & $k\!+\!2$ & $3$ & $-9$ & $3k\!+\!26$ & $-54$ & $-162$ 
\\
\hline
\hline
$\text{B}_{\rm \RomanNumeralCaps{4}}$
&
${12k\!+\!2} $ & ~$6k\!+\!1$~ & $1$ & $6k$ & $\phantom{-}4k\!-\!2$ &  $\phantom{-}6$ & $18$ 
\\
\cline{2-8}
&
${12k\!+\!2} $ & ~$6k\!+\!1$~ & $2$ & $6k\!-\!1$ & $-4k\!-\!6$ &  $\phantom{-}12$ & $36$ 
\\
\cline{2-8}
&
${12k\!+\!4} $ & $6k\!+\!4$ & $1$ & $6k\!+\!1$ & $-4k\!-\!4$ & $\phantom{-}6$ & $18$ 
\\
\cline{2-8}
&
${12k\!+\!8} $ & $6k\!+\!6$ & $1$ & $6k\!+\!3$ & $4k$ &  $\phantom{-}6$ & $18$ 
\\
\cline{2-8}
&
${12k\!+\!10} $ & $6k\!+\!5$ & $1$ & $6k\!+\!4$ & $-4k\!-\!6$ &  $\phantom{-}6$ & $18$ 
\\
\cline{2-8}
&
${12k\!+\!10} $ & $6k\!+\!5$ & $2$ & $6k\!+\!3$ & $\phantom{-}4k\!-\!2$ & $\phantom{-}12$ & $36$ 
\\
\cline{2-8}
&
${36k} $ & $6k\!+\!2$ & $3$ & $18k\!-\!3$ & $12k\!-\!8$ & $\phantom{-}18$ & $54$ 
\\
\cline{2-8}
&
${36k\!+\!6} $ & $6k\!+\!1$ & $3$ & $18k$ & $-12k\!-\!10$ & $\phantom{-}18$ & $54$ 
\\
\cline{2-8}
&
${36k\!+\!6} $ & $6k\!+\!1$ & $6$ & $18k\!-\!3$ & $\phantom{-}12k\!-\!14$ & $\phantom{-}36$ & $108$ 
\\
\cline{2-8}
&
${36k\!+\!12} $ & $6k\!+\!4$ & $3$ & $18k\!+\!3$ & $12k\!-\!4$ & $\phantom{-}18$ & $54$ 
\\
\cline{2-8}
&
${36k\!+\!18} $ & $6k\!+\!3$ & $3$ & $18k\!+\!6$ & $12k\!-\!2$ & $\phantom{-}18$ & $54$ 
\\
\cline{2-8}
&
${36k\!+\!18} $ & $6k\!+\!3$ & $6$ & $18k\!+\!3$ & $12k\!-\!10$ & $\phantom{-}36$ & $108$ 
\\
\cline{2-8}
&
${36k\!+\!24} $ & $6k\!+\!6$ & $3$ & $18k\!+\!9$ & $-12k\!-\!16$ & $\phantom{-}18$ & $54$ 
\\
\cline{2-8}
&
${36k\!+\!30} $ & $6k\!+\!5$ & $3$ & $18k\!+\!12$ & $12k\!+\!2$ & $\phantom{-}18$ & $54$ 
\\
\cline{2-8}
&
~${36k\!+\!30}$~ & $6k\!+\!5$ & $6$ & $18k\!+\!9$ & $-12k\!-\!26$ & $\phantom{-}36$ & $108$ 
\\
\hline
\end{tabular}
\end{center}
\end{table}

Despite this infinite number of possible $Z_n$ symmetries, the only base model cases that satisfy the additional condition $A_1 = 5 A_3$ (mod $n$) and have adequate suppression $(p \geq 7)$ of PQ-violating terms in the superpotential are $Z_{36}$  for model $\text{B}_{\rm \RomanNumeralCaps{3}}$ with $p = 12$ (with three distinct possibilities for the $H_u$ charge, labeled in the table by $m=0,1,2$), and the $Z_{36}$ for model $\text{B}_{\rm \RomanNumeralCaps{4}}$ with $p = 8$. 
In both cases, the Green-Schwarz mechanism is needed, as $\rho_{\text{GS}} = A_3 = 18$ (mod $36$).
 
If there are extra vectorlike chiral supermultiplets that owe their masses to $X$ and $Y$, 
then there are so many possible discrete symmetries that we cannot provide an exhaustive list. Some examples are given in Tables \ref{tab:Zn123extras} and \ref{tab:Zn4extras}, in which we have limited ourselves to cases with at most one ${\bf 5}+ {\bf \overline 5}$ or  ${\bf 10}+ {\bf \overline {10}}$ of $SU(5)$ at the TeV scale, and in which both anomaly-free constraints 
$A_2 = A_3$ (mod $n$) and $A_1 = 5 A_3$ (mod $n$) are required.
\begin{table}[p]
\caption{Some examples of non-$R$ discrete symmetries $Z_{n}$ satisfying the anomaly cancellation conditions $A_2 = A_3$ (mod $n$) and $A_1 = 5 A_3$ (mod $n$), obtained from base models
$\text{B}_{\rm \RomanNumeralCaps{1}}$, 
$\text{B}_{\rm \RomanNumeralCaps{2}}$, 
or
$\text{B}_{\rm \RomanNumeralCaps{3}}$
by adding up to one 
${\bf 5}+ {\bf \overline 5}$ or 
${\bf 10}+ {\bf \overline {10}}$ 
of $SU(5)$ at the TeV scale. The suppression of PQ violation $p$ is defined so that the lowest dimension PQ-violating superpotential term(s) allowed by the discrete symmetry and involving only $X$ and $Y$ and are of the form $X^j Y^{p-j}$. In all cases in this table, the $X$ and $Y$ charges are respectively $1$ and $-3$, and the $H_u$ charges are as listed with $m = 0,1,2$.  The $H_d$ and quark and lepton superfield charges are then determined as in Table \ref{tab:Zncharges}. The last three columns give the Green-Schwarz contribution $\rho_{\text{GS}}$ to the $Z_n$ anomalies and the PQ-QCD-QCD and PQ-EM-EM anomalies $N$ and $E$. The five cases with $\rho_{\text{GS}} = 0$ do not require a Green-Schwarz mechanism. However, of those, the two cases with $N=0$ have no PQ-QCD-QCD anomaly and therefore have an axion-like particle but do not provide a solution to the strong CP problem.
\label{tab:Zn123extras}}
\begin{center}
\begin{tabular}{| l  l | c r | 
c || c || c | c |}
\hline
~Base~&~Extension~ & ~~$n$~~ & 
~$p$~ & ~$H_u$~ & ~$\rho_{\text{GS}}$~ & ~$N$~ & ~$3E$~
\\[1pt]
\hline
\hline
~$\text{B}_{\rm \RomanNumeralCaps{1}}$~
  & ~$X^2 D \overline D + Y^2 L \overline L$~
  & $20$ & 
  ~$8$~ & ~$12$~ & $12$ & $4$ & $2$
\\
\cline{2-8}
~~
  & ~$Y^2 D \overline D + X^2 L \overline L$~
  & $36$ & 
  ~$12$~ & $12m$ & $0$ & $0$ & $18$
\\
\cline{2-8}
~~
  & ~$X^2 D \overline D + XY L \overline L$~
  & $36$ & 
  ~$12$~ & ~$4\!+\!12m$~ & $28$ & $4$ & $14$
\\
\cline{2-8}
~~
  & ~$Y^2 Q \overline Q + XY U \overline U + X^2 E \overline E$~
  & $36$ & 
  ~$12$~ & ~$4\!+\!12m$~ & $8$ & ~$-4$~ & $-14$
\\
\cline{2-8}
~~
  & ~$X^2 Q \overline Q + Y^2 U \overline U + Y^2 E \overline E$~
  & $20$ & 
  ~$8$~ & ~$0$~ & $16$ & $2$ & $-14$
\\
\cline{3-8}
~~
  & 
  & $30$ & 
  ~$10$~ & ~$10m$~ & $26$ & $2$ & $-14$
\\
\cline{3-8}
~~
  & 
  & $60$ & 
  ~$20$~ & ~$20m$~ & $56$ & $2$ & $-14$
\\
\hline
\hline
~$\text{B}_{\rm \RomanNumeralCaps{2}}$~ 
  & ~$Y^2 D \overline D + X^2 L \overline L$~
  & $20$ & 
  ~$8$~ & ~$8$~ & $12$ & $6$ & $18$
\\
\cline{2-8}
~~ 
  & ~$XY D \overline D + X^2 L \overline L$~
  & $36$ & 
  ~$12$~ & ~$8\!+\!12m$~ & $8$ & $4$ & $14$
\\
\cline{2-8}
 & ~$Y^2 Q \overline Q + XY U \overline U + X^2 E \overline  E$~
 & $20$ & 
 ~$8$~ & $12$ & 0 & 10 & 50
\\
\cline{2-8}
~~ 
  & ~$XY Q \overline Q + Y^2 U \overline U + X^2 E \overline  E$~
  & $36$ & 
  ~$12$~ & ~$8\!+\!12m$~ & $16$ & $8$ & $46$
\\
\cline{2-8}
~~ 
  & ~$Y^2 Q \overline Q + X^2 U \overline U + X^2 E \overline  E$~
  & $20$ & 
  ~$8$~ & ~$0$~ & $16$ & $8$ & $34$
\\
\cline{3-8}
~~ 
  & 
  & $30$ & 
  ~$10$~ & ~$10m$~ & $16$ & $8$ & $34$
\\
\cline{3-8}
~~
  & 
  & $60$ & 
  ~$20$~ & ~$20m$~ & $16$ & $8$ & $34$
\\
\hline\hline
~$\text{B}_{\rm \RomanNumeralCaps{3}}$~ 
 & ~none~
 & $36$ & 
 ~$12$~ & $8\!+\!12m$ & $18$ & 3 & 18
\\
\cline{2-8}
~~ 
 & ~$X^2 D \overline D + XY L \overline L$~
 & $20$ & 
 ~$8$~ & $16$ &  0 & $\frac{10}{3}$ & $\frac{50}{3}$
\\[2pt]
\cline{2-8}
~~ 
 & ~$XY D \overline D + Y^2 L \overline L$~
 & $20$ & 
 ~$8$~ & $16$ & 4 & $\frac{8}{3}$ & $\frac{34}{3}$
\\[2pt]
\cline{2-8}
~~ 
 & ~$X^2 D \overline D + Y^2 L \overline L$~
 & $28$ & 
 ~$12$~ & $20$ & 8 & $\frac{10}{3}$ & $\frac{38}{3}$
\\[2pt]
\cline{2-8}
~~ 
 & ~$X^2 D \overline D + X^2 L \overline L$~
 & $36$ & 
 ~$12$~ & $8\!+\!12m$ & 16 & $\frac{10}{3}$ & $\frac{62}{3}$
\\[2pt]
\cline{2-8}
~~
  & ~$X^2 Q \overline Q + XY U \overline U + XY E \overline  E$~~
  & $20$ & 
  ~$8$~ & $0$ & 0 & $\frac{10}{3}$ & $\frac{50}{3}$
\\[2pt]
\cline{2-8}
~~
  & ~$X^2 Q \overline Q + X^2 U \overline U + X^2 E \overline  E$~~
  & $36$ & 
  ~$12$~ & $8\!+\!12m$ & 12 & $4$ & $26$
\\
\cline{2-8}
~~
  & ~$Y^2 Q \overline Q + Y^2 U \overline U + Y^2 E \overline  E$~
  & $36$ & 
  ~$12$~ & $8\!+\!12m$ & 0 & $0$ & $-6$
\\
\cline{2-8}
~~
  & ~$X^2 Q \overline Q + X^2 U \overline U + Y^2 E \overline  E$~
  & $21$ & 
  ~$7$~ & $1\!+\!7m$ & 18 & $4$ & 18
\\
\cline{3-8}
~~
  & 
  & $28$ & 
  ~$12$~ & $8$ & 4 & $4$ & 18
\\
\cline{3-8}
~~
  & ~
  & $42$ & 
  ~$14$~ & $8\!+\!14m$ & 18 & $4$ & 18
\\
\cline{2-8}
~~
  & ~$X^2 Q \overline Q + X^2 U \overline U + XY E \overline  E$~
  & $20$ & 
  ~$8$~ & $8$ & 16 & $4$ & 22
\\
\cline{3-8}
~~
  & 
  & $30$ & 
  ~$10$~ & $8\!+\!10m$ & 6 & $4$ & 22
\\
\cline{3-8}
~~
  & 
  & $60$ & 
  ~$20$~ & $8\!+\!20m$ & 36 & $4$ & 22
\\
\hline
\end{tabular}
\end{center}
\end{table}
The order $n$, the $Z_n$ MSSM charges, and the suppression $p$ for these symmetries are the same
as found in Table \ref{tab:Znbase}, but the anomalies $N$ and $E$ that enter into the low-energy axion phenomenology are different because of the contributions of the vectorlike fields. We also note that there are several examples that have $\rho_{\text{GS}}= 0$ and therefore do not require the Green-Schwarz mechanism. The examples with $\rho_{\text{GS}}= 0$ and $N\not=0$ that have the smallest order $n$ include $Z_{20}$ for extensions of $\text{B}_{\rm \RomanNumeralCaps{2}}$ and $\text{B}_{\rm \RomanNumeralCaps{3}}$, and $Z_{12}$ and $Z_{36}$ for extensions of $\text{B}_{\rm \RomanNumeralCaps{4}}$. Each of these has $p=8$, which provides sufficient quality for the PQ symmetry for $f_A \lsim 4 \times 10^{9}$ GeV if the PQ-violating terms have order one couplings
with generic phases, and for larger $f_A$ otherwise. We note that when $n$ is not too large,
$\rho_{\text{GS}}\not=0$ tends to allow for higher $p$.

\begin{table}[t]
\caption{Some examples of non-$R$ discrete symmetries $Z_{n}$ satisfying the anomaly cancellation conditions $A_2 = A_3$ (mod $n$) and $A_1 = 5 A_3$ (mod $n$), obtained from base model $\text{B}_{\rm \RomanNumeralCaps{4}}$ by adding up to one ${\bf 5}+ {\bf \overline 5}$ or ${\bf 10}+ {\bf \overline {10}}$ of $SU(5)$ at the TeV scale. The meaning of $p$ is that the lowest dimension PQ-violating superpotential term(s) allowed by the discrete symmetry and involving only $X$ and $Y$ and are of the form $X^j Y^{p-j}$. The next three columns list the charges of $X$, $Y$, and $H_u$, with $m = 0,1,2$. The remaining MSSM charges are determined in each case as in Table \ref{tab:Zncharges}. The last three columns give the Green-Schwarz contribution $\rho_{\text{GS}}$ to the $Z_n$ anomalies and the PQ-QCD-QCD and PQ-EM-EM anomalies $N$ and $E$. The cases with $\rho_{\text{GS}} = 0$ do not require a Green-Schwarz mechanism. However, the last two examples shown, with $N=0$, have no PQ-QCD-QCD anomaly and therefore have an axion-like particle but do not provide a solution to the strong CP problem. 
\\{}
\label{tab:Zn4extras}}
\begin{center}
\begin{tabular}{| l  l | c r || c | c | c || c | c | c |}
\hline
~Base~&~Extension~ & ~$n$~ & ~$p$~ & ~$X$~ & ~$Y$~ & ~$H_u$~ & ~$\rho_{\text{GS}}$~ & ~$N$~ & ~$3E$~
\\[1pt]
\hline\hline
~$\text{B}_{\rm \RomanNumeralCaps{4}}$
 & ~none~
 & ~$36$~ & ~$8$~ & $3$ & $15$ & $4$ & 18 & 3 & 18
\\
\cline{2-10}
~
 & ~$XY D \overline D + Y^2 L \overline L$~
 & $12$ & ~$8$~ & $1$ & $5$ & $4m$ & $0$ & 3 & 24
\\
\cline{2-10}
 & ~$XY D \overline D + X^2 L \overline L$~
 & $36$ & ~$8$~ & $3$ & $15$ & $8$ &  0 & 3 & 12
\\
\cline{3-10}
 & ~~
 & $72$ & ~$14$~ & $3$ & $33$ & $2$ & 54 & 3 & 12
\\
\cline{2-10}
 & ~$XY D \overline D + XY L \overline L$~
 & $36$ & ~$8$~ & $3$ & $15$ & $4$ &  0 & 3 & 18
\\
\cline{2-10}
~
 & ~$X^2 D \overline D + X^2 L \overline L$~
 & $36$ & ~$8$~ & $3$ & $15$ & $4$ & 12 & 2 & 10
\\
\cline{2-10}
~
 & ~$Y^2 D \overline D + X^2 L \overline L$~
 & $12$ & ~$8$~ & $1$ & $5$ & $4m$ & 8 & 4 & 14
\\
\cline{3-10}
~
 & 
 & $18$ & ~$9$~ & $1$ & $8$ & $2\!+\!6m$ & 8 & 4 & 14
\\
\cline{3-10}
~
 & 
 & $36$ & ~$20$~ & $1$ & $17$ & ~$8\!+\!12m$~ & 8 & 4 & 14
\\

\cline{2-10}
~
 & ~$X^2 D \overline D + XY L \overline L$~
 & $12$ & ~$8$~ & $1$ & $5$ & $4m$ & 4 & 2 & 16
\\
\cline{3-10}
~
 & 
 & $14$ &  ~$7$~ & $1$ & $6$ & $5$ &4 & 2 & 16
\\
\cline{3-10}
~
 & 
 & $16$ & ~$10$~ & $1$ & $7$ & $6$ & 4 & 2 & 16
\\
\cline{3-10}
~
 & 
 & $18$ & ~$9$~ & $1$ & $8$ & $1\!+\!6m$ & 4 & 2 & 16
\\
\cline{3-10}
~
 & 
 & $20$ & ~$12$~ & $1$ & $9$ & $8$ & 4 & 2 & 16
\\
\cline{3-10}
~
 & 
 & $22$ & ~$11$~ & $1$ & $10$ & $9$ & 4 & 2 & 16
\\
\cline{3-10}
~
 & 
 & $24$ & ~$14$~ & $1$ & $11$ & $2\!+\!8m$ & 4 & 2 & 16
\\
\cline{3-10}
~
 & 
 & $28$ & ~$16$~ & $1$ & $13$ & $12$ & 4 & 2 & 16
\\
\cline{3-10}
~
 & 
 & $32$ & ~$18$~ & $1$ & $15$ & $14$ & 4 & 2 & 16
\\
\cline{3-10}
~
 & 
 & $36$ & ~$20$~ & $1$ & $17$ & $4\!+\!12m$ & 4 & 2 & 16
\\
\cline{2-10}
 & ~$Y^2 Q \overline Q + Y^2 U \overline U + Y^2 E \overline E$~
 & $36$ & ~$8$~ & $3$ & $15$ & $4$ & 0 & 6 & $42$
\\
\cline{2-10}
 & ~$X^2 Q \overline Q + X^2 U \overline U + X^2 E \overline E$~
 & $36$ & ~$8$~ & $3$ & $15$ & $4$ & 0 & 0 & $-6$
\\
\cline{2-10}
~ 
 & ~$X^2 Q \overline Q + X^2 U \overline U + XY E \overline E$~
 & $14$ & ~$7$~ & $1$ & $6$ & $2$ & 0 & 0 & 0
 \\
\hline
\end{tabular}
\end{center}
\end{table}
Also shown in Tables \ref{tab:Zn123extras} and \ref{tab:Zn4extras} are some examples that have $\rho_{\text{GS}}= 0$ but also $N=0$, so that there is no PQ-QCD-QCD anomaly. These examples will have a light axion-like particle (ALP), but will not provide a solution to the strong CP problem.  One peculiar case is that of base model $\text{B}_{\rm \RomanNumeralCaps{4}}$ extended by $X^2 Q \overline Q + X^2 U \overline U + XY E \overline E$. This particular vectorlike content implies a completely anomaly-free PQ symmetry with $E=N=0$, resulting in an infinite number of different $Z_n$ symmetries that protect it; only the first one (with $n=14$) is shown. However, all cases that give
$N=0$ are moot for our main purpose in this paper, as there is no reason to forbid high-dimension PQ-violating terms if the strong CP problem cannot be solved anyway.

The examples shown in Tables \ref{tab:Zn123extras} and \ref{tab:Zn4extras} are only some of the
many possibilities. There are similar $Z_n$ symmetries available for models that have other
combinations of vectorlike supermultiplets both at the TeV and intermediate scales. This is also true of the models with quixes, as for example in the last two rows of Table \ref{tab:NDW=1 models with two SU5 pairs}.

\subsection{Discrete $R$-symmetries $Z_n^R$\label{subsec:discreteR}}

Discrete $R$-symmetries provide even more possibilities for an accidental Peccei-Quinn symmetry protected to a high power $p$, including a larger number of cases that satisfy both  $A_2 = A_3$ (mod $n$) and $A_1 = 5 A_3$ (mod $n$). A non-exhaustive selection of such symmetries for the base models is shown in Table \ref{tab:ZnRbasemodels}. One of the discrete $Z_{24}^R$ symmetries had been previously proposed and studied for the MSSM in\footnote{Ref. \cite{Lee:2011dya} imposed a requirement that the discrete symmetry charges for MSSM quark and lepton superfields respect $SU(5)$ invariance, which limits the possibilities to only a few. We do not impose this requirement. We note that all of the symmetries we find can be made consistent with a partial unification with a Pati-Salam $SU(4)_{\rm PS} \times SU(2)_L \times U(1)_R$ \cite{Pati:1974yy} embedding, although this is not always immediately obvious with our discrete charge conventions.} ref.~\cite{Lee:2011dya}, and was found in ref.~\cite{Baer:2018avn,Bae:2019dgg} to extend to base models $\text{B}_{\rm \RomanNumeralCaps{2}}$ and $\text{B}_{\rm \RomanNumeralCaps{3}}$ with suppression $p=10$. There are actually several inequivalent $Z_{24}^R$ symmetries; the one found in
\cite{Lee:2011dya} and applied to $\text{B}_{\rm \RomanNumeralCaps{2}}$ and $\text{B}_{\rm \RomanNumeralCaps{3}}$ in \cite{Baer:2018avn,Bae:2019dgg} is equivalent (after shifting by a multiple of weak hypercharge) to what we have listed in Table \ref{tab:ZnRbasemodels} in the second row under $\text{B}_{\rm \RomanNumeralCaps{2}}$ and
the third row under $\text{B}_{\rm \RomanNumeralCaps{3}}$,  with $k=m=0$ and $\rho_{\text{GS}} = 18$ in each case. It was also pointed out in
\cite{Baer:2018avn,Bae:2019dgg} that when this $Z_{24}^R$ symmetry is realized in
base models  $\text{B}_{\rm \RomanNumeralCaps{1}}$ and $\text{B}_{\rm \RomanNumeralCaps{4}}$,
it only provides suppression $p=7$; we have chosen not to list these among the examples, although suppression $p=7$ may be good enough if $f_A$ is up to $4 \times 10^{9}$ GeV, since it can be promoted to $p=8$ by imposing an extra $Z_2$ symmetry that acts only on $X$ and $Y$. 

For model $\text{B}_{\rm \RomanNumeralCaps{4}}$, there are two distinct $Z_{12}^R$ symmetries that give $p=7$ (or $p=8$ by the trick just mentioned), one of which requires use of the Green-Schwarz mechanism, and one of which does not because $\rho_{\text{GS}} = A_1 = A_2 = A_3 = 0$ (mod 12). For $\rho_{\text{GS}}=0$ with protection up to $p=8$, one also has symmetries with
smallest order $n=54$ for models $\text{B}_{\rm \RomanNumeralCaps{1}}$, $\text{B}_{\rm \RomanNumeralCaps{2}}$ and $\text{B}_{\rm \RomanNumeralCaps{4}}$, 
and with protection up to $p=10$ for $n=54$ for models $\text{B}_{\rm \RomanNumeralCaps{3}}$
and $n=108$ for models $\text{B}_{\rm \RomanNumeralCaps{1}}$
and $\text{B}_{\rm \RomanNumeralCaps{4}}$. There is a plethora of other possibilities beyond those shown in the table.

Adding vectorlike supermultiplets adds to the possibilities for discrete $R$ symmetries.
A few examples are shown in table \ref{tab:examplesRextras}, limited for reasons of brevity to 
only a subset of the cases that have $\rho_{\text{GS}} =0$, so no Green-Schwarz mechanism necessary, and that have at most one $SU(5)$ multiplet at the TeV scale.

\begin{table}[p]
\caption{Some examples of discrete $R$-symmetries $Z_n^R$ that satisfy both discrete anomaly cancellation constraints $A_2 = A_3$ (mod $n$) and $A_1 = 5 A_3$ (mod $n$) for the base models. The superpotential has charge $2r$ and gauginos have charge $r$. The meaning of $p$ is that the lowest dimension PQ-violating superpotential term(s) allowed by the discrete symmetry and involving only $X$ and $Y$ and are of the form $X^j Y^{p-j}$. The next three columns list the charges of $X$, $Y$, and $H_u$ in terms of $k,k'=0,1$, and $m = 0,1,2$. The remaining MSSM charges are determined in each case as in Table \ref{tab:Zncharges}. The last column is the Green-Schwarz contribution $\rho_{\text{GS}}$ to the $Z_n^R$ anomalies. Cases with $\rho_{\text{GS}} = 0$ do not use the Green-Schwarz mechanism for anomaly cancellation. The PQ-QCD-QCD and PQ-EM-EM anomalies are $N=3$ and $E=6$ in each case.\label{tab:ZnRbasemodels}}
\begin{center}
\begin{tabular}{|c c c r||c|c|c||c|}
\hline
Model & $n$ & $r$ & ~$p$~ & ~$X$~ & $Y$ & ~$H_u$~ &  ~$\rho_{GS}$~ 
\\
\hline\hline
$\text{B}_{\rm \RomanNumeralCaps{1}}$  
& $15$ & ~$1$~ & ~$7$~ & $-1$ & $5$ & $3\!+\!5m$ & $12$  
\\
\cline{2-8}
~~
& $20$ & $1$ & ~$8$~ & $9\!+\!10k$ & $-5\!+\!10k$ & $13$ & $12$  
\\
\cline{2-8}
~~
& $23$ & $1$ & ~$9$~ & $4$ & $13$ & $14$ & $5$ 
\\
\cline{2-8}
~~
& $30$ & $1$ & ~$10$~ & $-1$ & $5$ & $3\!+\!10m$ & $12$  
\\
\cline{2-8}
~~
& $35$ & $1$ & ~$11$~ & $4$ & $25$ & $18$ & $17$  
\\
\cline{2-8}
~~
& $44$ & $2$ & ~$12$~ &$19$ & $35$ & $42$ &  $30$ 
\\
\cline{2-8}
~~
& $50$ & $1$ & ~$14$~ &$29$ & $15$ & $23$ &  $32$  
\\
\cline{2-8}
~~ & $54$ & 3 & ~$8$~ &$39$ & $-3$ & $1$ &  $0$ 
\\
\cline{2-8}
~~ & $60$ & 2 & ~$16$~ &$13$ & $25$ & ~$6\!+\!20m$~ &  $54$  
\\
\cline{2-8}
~~ & $108$ & 3 & ~$10$~ &~$21\!+\!18k\!+\!54k'$~ & ~$51\!+\!54k\!+\!54k'$~ & $7\!+\!12k$ &  $0$ 
\\
\hline
\hline
$\text{B}_{\rm \RomanNumeralCaps{2}}$  
& $19$ & 1 & ~$7$~ &$-3$ & $11$ & $0$ &  $1$  
\\
\cline{2-8}
~~ & $24$ & 1 & ~$10$~ &$-1\!+\!12k$ & $5\!+\!12k$ & $1\!+\!8m$ &  $18$  
\\
\cline{2-8}
~~ & $24$ & 1 & ~$10$~ &$5\!+\!12k$ & $11\!+\!12k$ & $1\!+\!8m$ &  $6$ 
\\
\cline{2-8}
~~ & $34$ & 1 & ~$12$~ &$-3$ & $11$ & $-5$ &  $16$ 
\\
\cline{2-8}
~~ & $54$ & 3 & ~$8$~ &$9$ & $33$ & $7$ &  $0$  
\\
\cline{2-8}
~~ & $56$ & 1 & ~$18$~ &$11\!+\!28k$ & $25\!+\!28k$ & $25$ &  $10$  
\\
\hline
\hline
$\text{B}_{\rm \RomanNumeralCaps{3}}$  
& $13$ & 1 & ~$7$~ &$6$ & $-3$ & $2$ &  $8$  
\\
\cline{2-8}
~~ & $20$ & 2 & ~$8$~ &$5$ & $9$ & $6$ &  $14$ 
\\
\cline{2-8}
~~ & $24$ & 1 & ~$10$~ &$5\!+\!12k$ & $11\!+\!12k$ & $1\!+\!8m$ &  $18$ 
\\
\cline{2-8}
~~ & $24$ & 1 & ~$10$~ &$-1\!+\!12k$ & $5\!+\!12k$ & $1\!+\!8m$ &  $6$ 
\\
\cline{2-8}
~~ & $35$ & 1 & ~$11$~ &$25$ & $-3$ & $18$ &  $17$
\\
\cline{2-8}
~~ & $48$ & 1 & ~$12$~ & $19\!+\!24k$ & $41\!-\!24k$ & $1\!+\!16m$ &  $6$  
\\
\cline{2-8}
~~ & $54$ & 3 & ~$10$~ & $5\!+\!12k$ & $-9\!+\!18k$ & $1\!+\!6k\!+\!18m$ &  $0$  
\\
\hline
\hline
$\text{B}_{\rm \RomanNumeralCaps{4}}$  
&
$12$ & $1$ & ~$7$~ &$-4$ & $5\!+\!6k$ & $1\!+\!4m$ &  $0$  
\\
\cline{2-8}
~~
&
$12$ & $1$ & ~$7$~ &$5\!+\!6k$ & $-4$ & $1\!+\!4m$ &  $6$ 
\\
\cline{2-8}
~~
&
$16$ & $1$ & ~$8$~ &$5\!+\!8k$ & $-4$ & $1$ &  $14$  
\\
\cline{2-8}
~~
&
$20$ & $1$ & ~$9$~ &$12$ & $9\!+\!10k$ & $13$ &  $12$  
\\
\cline{2-8}
~~
&
$25$ & $1$ & ~$10$~ &$-3$ & $4$ & $-2$ &  $7$  
\\
\cline{2-8}
~~
&
$28$ & $1$ & ~$11$~ &$\phantom{-}4\!+\!14k$ & $11\!+\!14k'$ & $-3$ & $24$  
\\
\cline{2-8}
~~
&
$32$ & $1$ & ~$12$~ &$\phantom{-}5\!+\!16k$ & $-4$ & $17$ &  $30$ 
\\
\cline{2-8}
~~ & $54$ & $3$ & ~$8$~ &$-9\!+\!18k$ & $39\!+\!9k$ & $1\!+\!6k$ &  $0$  
\\
\cline{2-8}
~~ & $108$ & $3$ & ~$10$~ &$90$ & $21\!+\!54k$ & $7$ &  $0$  
\\
\hline
\end{tabular}
\end{center}
\end{table}

\begin{table}[p]
\caption{Some examples of discrete $R$-symmetries $Z_{n}^R$ that satisfy the anomaly cancellation conditions $A_1 = A_2 = A_3 = 0$ (mod $n$) without the Green-Schwarz mechanism (so $\rho_{\text{GS}}=0$). The superpotential has charge $2r$ and gauginos have charge $r$. The meaning of $p$ is that the lowest dimension PQ-violating superpotential term(s) allowed by the discrete symmetry and involving only $X$ and $Y$ and are of the form $X^j Y^{p-j}$. The next three columns list the charges of $X,Y,H_u$, in terms of $k=0,1$, and $m = 0,1,2$, and $s=0,1,2,3$. The remaining MSSM charges are determined in each case as in Table \ref{tab:Zncharges}. The last two columns give the PQ-QCD-QCD and PQ-EM-EM anomalies $N$ and $E$.  
\label{tab:examplesRextras}}
\begin{center}
\begin{tabular}{|l l|c c r|| c | c | c || c | c |}
\hline
~Base~&~Extension~ & ~$n$~ & ~$r$~ & ~$p$~ & ~$X$~ & ~$Y$~ & ~$H_u$~ & ~$N$~ & ~$3E$~\\[1pt]
\hline\hline
~$\text{B}_{\rm \RomanNumeralCaps{1}}$~
  & ~none~
  & $54$ & $3$ & ~$8$~ & $39$ & ~$-3$~ & $1$ & $3$ & $18$
\\
\cline{2-10}
~~
  & ~$X^2 D \overline D + XY L \overline L$~
  & $22$ & $1$ &  ~$8$~ & $9$ & ~$-3$~ & $9$ & $4$ & $14$
\\
\cline{2-10}
~~
  & ~$XY D \overline D + Y^2 L \overline L$~
  & $24$ & $1$ & ~$8$~ & ~$7\!+\!12k$~ & $5\!+\!12k$ & $7\!+\!8m$ & $2$ & $-2$
\\
\hline
\hline
~$\text{B}_{\rm \RomanNumeralCaps{2}}$~ 
  & ~none~
  & $54$ & $3$ & ~$8$~ & $9$ & ~$33$~ & $7$ & $3$ & $18$
\\
\cline{2-10}
~~ 
  & ~$X^2 D \overline D + Y^2 L \overline L$~
  & $20$ & $2$ & ~$8$~ & $5$ & ~$9$~ & $14$ & $2$ & $34$
\\
\cline{2-10}
~~ 
  & ~$Y^2 D \overline D + X^2 L \overline L$~
  & $32$ & $1$ & ~$10$~ & $3\!+\!16k$ & ~$-7\!+\!16k$~ & $5$ & $6$ & $18$
\\
\cline{2-10}
~~ 
  & ~$Y^2 D \overline D + XY L \overline L$~
  & $108$~ & $6$ & ~$20$~ & $11$ & ~$87$~ & $22\!+\!36m$ & $6$ & $30$
\\
\cline{2-10}
~~ 
  & ~$Y^2 Q \overline Q + X^2 U \overline U + X^2 E \overline E$~
  & $24$ & $1$ & ~$10$~ & $5\!+\!6s$ & ~$11 \!+\! 6s$~ & $5\!+\!8m$ & $8$ & $34$
\\
\cline{2-10}
~~ 
  & ~$Y^2 Q \overline Q + X^2 U \overline U + Y^2 E \overline E$~
  & $56$ & $1$ & ~$18$~ & $11\!+\!14s$ & ~$25 \!+\! 14s$~ & $-3$ & $8$ & $58$
\\
\hline\hline
~$\text{B}_{\rm \RomanNumeralCaps{3}}$~ 
 & ~none~
 & $54$ & $3$ & ~$10$~ & $5\!+\!12k$ & $-9\!+\!18k$ & $1\!+\!6k\!+\!18m$ & 3 & 18
\\
\cline{2-10}
~~ 
 & ~$XY D \overline D + XY L \overline L$~
 & $20$ & $2$ & ~$8$~ & $5$ & $9$ & $6$ &  $\frac{8}{3}$ & $\frac{46}{3}$
\\[2.5pt]
\cline{2-10}
~~ 
 & ~$X^2 D \overline D + X^2 L \overline L$~
 & $24$ & $1$ & ~$8$~ & $9\!+\!12k$ & $-1\!+\!12k$ & $1\!+\!8m$ &  $\frac{10}{3}$ & $\frac{62}{3}$
\\[2.5pt]
\cline{2-10}
~~ 
  & ~$X^2 Q \overline Q + Y^2 U \overline U + X^2 E \overline E$~
  & $24$ & $1$ & ~$10$~ & $5\!+\!6s$ & ~$11 \!+\! 6s$~ & $5\!+\!8m$ & $\frac{8}{3}$ & $\frac{46}{3}$
\\[2.5pt]
\cline{2-10}
~~ 
  & ~$X^2 Q \overline Q + Y^2 U \overline U + Y^2 E \overline E$~
  & $56$ & $1$ & ~$18$~ &$11\!+\!14s$ & ~$25 \!+\! 14s$~ & $-3$ &  $\frac{8}{3}$ & $\frac{22}{3}$
\\[2.5pt]
\hline\hline
~$\text{B}_{\rm \RomanNumeralCaps{4}}$~ 
 & ~none~
 & $12$ & $1$ & ~$7$~ &$-4$ & $5\!+\!6k$ & $1\!+\!4m$ &  3 & 18
\\
\cline{3-10}
~~ 
 & & $54$ & $3$ & ~$8$~ & $-9\!+\!18k$ & $39\!+\!9k$ & $1\!+\!6k$ & 3 & 18
\\
\cline{3-10}
~~ 
 & & $108$~ & $3$ & ~$10$~ & $90$ & $21\!+\!54k$ & $7$ & 3 & 18
\\
\cline{2-10}
~~ 
 & ~$Y^2 D \overline D + X^2 L \overline L$~
 & $12$ & $2$ & ~$8$~ & $5$ & $9$ & $2\!+\!4m$ & 4 & 14
\\
\cline{2-10}
~~ 
 & ~$X^2 D \overline D + Y^2 L \overline L$~
 & $16$ & $1$ & ~$8$~ &$-4$ & $5\!+\!8k$ & $3$ &  2 & 22
\\
\cline{2-10}
~~ 
  & ~$XY Q \overline Q + Y^2 U \overline U + Y^2 E \overline E$~
  & $24$ & $4$ & ~$10$~ & $5$ & ~$-1 \!+\! 12k$~ & $2\!+\!4k\!+\!8m$ & $4$ & $32$
\\
\cline{2-10}
~~ 
  & ~$Y^2 Q \overline Q + X^2 U \overline U + X^2 E \overline E$~
  & $28$ & $1$ & ~$11$~ &$11\!+\!7s$ & $4 \!-\! 7s \!+\! 14k$ & $11$ &  $4$ & $14$
\\
\cline{2-10}
~~ 
  & ~$Y^2 Q \overline Q + XY U \overline U + X^2 E \overline E$~
  & $60$ & $2$ & ~$16$~ &$7$ & ~$-5$~ & $2\!+\!20m$ &  $5$ & $22$
\\
\hline
\end{tabular}
\end{center}
\end{table}

We have checked that for every available pair of anomaly coefficients $N$ and $E$ that determine the
low-energy axion phenomenology as discussed in the next section, one can find a variety of corresponding discrete symmetries. From the point of view of low-energy phenomenology, the exact identity of the discrete symmetry may be of limited interest, since it is not possible to determine whether the discrete symmetry is an $R$-symmetry, or its order $n$, or whether it may have $\rho_{\text{GS}} =0$. The more important point seems to be the existence proof that it is always possible to find such discrete symmetries, so that the global PQ symmetry is consistent. The correlation of $N$, $E$, and the possible presence of TeV-scale vectorlike quarks or leptons could eventually point the way to specific ultraviolet completions. 

\subsection{Impact on baryon number and lepton number violation}

\vspace{-0.2cm}

Recall that in the MSSM there are renormalizable superpotential terms that violate lepton number L and baryon number B, schematically:
\beq
W_{\text{L-violating}} &=& 
H_u \ell 
+ q \ell \overline d 
+ \ell\ell\overline e 
,
\qquad\quad
W_{\text{B-violating}} \>=\>  \overline u\hspace{0.7pt} \overline d\hspace{0.7pt} \overline d
.
\eeq
\vspace{-0.75cm}

\noindent
Taken together, these would predict very rapid proton decay. The most common way of avoiding this is to impose the $Z_2$ matter parity $= (-1)^{3(\text{B}-\text{L})}$ discrete symmetry  (or equivalently $R$-parity) \cite{Farrar:1978xj}-\cite{Weinberg:1981wj}. There are also non-renormalizable operators that suppressed by the cutoff scale but violate both B and L, and so could directly mediate proton decay in violation of current bounds:
\beq
W &=& \frac{1}{M_P} qqq\ell + 
\frac{1}{M_P} 
\overline u\hspace{0.7pt} \overline u\hspace{0.7pt} 
\overline d\hspace{0.7pt} \overline e
.
\eeq
\vspace{-0.75cm}

\noindent
As was noted in \cite{Babu:2002ic} for the $Z_{22}$ discrete symmetry example found there, and in \cite{Lee:2011dya} for their $Z_{24}^R$ symmetry, the discrete symmetry that protects the PQ symmetry can also help by forbidding these dangerous baryon number and lepton number violating operators.

In Table \ref{BLviolation}, we show the discrete symmetry charges $z_{\cal O} - 2r$ of the above superpotential operators ${\cal O}$. The operator is allowed if the entry in the table vanishes. Note that the lepton-number violating operator $H_u \ell$ is always forbidden if the discrete symmetry is an $R$ symmetry, but is always allowed if it is a non-$R$ symmetry (with $r=0$). This is because we required that the square of this operator is allowed to provide neutrinos masses. Also, the discrete symmetry always forbids the operators $q\ell \overline d$ and $\ell \ell \overline e$. This is because if they are allowed, one can check that the  term $Y^4$ would also be allowed, and would violate the PQ symmetry, against our defining requirement for the discrete symmetry. Therefore, the PQ-protecting $Z_n^R$ discrete symmetry never allows renormalizable L violation, and $Z_n$ only allows soft L violation. Regarding the other B and L violating operators, the charges in the table have no particular reason to vanish, and we find that in the vast majority of discrete symmetry cases they do not. A catalog of specific cases will not be attempted here, but the general lesson is that quite often the  discrete symmetries are powerful enough to automatically suppress proton decay sufficiently to satisfy present bounds, in addition to providing the high-quality PQ symmetry. One can also always supplement the PQ-protecting discrete symmetry with either matter parity, or the $Z_3$ baryon triality\cite{Ibanez:1991pr}, which forbids all proton decay.
\begin{table}
\begin{center}
\begin{minipage}[]{0.95\linewidth}\caption{The discrete charges $z_{\cal O} - 2r$ for B and/or L violating operators ${\cal O}$ in the MSSM superpotential, for each of the base models (and their extensions), in our conventions. Each operator is allowed only if the entry vanishes. The terms $\ell\ell\overline e$ and $q\ell\overline d$ are therefore always forbidden, because otherwise the PQ-violating term $Y^4$ would be allowed. The term $H_u \ell$ is forbidden for discrete $R$ symmetries ($r\not=0$).\label{BLviolation}}
\end{minipage}

\vspace{0.3cm}

\begin{tabular}{|c || c | c | c |}
\hline
${\cal O}$ & ~~$\text{B}_{\rm \RomanNumeralCaps{1}}$~~ & 
~~$\text{B}_{\rm \RomanNumeralCaps{2}}$, $\text{B}_{\rm \RomanNumeralCaps{4}}$~~ &
~~$\text{B}_{\rm \RomanNumeralCaps{3}}$~~
\\
\hline\hline
~$H_u \ell$~ & $-r$ & $-r$ & $-r$ 
\\
\hline
~$\ell\ell\overline e$, ~$q\ell\overline d$ ~ & $-2x+r$ & $2x-r$ & $-6x+3r$ 
\\
\hline
~$\overline u\hspace{0.7pt} \overline d\hspace{0.7pt} \overline d$ ~ & ~$h-4x+4r$~ & ~$h+4x$~ & ~$h-12x+8r$~ 
\\
\hline
~$qqq\ell$ ~ & ~$-h-r$~ & ~$-h-r$~ & ~$-h-r$~ 
\\
\hline
~$\overline u\hspace{0.7pt} \overline u\hspace{0.7pt} 
\overline d\hspace{0.7pt} \overline e$ ~ & ~$h-4x+5r$~ & ~$h+4x+r$~ & ~$h-12x+9r$~ 
\\
\hline
\end{tabular}
\end{center}
\end{table}

\newpage 

\section{Axion signals and detection prospects\label{sec:results}}
\setcounter{equation}{0}
\setcounter{figure}{0}
\setcounter{table}{0}
\setcounter{footnote}{1}

In this section, we discuss the low-energy axion couplings to photons, electrons, and nucleons,
and their impact on limits and detection prospects, for the supersymmetric base models and their extensions discussed above. For comparison, we will 
also give results for the standard non-supersymmetric benchmark QCD
axion models of Kim-Shifman-Vainshtein-Zakharov (KSVZ) \cite{Kim:1979if, Shifman:1979if},
and DFSZ \cite{Dine:1981rt, Zhitnitsky:1980tq} of types I and II.

In the non-supersymmetric benchmark DFSZ-I model, the Standard Model leptons and down-type quarks get their mass from one Higgs doublet, and the up-type quarks couple to the other Higgs doublet, as is the case with the MSSM. In DFSZ-II, the Standard Model leptons instead couple to the Higgs doublet that provides for the mass of the up-type quarks. In the KSVZ models, the Standard Model quarks and leptons are not charged under $U(1)_\textrm{PQ}$, unlike the DFSZ models. We consider several KSVZ constructions in which the heavy vectorlike quarks mediating the PQ anomaly  have different electroweak quantum numbers. The original Kim model \cite{Kim:1979if} in which the vectorlike quark transforms under the Standard Model gauge group as $({\bf 3}, {\bf 1}, 0)$ + $({\bf \overline{3}}, {\bf 1}, 0)$, will be called $\textrm{KSVZ}_0$, while $\textrm{KSVZ}_D$, $\textrm{KSVZ}_U$, $\textrm{KSVZ}_Q$ have vectorlike quarks with gauge quantum numbers as that of $D + \overline{D}$, $U + \overline{U}$, $Q + \overline{Q}$, respectively. Table~\ref{tab:benchmarks} summarizes these benchmarks, with their PQ-QCD-QCD anomaly coefficients $N$, and the coupling coefficients defined in eqs.~(\ref{eq:coeff_photon}) and (\ref{eq:coeff_fermion_a}) and appearing in eqs.~(\ref{eq:axion_couplings_begin})-(\ref{eq:DeltaCae}) and (\ref{eq:low-energy axion couplings}).%
\renewcommand{\arraystretch}{1.45}
\begin{table}[b]
\begin{center}
\begin{minipage}[]{0.95\linewidth}\caption{Benchmark non-supersymmetric QCD axion models and the fermions that mediate the $U(1)_{\rm PQ}$ anomaly. $N$ is the $U(1)_{\text{PQ}}$-$[SU(3)_c]^2$ anomaly coefficient and $c_\gamma$, $c_u$, $c_d$, and $c_e$ are low-energy axion coupling coefficients for photons, up-type quarks, down-type quarks, and electrons, respectively, appearing in eqs.~(\ref{eq:axion_couplings_begin})-(\ref{eq:DeltaCae}) and (\ref{eq:low-energy axion couplings}). In KSVZ constructions, only the new vectorlike quarks carry a PQ charge. In both DFSZ models, $\tan \beta = s_\beta/c_\beta$ is the ratio of the Higgs VEVs analogous to the MSSM. The domain wall number is $N_\textrm{DW} = 2 N$ in all of these benchmark cases.\label{tab:benchmarks}}
\end{minipage}

\vspace{0.3cm}

\begin{tabular}{|c | c | c | c | c | c | c |}
\hline
~Benchmark~ & ~PQ charged fermions~ & 
~$N$~ & ~$c_\gamma$~ & ~$c_u$~ & ~$c_d$~ & ~$c_e$~
\\
\hline\hline
~$\textrm{KSVZ}_0$~ & ~$({\bf 3}, {\bf 1}, 0)$ + $({\bf \overline 3}, {\bf 1}, 0)$~ & 
~$\frac{1}{2}$~ & ~$0$~ & ~$0$~ & ~$0$~ & ~$0$~
\\
~$\textrm{KSVZ}_D$~ & ~$D + \overline{D}$~ & 
~$\frac{1}{2}$~ & ~$2\over 3$~ & ~$0$~ & ~$0$~ & ~$0$~
\\
~$\textrm{KSVZ}_U$~ & ~$U + \overline{U}$~ & 
~$\frac{1}{2}$~ & ~$8\over 3$~ & ~$0$~ & ~$0$~ & ~$0$~
\\
~$\textrm{KSVZ}_Q$~ & ~$Q + \overline{Q}$~ & 
~$1$~ & ~$\frac{5}{3}$~ & ~$0$~ & ~$0$~ & ~$0$~
\\
~DFSZ-I~ & ~SM fermions~ & 
~$3$~ & ~$\frac{8}{3}$~ & ~$\frac{c_\beta^2}{3}$~ & ~$\frac{s_\beta^2}{3}$~ & ~$\frac{s_\beta^2}{3}$~
\\
~DFSZ-II~ & ~SM fermions~ & 
~$3$~ & ~$\frac{2}{3}$~ & ~$\frac{c_\beta^2}{3}$~ & ~$\frac{s_\beta^2}{3}$~ & ~$-\frac{c_\beta^2}{3}$~
\\
\hline
\end{tabular}
\end{center}
\end{table}
Among these benchmarks models, only $\textrm{KSVZ}_0$, $\textrm{KSVZ}_D$, and $\textrm{KSVZ}_U$ models have domain wall number $N_\textrm{DW} = 1$.

As noted in ref.~\cite{Bae:2017hlp}, all four supersymmetric DFSZ base models 
$\text{B}_{\rm \RomanNumeralCaps{1}}, \text{B}_{\rm \RomanNumeralCaps{2}}, \text{B}_{\rm \RomanNumeralCaps{3}}$, and $\text{B}_{\rm \RomanNumeralCaps{4}}$ have $E/N = 2$, which is close to cancelling the model-independent contribution $-1.92\pm 0.04$ in eqs.~(\ref{eq:axion_couplings_begin}) and (\ref{eq:gAgamma}), and so have suppressed axion-photon couplings compared to the standard non-supersymmetric benchmark cases. However, as we will see below, adding extra vectorlike supermultiplets that get their masses from the VEVs of the $X$ and $Y$ scalar fields can enhance the axion couplings. This is because they have non-zero net PQ charges contributing to the anomaly coefficients $N$ and $E$. Interestingly, the extensions with $N_\textrm{DW} = 1$ that avoid the cosmological domain wall problem have the smallest $|N|$ and therefore can give rise to the most enhanced low-energy axion couplings. As discussed in Section~\ref{sec:discrete}, for every $(N, E)$ found in the base models and their extensions, the PQ symmetry can be protected to a high degree (generally in a variety of different ways). Consequently, high-quality QCD axions in many of these extensions, including some $N_\textrm{DW} > 1$ cases, should be accessible to future axion searches.

\begin{figure}
  \begin{minipage}[]{0.995\linewidth}
    \includegraphics[width=14.5cm,angle=0]{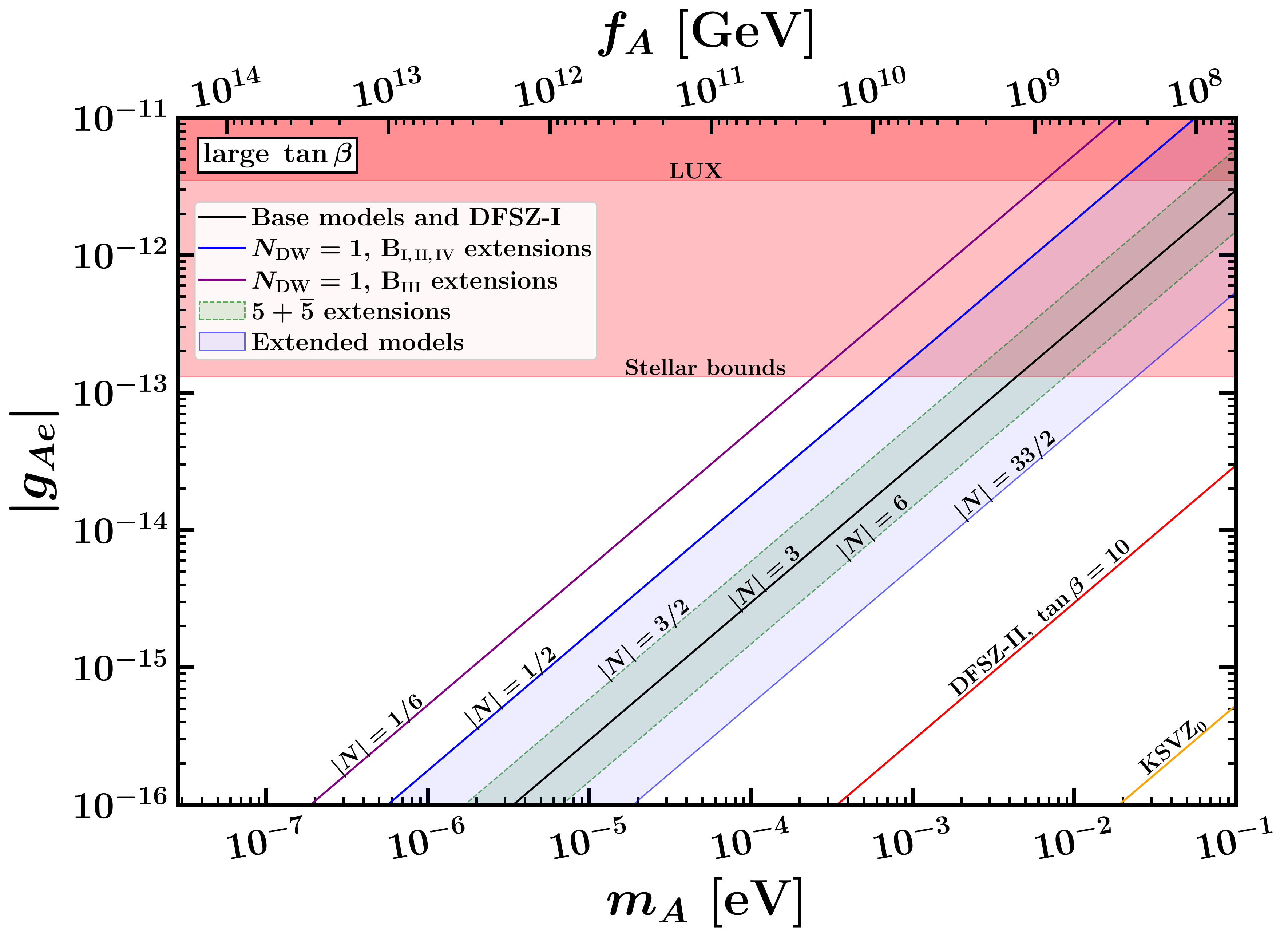}
  \end{minipage}
\begin{center}\begin{minipage}[]{0.98\linewidth}
\caption{The axion-electron coupling $|g_{Ae}|$ as a function of the axion mass $m_A$ and the axion decay constant $f_A$. The results at large $\tan\beta$ for the four supersymmetric base models, and the DFSZ-I benchmark, all lie on the $|N|=3$ line, where $N$ is the PQ-QCD-QCD anomaly. The range for base models extended by only a single ${\bf 5}+ {\bf \overline 5}$ occupy the green shaded band bounded by $|N|=3/2$ and $|N|=6$. The light blue region is the allowed range for extensions of base models that include a ${\bf 5}+ {\bf \overline 5}$ or ${\bf 10}+ {\bf \overline {10}}$ at the TeV scale and/or ${\bf 5}+ {\bf \overline 5}$ or  ${\bf 10}+ {\bf \overline {10}}$ at $M_\textrm{int}$, or $2 \times (L + \overline L)$ at the TeV scale and an exotic quix pair $D_6+ \overline{D}_6$ at $M_\textrm{int}$. The cases with $N_\textrm{DW} = 1$ have either $|N| = 1/2$ (extensions of $\text{B}_{\rm \RomanNumeralCaps{1}, \RomanNumeralCaps{2}, \RomanNumeralCaps{4}}$, heavier solid blue line)  or $1/6$ (extensions of $\text{B}_{\rm \RomanNumeralCaps{3}}$, solid purple line), and saturate the upper limit on $|g_{Ae}|$ in these extensions. The shading for allowed supersymmetric models excludes $1/6 < |N| < 1/2$, where there are no cases. The much smaller predictions of the benchmarks DFSZ-II with $\tan \beta = 10$ and $\textrm{KSVZ}_0$ are also shown for comparison. The current experimental limits on $|g_{Ae}|$ from the LUX experiment and from stellar cooling are also shown on the plot, as labeled.\label{fig:axion-electron}}
\end{minipage}\end{center}
\end{figure}

We begin with the axion coupling to electrons, because this gives the most robust bound on $f_A$ from present astrophysical data. Figure~\ref{fig:axion-electron} shows the axion-electron coupling $|g_{A e}|$ as a function of the axion mass $m_A$ [and equivalently $f_A$, through eq.~(\ref{eq:mAintermsoffA})], for large $\tan \beta$. 
For the DFSZ-I benchmark model and all supersymmetric models, the results are not very sensitive to the value of $\tan\beta$ as long as it is large, and $|g_{Ae}| = m_e/|N| f_A$ in the large $\tan\beta$ limit, so the lines can be simply labeled by $|N|$. The line $|N|=3$ is thus the common result for DFSZ-I and for all four base models. It is surrounded by a green shaded region bounded by the labels $|N|=3/2$ and $|N|=6$, which is the range of possibilities for all extensions of the base models with only one ${\bf 5}+ {\bf \overline 5}$ at either the intermediate or TeV scale. The larger blue shaded band between $|N| = 1/2$ and $|N| = 33/2$ is the range of results for the more general class of models obtained by extending the base models to include a ${\bf 5}+ {\bf \overline 5}$ or ${\bf 10}+ {\bf \overline {10}}$ at the TeV scale and/or ${\bf 5}+ {\bf \overline 5}$ or  ${\bf 10}+ {\bf \overline {10}}$ at $M_\textrm{int}$, or $2 \times (L + \overline L)$ at the TeV scale and an exotic quix pair $D_6+ \overline{D}_6$ at $M_\textrm{int}$. The largest possible couplings $g_{Ae}$ for a given $f_A$ are obtained for $|N|=1/6$ for extensions of $\text{B}_{\rm \RomanNumeralCaps{3}}$, and $|N| = 1/2$ for extensions of $\text{B}_{\rm \RomanNumeralCaps{1},\RomanNumeralCaps{2},\RomanNumeralCaps{4}}$. There are no extensions that have $|N|$ in between the allowed values 1/2 and 1/6, so we did not include that range in the blue shaded band. We also show the results for KSVZ$_0$ and DFSZ-II with $\tan\beta = 10$. Note that $|g_{Ae}|$ is much smaller, and more sensitive to large $\tan\beta$, in DFSZ-II than in DFSZ-I. In these cases, since the tree-level $g_{Ae}$ is zero and very small respectively, we include the leading log renormalization contribution from eq.~(\ref{eq:DeltaCae}).

The models with $N_{\rm DW} = 1$ saturate the upper limit on the axion-electron coupling as shown in the figure. These include extensions of base models $\text{B}_{\rm \RomanNumeralCaps{1}}$, $\text{B}_{\rm \RomanNumeralCaps{2}}$, and $\text{B}_{\rm \RomanNumeralCaps{4}}$ with $|N| = 1/2$ and $\text{B}_{\rm \RomanNumeralCaps{3}}$ with $|N| = 1/6$ shown in Table \ref{tab:NDW=1 models with two SU5 pairs}. There are also extensions of base model $\text{B}_{\rm \RomanNumeralCaps{3}}$ that have $|N| = 1/2$ but with $N_\textrm{DW} \ne 1$. The lower limits in all extensions correspond to the models with largest $|N|$ that can occur. In all base model extensions with $SU(5)$ pairs, or $D_6 + \overline{D}_6$ quixes, the largest $|N| = 33/2$ occurs in $\text{B}_{\rm \RomanNumeralCaps{2}}$ extended with a ${\bf 10}+ {\bf \overline {10}}$ at the TeV scale and a ${\bf 10}+ {\bf \overline {10}}$ at $M_{\rm int}$. And, in the extensions with exactly one ${\bf 5}+ {\bf \overline {5}}$, the largest $|N|=6$ occurs in the extension of $\text{B}_{\rm \RomanNumeralCaps{2}}$ with  a ${\bf 5}+ {\bf \overline {5}}$ at the TeV scale.

Current experimental limits from the LUX experiment \cite{Akerib:2017uem}, and much stronger bounds from the brightness of the tip of the red-giant branch \cite{Capozzi:2020cbu, Straniero:2020iyi} are shown as shaded regions in Figure~\ref{fig:axion-electron}. There are also bounds \cite{Isern:2008nt}-\cite{Bertolami:2014wua} from the cooling of white dwarfs, which are somewhat less strong than the red-giant bounds, and some hints of stellar cooling that might be ascribed to axions \cite{Giannotti:2017hny}. The red giant bound on the axion-electron coupling $g_{Ae} < 1.3 \times 10^{-13}$ sets the most stringent astrophysical constraint throughout our supersymmetric DFSZ axion model space:
\beq
f_A &>& \frac{\sin^2\hspace{-1.8pt}\beta}{|N|}\, (\mbox{$3.9 \times 10^9$ GeV}),
\label{eq:citefAboundgAe}
\eeq
which can also be written directly in terms of the $X,Y$ scalar field VEVs [see eq.~(\ref{eq:VintermsoffA})] as
\beq
|v_A| &>& \sin^2\hspace{-1.8pt}\beta\> (\mbox{$7.8 \times 10^9$ GeV}). 
\label{eq:citevAboundgAe}
\eeq
The lower bound on the axion decay constant for $|N| = (1/6, 1/2, 3)$ for large $\tan \beta$ is $f_A \gtrsim (2.3 \times 10^{10},\> 7.8 \times 10^{9},\> 1.3 \times 10^{9})$ GeV which corresponds to an upper bound on the axion mass $m_A \lesssim (2.4 \times 10^{-4},\> 7.3 \times 10^{-4},\> 4.4 \times 10^{-3})$ eV.

\begin{figure}[!tb]
  \begin{minipage}[]{\linewidth}
    \includegraphics[width=12.7cm,angle=0]{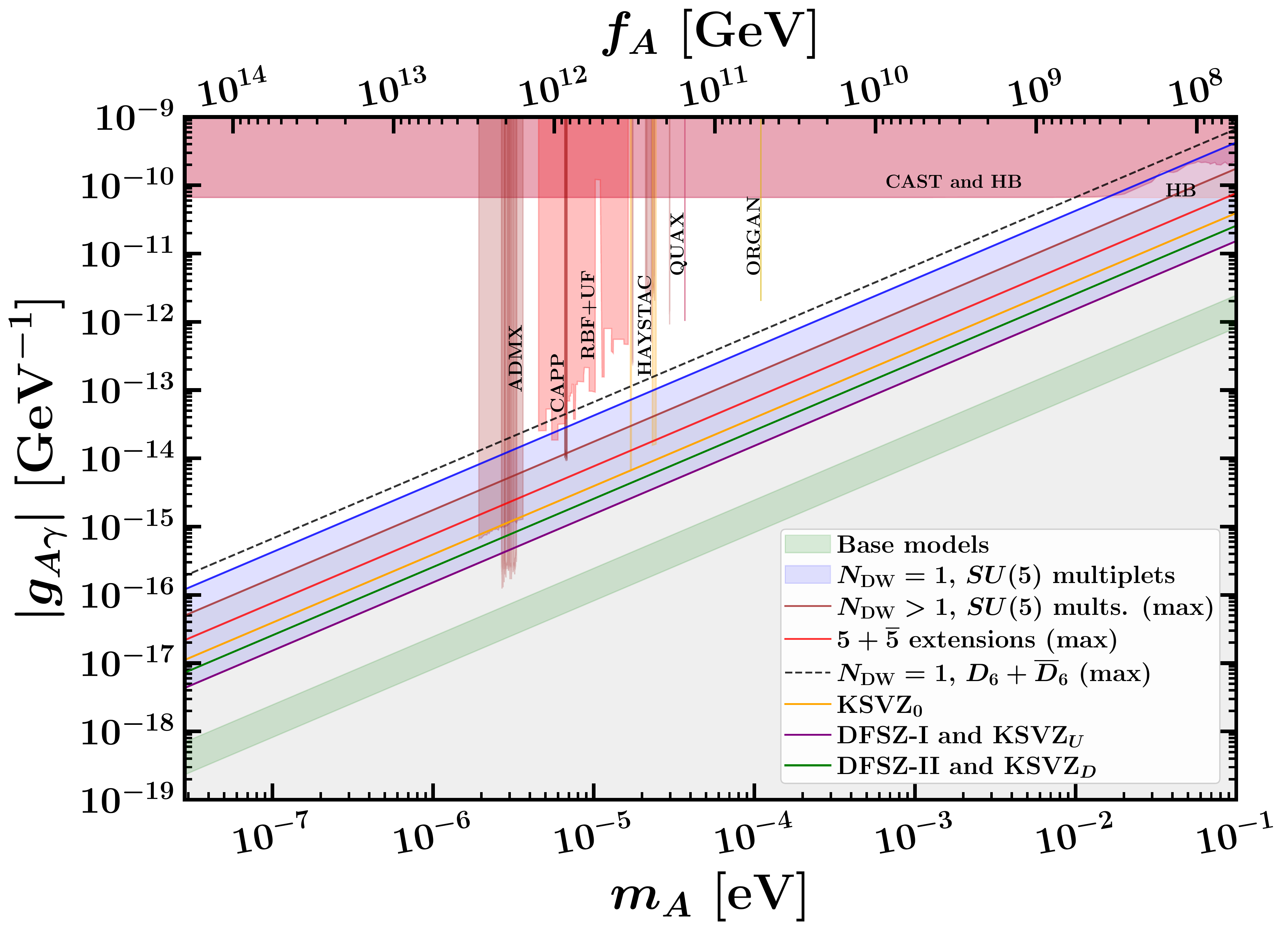}
  \end{minipage}
  \begin{minipage}[]{\linewidth}
  \includegraphics[width=12.7cm,angle=0]{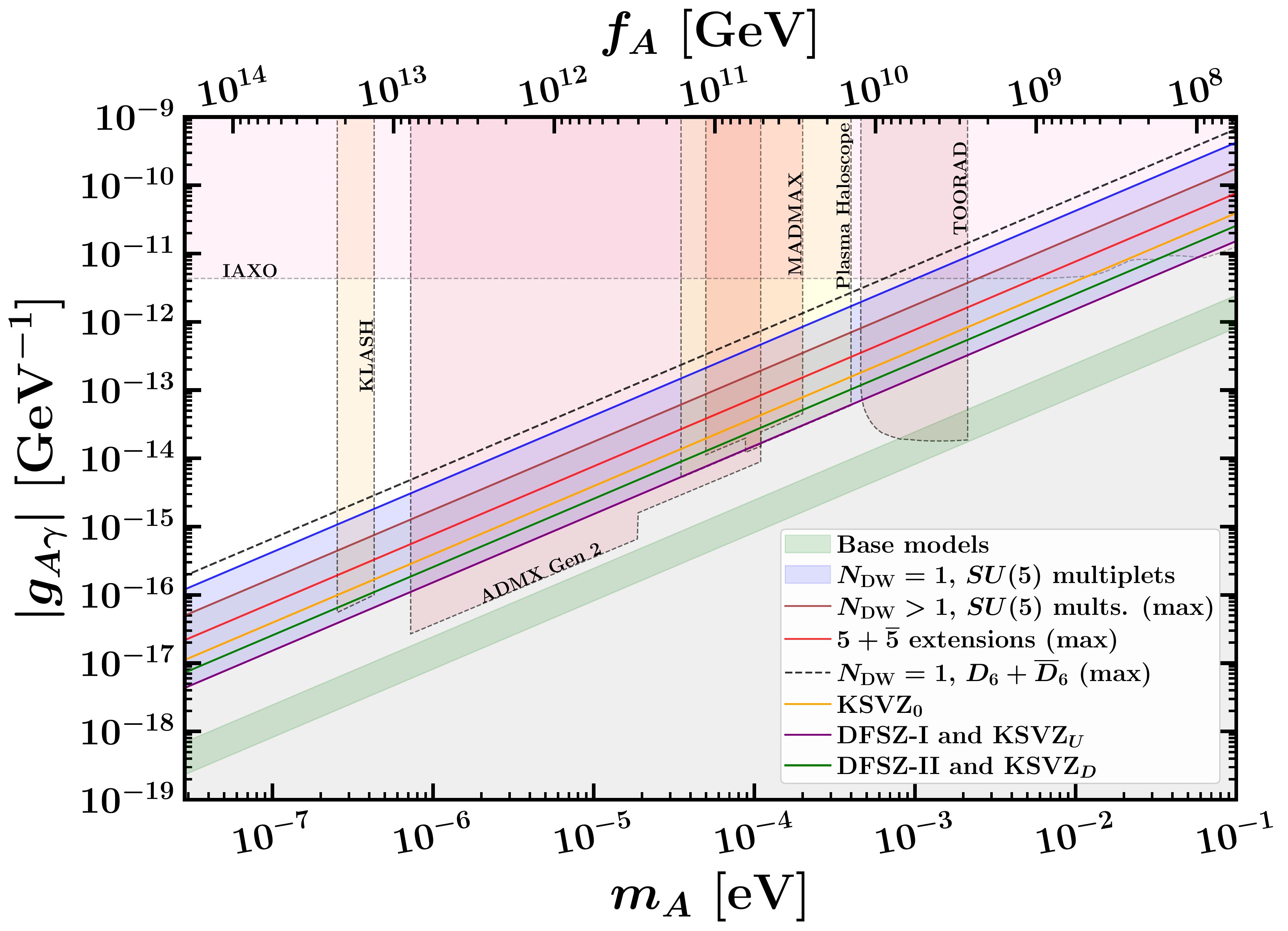}
  \end{minipage}
\begin{center}

\vspace{-0.5cm}

\caption{Model predictions for axion-photon coupling $g_{A\gamma}$ as a function of the axion mass and the axion decay constant, with current bounds from various axion searches (top panel), and projections for future experiments (bottom panel). The prediction, including uncertainty, for supersymmetric DFSZ base models is the shaded green band. Cases with $N_\textrm{DW} = 1$ that occur in base models extended to include $SU(5)$ multiplets are shown as the shaded blue region. The largest $|g_{A\gamma}|$ that have $N_\textrm{DW} > 1$ in such extensions is the solid brown line, and the maximum in extensions that include only a single ${\bf 5}+ {\bf \overline 5}$ is the solid red line. The largest coupling found in models with $2 \times (L + \overline L)$ at the TeV scale and a quix pair $D_6+ \overline{D}_6$ at $M_\textrm{int}$ is the dashed black line. The light gray shading everywhere below the brown line is to emphasize that the axion can accidentally decouple from the photon in some extensions with $N_\textrm{DW} > 1$. For comparison, the lines for the standard benchmark axion models $\textrm{KSVZ}_0$, DFSZ-I (same as $\textrm{KSVZ}_U$), and DFSZ-II (same as $\textrm{KSVZ}_D$) are shown.\label{fig:axion-photon}}
\end{center}
\end{figure}

We now turn to the axion-photon coupling and direct detection prospects for axions. In Figure~\ref{fig:axion-photon}, we show the axion-photon coupling $|g_{A \gamma}|$ as a function of the axion mass $m_A$ and the axion decay constant $f_A$. The axion-photon coupling depends only on the ratio $E/N$ and  $f_A$, so each line in this plot corresponds to a value of $|E/N - 1.92(4)|$. The results for the base models (which have $E/N=2$), including the uncertainty on the model-independent contribution, are shown as the green shaded band. The blue shaded region shows the range of much larger couplings obtained for extended models made from $SU(5)$ multiplet combinations that give $N_{\rm DW} = 1$.  The $E/N$ values that occur in these extensions  were listed in Table~\ref{tab:NDW=1 models with two SU5 pairs}. This blue shaded region is bounded from above by a solid blue line that corresponds to the extensions that have $E/N = 68/3$, and the lower bound for these models is $E/N = 8/3$ (which is the same as DFSZ-I). This shows that requiring $N_\textrm{DW} = 1$ guarantees that the axion-photon coupling is at least as large as for the DFSZ-I model, and is usually considerably larger, for a given axion mass. The extensions of base models with $2 \times (L + \overline L)$ at the TeV scale and a quix pair $D_6+ \overline{D}_6$ at $M_\textrm{int}$ can give an even larger axion-photon coupling compared to the extensions with $SU(5)$ multiplets. This is shown by the dashed black line that corresponds to $E/N = 104/3$, which occurs in the extension of $\text{B}_{\rm \RomanNumeralCaps{2}}$ with $N_\textrm{DW} = 1$.

\begin{figure}[!tb]
  \begin{minipage}[]{0.995\linewidth}
    \includegraphics[width=13.5cm,angle=0]{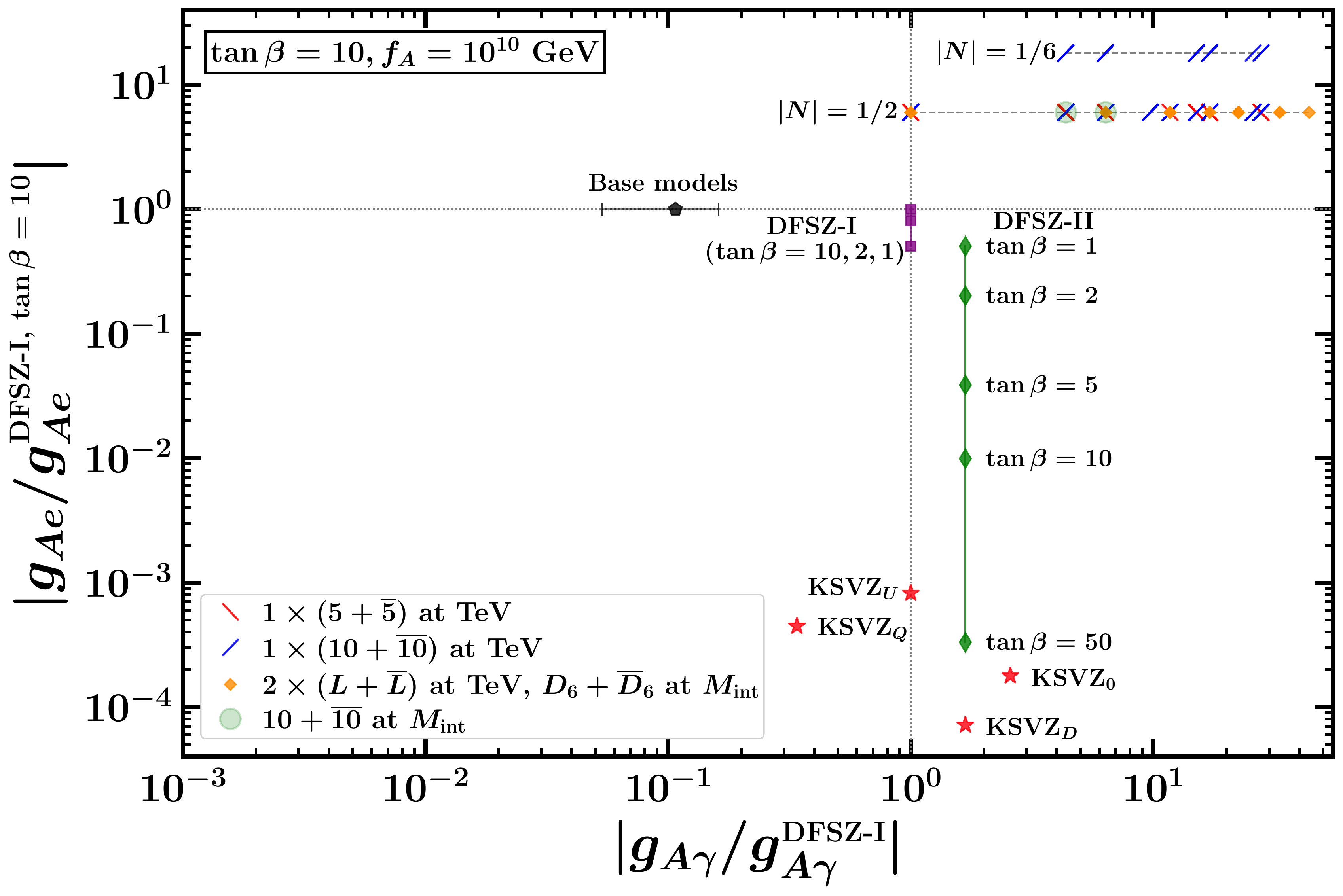}
  \end{minipage}
\vspace{0.2cm}
  \begin{minipage}[]{0.995\linewidth}
  \includegraphics[width=13.5cm,angle=0]{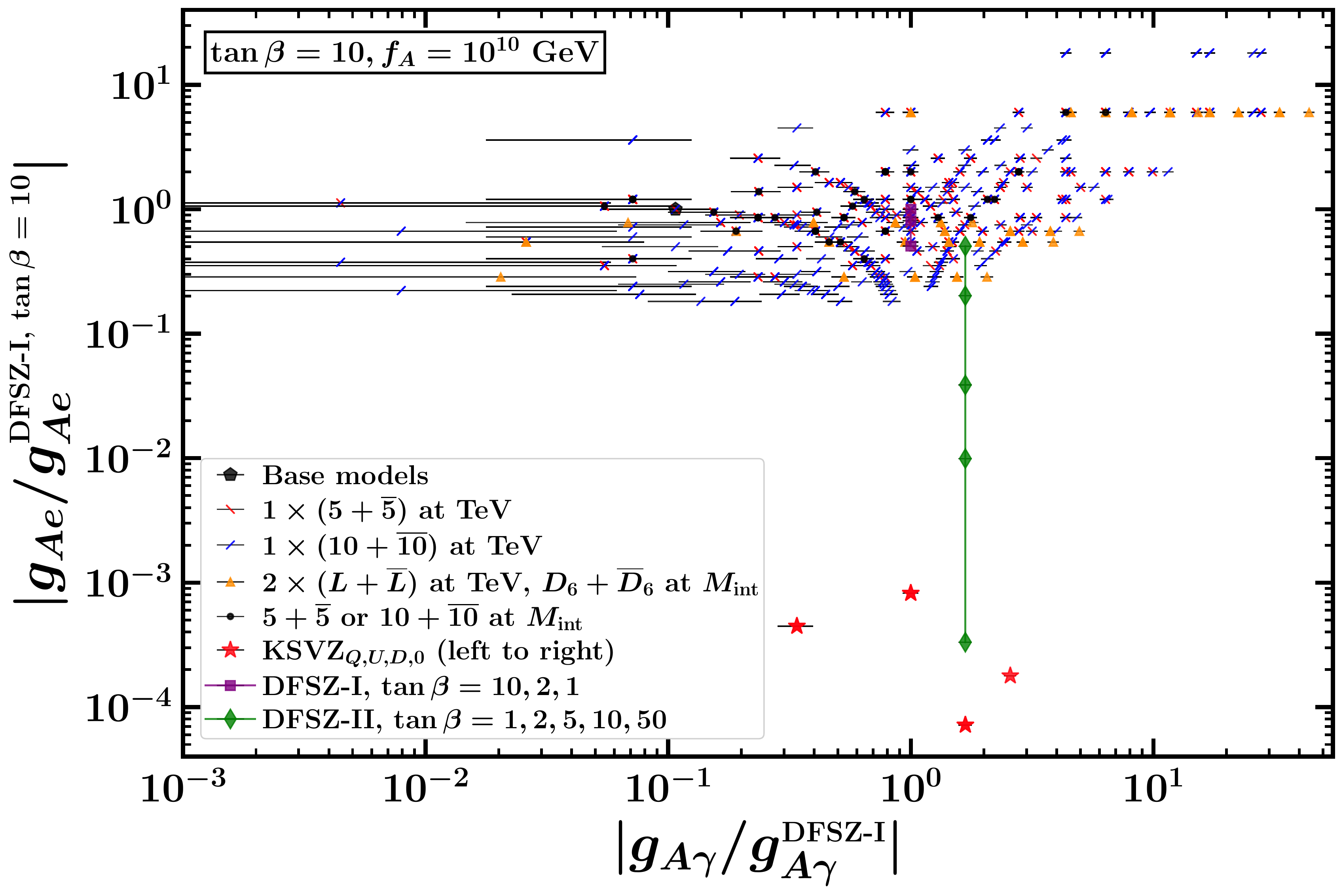}
  \end{minipage}
  
\vspace{-0.7cm}
  
\begin{center}\begin{minipage}[]{0.999\linewidth}
\caption{The axion-electron coupling plotted against the axion-photon coupling, both rescaled by the corresponding couplings of the benchmark non-supersymmetric DFSZ-I model with $\tan \beta = 10$, $f_A = 10^{10}$ GeV. Different points on the plot correspond to the base models and their extensions, categorized by the additional particle content at the TeV scale as discussed in the text, again with fixed $\tan \beta = 10$. The top panel shows only the models with $N_\textrm{DW} = 1$ in these extensions, while the bottom panel shows all possible combinations of mass terms for extra vectorlike supermultiplets in each model extension, as labeled. Extensions with $N_{\rm DW} = 1$ (in the top-right corner of both panels) have axion-photon and axion-electron couplings both enhanced. Also shown in both panels are the points for the benchmarks $\textrm{KSVZ}_{Q, U, D, 0}$ (independent of $\tan \beta$), and DFSZ-I and DFSZ-II for various choices of $\tan \beta$, as labeled. The uncertainty bars on $g_{A\gamma}$ are shown for the base models in the top panel, and for all points in the bottom panel. \label{fig:AeAgamma}}
\end{minipage}\end{center}
\end{figure}

Also shown in Figure~\ref{fig:axion-photon} is a light gray shaded region obtained for models with $N_{\rm DW} > 1$, obtained by extending the base models to include a
${\bf 5}+ {\bf \overline 5}$ or 
${\bf 10}+ {\bf \overline {10}}$
at the TeV scale and/or
${\bf 5}+ {\bf \overline 5}$ or 
${\bf 10}+ {\bf \overline {10}}$
at $M_\textrm{int}$. The upper limit in these models with $N_{\rm DW}>1$, shown with a solid brown line, is realized by an extension with a ${\bf 10}+ {\bf \overline {10}}$ at the TeV scale and another ${\bf 10}+ {\bf \overline {10}}$ at $M_\textrm{int}$ with $E/N = -20/3, N_\textrm{DW} = 3$.  The base models extended with only a ${\bf 5 + \overline{5}}$ at the TeV scale or $M_{\rm int}$ cannot have $N_{\rm DM}=1$, and have an upper limit for $|g_{A\gamma}|$ shown as the solid red line in Figure~\ref{fig:axion-photon}. This corresponds to $E/N = 17/3$, realized in an extension of the  model $\text{B}_{\rm \RomanNumeralCaps{2}}$ with a ${\bf 5 + \overline{5}}$ at the TeV scale. The axion-photon coupling can vanish within errors in some of the extensions (including quixotic extensions) of base models when $E/N$ happens to be close to the imperfectly known model-independent contribution $-1.92 \pm 0.04$, as will be further illustrated below in the bottom panel of Figure~\ref{fig:AeAgamma}. This is the reason for shading the entire region in light gray below the solid brown line. Also shown in Figure~\ref{fig:axion-photon} are the lines for the non-supersymmetric DFSZ-I, DFSZ-II, and $\textrm{KSVZ}_0$ benchmark models. Note that, as can be seen in Table~\ref{tab:benchmarks}, DFSZ-I and DFSZ-II have the same axion-photon couplings as $\textrm{KSVZ}_U$ and $\textrm{KSVZ}_D$, respectively.

\begin{figure}[!tb]
  \begin{minipage}[]{0.995\linewidth}
    \includegraphics[width=14.0cm,angle=0]{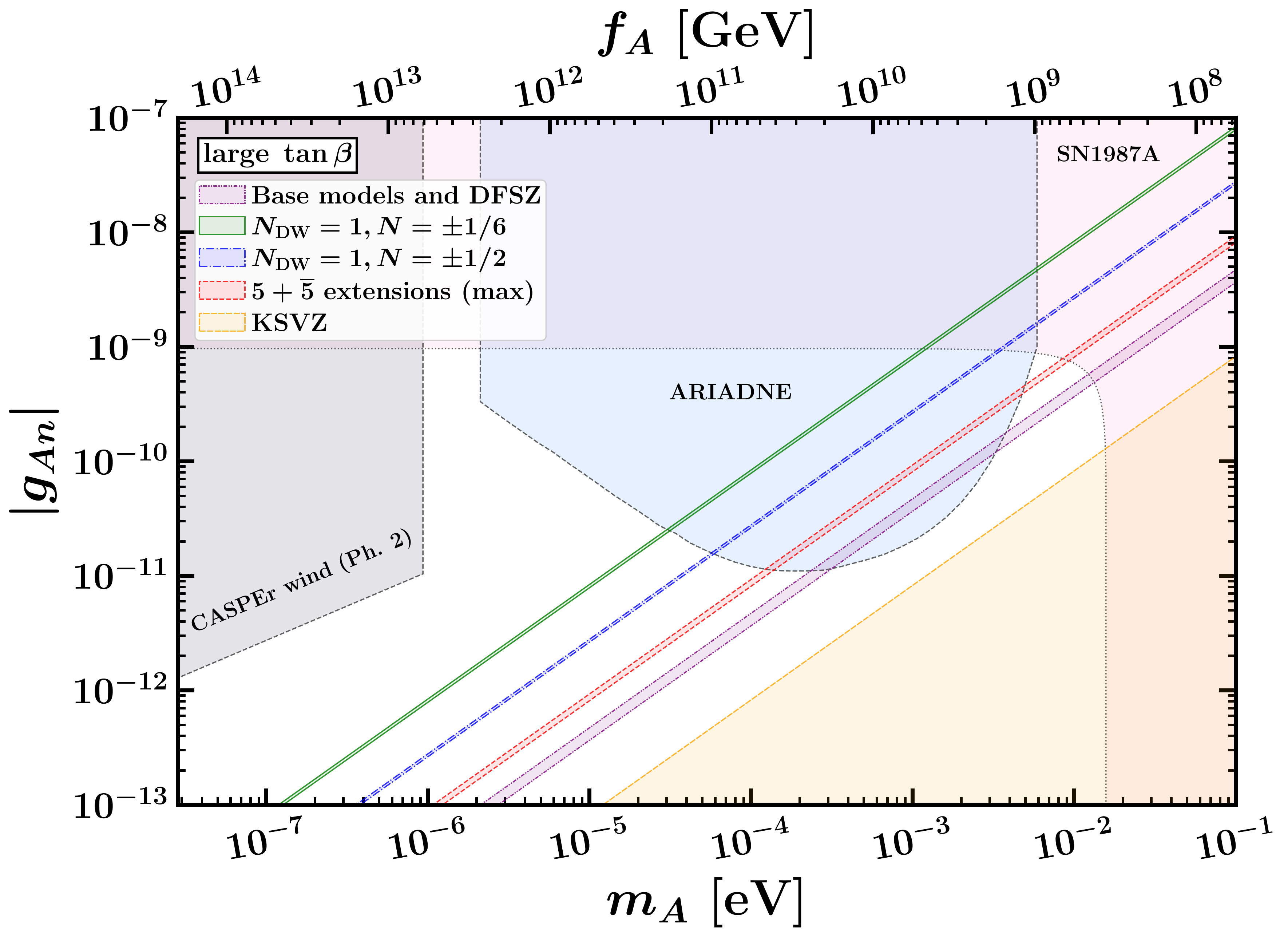}
  \end{minipage}
  
\vspace{-0.4cm}
  
\begin{center}\begin{minipage}[]{0.97\linewidth}
\caption{
The axion-neutron coupling as a function of the axion mass and the axion decay constant. The model predictions for large $\tan\beta$ are diagonal bands including uncertainties in the coupling. As labeled, they correspond to the supersymmetric base models (same as non-supersymmetric DFSZ-I and DFSZ-II benchmarks), the extensions with $N_\textrm{DW} = 1$ that can have enhanced axion-nucleon couplings, and the case that gives rise to the largest axion-nucleon coupling in the extension with a single ${\bf 5} + {\bf \overline{5}}$. Within uncertainties, the neutron may decouple from KSVZ axions, as emphasized by the orange shading extending to very low couplings. The experimental bound from supernova SN 1987A is also shown
as it applies to the supersymmetric DFSZ models as parameterized by eqs.~(\ref{eq:gAn}) and (\ref{eq:gAp}) with large $\tan\beta$, taking into account the correlation with $g_{Ap}$. Also shown are future projections for CASPEr wind (phase 2) and ARIADNE. The projection for the ARIADNE experiment is based on an optimistic choice regarding the CP-violating coupling of axions to neutrons, as discussed in the text. \label{fig:axion-nucleon}}
\end{minipage}\end{center}
\end{figure}

The current experimental bounds on $g_{A\gamma}$ as a function of $m_A$ are shown as shaded regions
in the top panel of Figure~\ref{fig:axion-photon}. These include limits
from the helioscope CAST \cite{Andriamonje:2007ew, Anastassopoulos:2017ftl}
and from evolution of Horizontal Branch (HB) stars \cite{Ayala:2014pea}. By coincidence these
happen to give almost the same limit $g_{A\gamma} < 6.5 \times 10^{-11}$ GeV$^{-1}$ over a very wide range of $f_A$, although the CAST bound becomes weaker for $f_A \lsim 4 \times 10^{8}$ GeV.
Also shown are the results of searches over much narrower bands in $f_A$ by haloscopes, under the assumption that axions are the dark matter: ADMX \cite{Asztalos:2009yp, Du:2018uak, Boutan:2018uoc, Bartram:2020ysy}, CAPP \cite{Lee:2020cfj}, RBF \cite{DePanfilis:1987dk}, UF \cite{Hagmann:1990tj}, HAYSTAC \cite{Backes:2020ajv, Zhong:2018rsr}, QUAX \cite{Alesini:2019ajt}, and ORGAN \cite{McAllister:2017lkb}. (In making Figures \ref{fig:axion-photon} and \ref{fig:axion-nucleon}, we made substantial use of the axion limit data collected at ref.~\cite{cajoharegithub}.) This shows that some limited ranges of $f_A$ for extended supersymmetric models with $N_{\rm DW}=1$ have already been probed. In the lower panel of Figure~\ref{fig:axion-photon}, the shaded regions with dashed borders show the projections for future sensitivity from the helioscope IAXO \cite{Shilon:2012te}, and haloscopes (ADMX \cite{Stern:2016bbw}, KLASH \cite{Alesini:2017ifp}, MADMAX \cite{Beurthey:2020yuq}, Plasma haloscope \cite{Lawson:2019brd}, and TOORAD \cite{Schutte-Engel:2021bqm}). Except for a narrow range in the case of TOORAD, these do probe the supersymmetric DFSZ base models, but it is encouraging that the haloscopes do cover many of the possibilities for the model extensions, including essentially all of the allowed range predicted by our models with $N_{\rm DW}=1$. However, it is again important to note that the haloscope search projections assume that axions are the main component of the dark matter.

The axion couplings to electrons and photons both scale with $1/f_A$. It is therefore interesting to compare the ratios $g_{A e}/g^{{\rm DFSZ-I}, \tan \beta=10}_{A e}$ and $g_{A \gamma}/g^{{\rm DFSZ-I}}_{A \gamma}$, in which the dependence on the scale $f_A$ (and the axion mass) very nearly cancels. We choose the normalizing denominators to be the results for the non-supersymmetric benchmark DFSZ-I (with $\tan \beta = 10$). Figure~\ref{fig:AeAgamma} shows a scatterplot of these ratios for the base models and their extensions along with the benchmarks DFSZ-I, DFSZ-II, and $\textrm{KSVZ}_{Q, U, D, 0}$, as labeled. In the plot, the extensions of base models are categorized based on the additional particle content at the TeV scale, which could eventually be discovered in collider experiments. First, there are models with a ${\bf 5 + \overline{5}}$ or ${\bf 10 + \overline{10}}$ at the TeV scale with or without a ${\bf 5 + \overline{5}}$ or ${\bf 10 + \overline{10}}$ at $M_\textrm{int}$, then there are models with 2 pairs of vectorlike leptons $L + \overline{L}$ at the TeV scale and $D_6 + \overline{D}_6$ quixes at $M_\textrm{int}$, and lastly there are models with no new particle content at the TeV scale which includes the extensions with a ${\bf 5 + \overline{5}}$ or ${\bf 10 + \overline{10}}$ at $M_\textrm{int}$. In these extensions, different combinations of the superpotential mass terms involving the $X$ and $Y$ fields for the additional vectorlike supermultiplets give rise to different points in the figure. In the top panel of Figure~\ref{fig:AeAgamma}, we only show cases with $N_{\rm DW}=1$, whereas in the bottom panel we show all possibilities (i.e. $N_\textrm{DW} \ge 1$) that occur in these extensions. Also, in the top panel the uncertainty bars are only shown for the base models, for which the uncertainties are significant due to the proximity of $E/N = 2$ to $1.92(4)$. In the bottom panel, the $g_{A\gamma}$ uncertainty bars are shown for all points.

Except for the $f_A$ dependence that enters the ratio $g_{A e}/g^{{\rm DFSZ-I}, \tan \beta=10}_{A e}$ through $\Delta C_{Ae}$ for models in which $C_{Ae}$ vanishes at tree level, the plots in Figure \ref{fig:AeAgamma} do not otherwise depend on the axion decay constant (or equivalently axion mass). For plotting purposes, we fixed $\tan \beta = 10$ for the base models and their extensions. (Results for other large values of $\tan\beta$ would not be visually distinguishable.) In DFSZ-I and the base models and their extensions, for a fixed $\tan \beta$, horizontal lines in these plots have constant $|N|$, while vertical lines have constant $E/N$. The horizontal dotted line in the top panel corresponds to $|N| = 3$ for fixed $\tan \beta = 10$ (in DFSZ-I type models), the vertical dotted line corresponds to $E/N = 8/3$, and the intersection point at $(1, 1)$ of course corresponds to the benchmark DFSZ-I with $\tan \beta = 10$. In the top-right corner of the plot in the top panel, the horizontal dashed lines correspond to cases with $N_\textrm{DW} = 1$ in the extensions of base models with $|N| = 1/6$ and $|N| = 1/2$ for a fixed $\tan \beta = 10$. As noted earlier, one or more cases with $N_{\rm DW} = 1$ provide for the largest axion-photon coupling (when $E/N$ is far away from $1.92(4)$) and axion-electron coupling (for smallest $|N|$ when tree-level contributions dominate), as is reflected in the figure. In the bottom panel, there are additional points on the horizontal line with $|N| = 1/2$ that are not present in the top panel; these points are from the extensions of base model $\text{B}_{\rm \RomanNumeralCaps{3}}$ with $|N| = 1/2$, $N_{\rm DW} = 3$. Also, in the figure, the DFSZ-I and DFSZ-II benchmark models are shown for various values of $\tan \beta$. We can therefore understand the impact of $\tan \beta$ on the points for base models and its extensions, as they have a similar $\tan \beta$ dependence as that of DFSZ-I.

Finally, we turn to the axion-nucleon couplings, for which the neutron coupling is most accessible to constraint. Figure~\ref{fig:axion-nucleon} shows the axion-neutron coupling as a function of the axion mass $m_A$ and the decay constant $f_A$, choosing $\tan \beta = 10$ for plotting purposes. Due to significant uncertainties in the axion coupling predictions, each case in the figure is a shaded band instead of a line. First, bands are shown for the $N_{\rm DW}=1$ cases that occur in $N=\pm 1/2$ and $\pm 1/6$ extensions of the base models. The axion-nucleon coupling, in general, depends on the sign of $N$, but in the case of the neutron the shaded bands for $N=1/2$ and $N=1/6$ very nearly overlap with the shaded bands for $N=-1/2$ and $N=-1/6$, respectively, so we have combined them. Also shown is the band of maximum allowed axion-neutron coupling in the extensions with only a ${\bf 5 + \overline{5}}$, which occurs for $N = 3/2$ in base model $\text{B}_{\rm \RomanNumeralCaps{1}}$ extended with a ${\bf 5 + \overline{5}}$ at $M_\textrm{int}$. In addition, ranges are shown for the supersymmetric base models and the DFSZ and KSVZ models, as labeled. Note that for a fixed $\tan \beta$ and $N$, the base models and DFSZ-I and DFSZ-II all have the same axion-nucleon couplings. Within uncertainties, the KSVZ axion can accidentally decouple from the neutron, so the shading
in that case extends to arbitrarily low $g_{An}$. 

The existence of the neutrino signal from Supernova SN 1987A has been interpreted by ref.~\cite{Carenza:2019pxu} to put the following constraint on axion-nucleon couplings:
\beq
g_{A n}^2 + 0.61 g_{A p}^2 +0.53 g_{A n} g_{A p} &\lsim& 8.26 \times 10^{-19} 
.
\label{eq:nucleoncouplingSN1987Abound}
\eeq
However, it has been suggested e.g.~in ref.~\cite{Fischer:2016cyd} that such bounds should be taken as a guide rather than a sharp bound, given the uncertainties involved. Taken at face value the bound implies, for large $\tan\beta$,
\beq
f_A \gsim \sqrt{0.15 + 0.66/N^2 }\> (10^{9}\>\mbox{GeV}),
\eeq
for our models, by using eqs.~(\ref{eq:gAn}) and (\ref{eq:gAp}). (There is also a term proportional to $1/N$ under the square root, but its coefficient is consistent with 0 within uncertainties.) For the critical case of small $|N|$, this is not quite as strong as eq.~(\ref{eq:citefAboundgAe}) obtained above from the constraints on $g_{Ae}$ from red-giant and white dwarf cooling. In Figure \ref{fig:axion-nucleon}, we show the bound as it applies to the supersymmetric DFSZ models as parameterized by eqs.~(\ref{eq:gAn}) and (\ref{eq:gAp}) with large $\tan\beta$, recognizing that it is not a model-independent bound on $g_{An}$ (because other models can have different correlations between $g_{An}$ and $g_{Ap}$) and in any case may not be robust in detail.

There is also an even stronger candidate bound \cite{Beznogov:2018fda} of $|g_{An}| < 2.8 \times 10^{-10}$  from avoiding too much cooling of the hot neutron star HESS J1731-347. However, since it is difficult to account for the temperature of this particular object even without axionic cooling \cite{Sedrakian:2018kdm}, we do not include this bound in Figure \ref{fig:axion-nucleon}.

Also shown in Figure \ref{fig:axion-nucleon} are the experimental reach for the axion-neutron coupling of Phase-II of the proposed CASPEr wind experiment \cite{JacksonKimball:2017elr}, and a projection for the ARIADNE experiment \cite{Arvanitaki:2014dfa,Geraci:2017bmq}. ARIADNE would be sensitive to
the product of $g_{A n}$ and the CP-odd scalar coupling $g^{s}_{A n}$
(as in $\mathcal{L}^{\rm CP-odd} \supset -A g^{s}_{A n} \overline{\Psi}_n \Psi_n$), and the region shown in Figure \ref{fig:axion-nucleon} is based on an optimistic choice of $g^{s}_{A n} = 10^{-12} \textrm{ GeV}/f_A$ \cite{Arvanitaki:2014dfa,DiLuzio:2020wdo,OHare:2020wah}. The coupling $g^{s}_{A n}$ could arise from a small non-zero CP violation in $\theta_{\rm eff}$, and so cannot be uniquely predicted by the other relevant model parameters $N$ and $\tan\beta$. We note that, at least with this optimistic assumption, ARIADNE can probe $g_{An}$ for $f_A$ up to about $2.5\times 10^{10}$ GeV for the supersymmetric DFSZ base models, which are invisible to the projected searches for $g_{A\gamma}$.

\section{Conclusion\label{sec:conclusion}}
\setcounter{equation}{0}
\setcounter{figure}{0}
\setcounter{table}{0}
\setcounter{footnote}{1}

In this paper, we considered supersymmetric DFSZ type axion models with the field content of the MSSM plus two gauge-singlet fields $X,Y$ that spontaneously break the PQ symmetry, and some extra vectorlike quark and lepton supermultiplets. These models simultaneously give a solution to the $\mu$ problem and the strong CP problem. The extra vectorlike content is chosen such that the perturbative gauge coupling unification is maintained. 
The PQ symmetry is spontaneously broken by the scalar components of $X$ and $Y$, which acquire intermediate scale vacuum expectation values, simultaneously giving rise to a high-quality nearly invisible axion with a decay constant within the current astrophysical limits and a $\mu$ term around the TeV scale. The axino and saxion also have masses of order the TeV scale.
Different combinations of mass terms for the additional vectorlike supermultiplets involving $X$ and $Y$ result in different combinations of the PQ anomaly coefficients $(N, E)$, which in turn determines the low-energy axion couplings.

For the base models and their extensions, we studied how to obtain the PQ symmetry as an accidental symmetry that is guarded against the dangerous PQ violating superpotential terms of the form $X^j Y^{p-j}/M_P^{p-3}$ to an extent that is compatible with the experimental constraint on the QCD $\theta$ parameter, by imposing anomaly-free discrete non-$R$ or $R$ $Z_n$ symmetries with or without the Green-Schwarz mechanism. If the axion decay constant $f_A$ is as low as $10^9$ GeV, we can allow $p = 7$, and for
$f_A \lesssim 10^{12}$ GeV, we may instead need a higher suppression of up to $p = 12$.
In order for the $Z^R_n$ symmetries (which reduces to non-$R$ $Z_n$ symmetry if the $Z^R_n$ charge of the gauginos $r = 0$) to be anomaly-free, we impose a weaker constraint
along with the additional stronger constraint on the $Z^R_n \times G \times G$ anomalies for
$G = SU(3)_c$, $SU(2)_L$, and $U(1)_Y$ as discussed in Section~\ref{sec:discrete}.
Out of all possible non-$R$ $Z_n$ symmetries for the base models, the only cases with adequate suppression
($p \ge 7$) that satisfy the stronger set of anomaly constraints are $Z_{36}$ for base model
$\text{B}_{\rm \RomanNumeralCaps{3}}$ with $p = 12$, and a $Z_{36}$ for
$\text{B}_{\rm \RomanNumeralCaps{4}}$ with $p = 8$. Both of these cases require the Green-Schwarz mechanism.
On the other hand, there are a lot more anomaly-free $Z_n$ $R$-symmetries for each base model,
some of which do not require the Green-Schwarz mechanism.
For example, there is a $Z^R_{54}$ for base model $\text{B}_{\rm \RomanNumeralCaps{3}}$ with $p = 10$, $Z^R_{12}$ for base model $\text{B}_{\rm \RomanNumeralCaps{4}}$ with only $p = 7$, both of which do not require the Green-Schwarz mechanism. With the Green-Schwarz mechanism, there are many more $Z^R_n$ symmetries that have lower $n$ and higher suppression $p$.
In our approach, the discrete symmetry is fundamental and exact (being anomaly free) and the PQ symmetry is not fundamental and approximate (being an accidental consequence of the discrete symmetry).

Adding vectorlike supermultiplets adds greatly to the possibilities for both discrete non-$R$ and $R$ symmetries. For each possible pair of PQ anomaly coefficients $(N, E)$ in the extensions, we find that there are always anomaly-free discrete symmetries that protect the PQ symmetry to a high degree of accuracy. Not only do these discrete symmetries provide for an accidental $U(1)_{\rm PQ}$ symmetry, but they also can forbid dangerous baryon number and lepton number violating operators, in most cases, that could mediate dangerous proton decay. There are so many available discrete symmetries that we did not attempt a complete categorization.
The additional fields and the discrete symmetry of the models we have proposed increase the complexity of the MSSM. The discrete symmetry could arise from an additional U(1) symmetry in the ultraviolet, but other than that we do not have any insight to offer as to the reasons for this additional complexity, other than the problems that it solves.

In our models, the vectorlike supermultiplets are coupled to Standard Model quark and lepton superfields to avoid cosmological problems. These couplings could be very small in magnitude while still allowing the vectorlike particles to decay promptly. Thus, while they would in general include additional CP-violating phases, the resulting CP-violating effects on the Standard Model could easily be negligible if the magnitudes of the couplings are sufficiently small. We also note that in the MSSM there are potential CP-violating effects from gaugino masses and the $\mu$ parameter. Our framework has nothing to say about these effects, except that they are not worse than in the usual MSSM, and as usual they are ameliorated for superpartners in the TeV range.

Due to the soft supersymmetry-breaking terms, the scalar components of the vectorlike chiral supermultiplets tend to be heavier than their corresponding fermions, as studied for example in ref.~\cite{Martin:2009bg}. Therefore, the constraints on the masses of the extra vectorlike supermultiplets come from the searches for pair-production of vectorlike quarks and leptons at the particle colliders. From the most recent LHC searches by the ATLAS and CMS collaborations for vectorlike quarks \cite{ATLAS:VLQlimits, CMS:VLQlimits}, the up-type (down-type) vectorlike quarks that are assumed to decay to the top (bottom) quarks are excluded at 95\% confidence level up to 1310 GeV - 1600 GeV (1200 GeV - 1570 GeV), depending on its branching ratios.
And, the weak isodoublet vectorlike leptons that decay to the tau leptons are excluded at 95\% confidence level from 120 GeV to 790 GeV by the CMS collaboration in ref.~\cite{CMS:VLLlimits}.
Lastly, as it was pointed out in the refs.~\cite{Kumar:2015tna, Bhattiprolu:2019vdu}, the weak isosinglet charged vectorlike leptons that mix with the tau lepton have essentially no reach prospects at the current and the future proton-proton colliders due to low pair-production cross-sections and unfavorable branching ratios.

After the PQ breaking the axion potential typically acquires more than one inequivalent degenerate minimum, given by the domain wall number $N_\textrm{DW}$, leading to a cosmological domain wall problem if the symmetry is broken in the post-inflationary era. Models with $N_\textrm{DW} = 1$ evade this problem.
The four base models without extra vectorlike supermultiplets have $N_\textrm{DW} \ne 1$, and therefore may suffer from the domain wall
problem. They also have suppressed axion-photon couplings compared to the standard benchmark QCD axion models, and so may be invisible to future proposed direct axion searches. However, in the extensions of base models that include extra vectorlike supermultiplets, we obtained a wide variety of larger axion couplings.  It is notable that the extensions with $N_\textrm{DW} = 1$, which always have at least one strongly interacting vectorlike supermultiplet at the intermediate scale, 
have the smallest $|N|$ and therefore give rise to enhanced axion couplings, which are likely to be within reach of future axion searches.

{\it Acknowledgments:}
This work is supported in part by the National Science Foundation
under grant number 2013340.


\end{document}